\documentclass{article}

\usepackage{arxiv}

\usepackage[utf8]{inputenc} 
\usepackage[T1]{fontenc}    
\usepackage{hyperref}       
\usepackage{url}            
\usepackage{booktabs}       
\usepackage{amsfonts}       
\usepackage{nicefrac}       
\usepackage{microtype}      
\usepackage{lipsum}
\usepackage{graphicx}
\usepackage{dirtytalk}
\usepackage{sectsty}
\usepackage{graphicx}
\usepackage{xcolor}
\usepackage{amsmath}
\usepackage{amssymb}
\usepackage{dirtytalk}
\usepackage{subcaption}
\usepackage{longtable}
\graphicspath{ {./images/} }

\title{SEC Form 13F-HR: Statistical investigation of trading imbalances and profitability analysis}

\author{
 Deborah Miori \\
  University of Oxford\\
  Oxford, United Kingdom\\
  \texttt{deborah.miori@maths.ox.ac.uk} \\
   \And
 Mihai Cucuringu \\
  University of Oxford\\
  Oxford, United Kingdom\\
  \texttt{mihai.cucuringu@stats.ox.ac.uk} \\
}

\newcounter{noteMCctr} 

\definecolor{colour3}{RGB}{178,55,250} 
\newcommand{\mc}[1]{\textcolor{black}{{{}}#1}}

\newcommand{\dm}[1]{\textcolor{black}{{{}}#1}}

\newcommand{\imp}[1]{\textcolor{black}{{{}}#1}}

\begin{document}
\maketitle
\begin{abstract}
US Institutions with more than $\$ 100$ million assets under management must disclose part of their long positions into the SEC Form 13F-HR on a quarterly basis. We consider the number of variations in holdings between consecutive reporting periods, and compute imbalances in buying versus selling behaviour for the assets under consideration. A significant opportunity for profit arises if  an external investor is willing to trade \textit{contrarian} to the 13F filings imbalances. Indeed, imbalances capture the amount of information already consumed in the market and the related trades tend to be inflated by crowding and herding. Betting on a relatively short-term movement of prices against the sign of imbalances results in a profitable strategy especially when using a time horizon between 21 and 42 trading 
days (corresponding to 1-2 calendar months) after each financial quarter ends.
\end{abstract}


\tableofcontents

\section{\dm{Introduction}}

\dm{Diversification is a double-hedged sword in financial markets. It helps with offsetting risks in quiet market times but significantly increases the likelihood of being affected by financial contagion during periods of sudden stress \cite{CACCIOLI2014233}. After the financial crisis of 2007-08, regulators strengthened their control on systemically important institutions \cite{thomson2009systemically} and promoted higher transparency of markets by increasing requirements on disclosures of holdings. The behaviour and performance of banks and non-bank financial institutions (such as mutual funds, pension funds, insurance companies, hedge funds, etc.) has become an active area of research since then. Examples are the investigation of corporate and Treasury bond funds in \cite{corporate-bond-funds}, \cite{treasury-bond-funds}, \cite{connected-stocks}, of insurance companies in \cite{portf-simil-insurance}, \cite{Ellul-Insurers}, and of hedge funds in \cite{HF-covid}, \cite{HF-covid-basistrade} and \cite{GuoMincaWang+2016+139+155}. Interconnections between the returns of hedge funds, banks, brokers and insurance companies are further analysed in \cite{billio2010} and \cite{BillioInterconnectedness}, for the US and European markets respectively.
Many studies (e.g. \cite{CommonHoldings}, \cite{volpati2020zooming}, \cite{coimpact}, \cite{sardines}, \cite{Brown2019CrowdedTA}) also address the importance of studying crowding, which is the tendency of different investors to focus on a similar set of factors, strategies, or securities. This happens both inside and across institutions' categories, and results in overlapping positions with generally lower returns but for hedge funds.}

\dm{The information contained in the Security and Exchanges Commission (SEC) Form 13F-HR can be of interest to any study spanning diverse agents' classes, since this filing publicly provides updates on the holdings of largest US institutional investors on a quarterly basis.
Despite being released with a lag of $45$ calendar days after the financial quarter ends and reporting only long positions, Form 13F allows insights on an extremely eclectic set of market participants and has been published with satisfactory quality for at least a decade.
Seldom instances of research have been focusing on it so far, likely due to difficulties in systematically accessing and pre-processing these reports, which are singularly stored on a SEC platform and appear at different times.
An example is \cite{10.1093/rfs/hhs111}, which leverages on these filings to forecast stock returns in mutual funds by \say{trusting} managers of companies that demonstrated a track record of profitability in the past. Otherwise, \cite{disclosures-copycat} shows that 13F filings are strongly used by plain copycat investors even if this has little evidence of long-term benefits. The research in \cite{10K-13F} relates SEC Form 13F to SEC Form 10K, which must be filed annually by companies with registered shares and provides a comprehensive summary of their financial performance. As conceivable, the author finds that pessimism in Form 10K leads to a decrease in related institutional holdings.
Our study considers a different investment approach from the above ones, and employs Form 13F data to aggregate changes in investors' positions, thus computing the flow of money in and out of different assets.}

\textbf{Main contribution.} We define a measure of trading imbalance for stocks, which weights the strength of related buying versus selling behaviour between consecutive reporting periods.
\dm{In this way, we are able to investigate the global view that our diverse set of institutional investors develops on each stock. Our main contribution lies in the extraction of a statistically significant signal from the imbalances computed via 13F filings, which suggests that asset prices move in the opposite direction to their imbalance sign at the end of each quarter.}

\textbf{Structure of the paper.}
Section  \ref{sec:data} describes in detail the data used and the pre-processing needed. Next, the methodology to compute imbalances and trading signals is provided in Section  \ref{sec:methodology}. \imp{Preliminary considerations on imbalances and their price impact on contemporaneous returns are presented in Section \ref{sec:preliminary}.} Section \ref{sec:results} shows our profits when betting contrarian to imbalances and it describes experiments of conditioning positions on stocks' past prices or to the sector membership of each asset. Finally, Section \ref{sec:conclusions} concludes our work with some final remarks.

\section{Data}
\label{sec:data}


\subsection{Form 13F-HR}

The SEC introduced Form 13F-HR (\say{Holdings Report}) to provide some publicly available disclosure of institutions' investments back in 1975.
Following its updated rules, the related managers need now to report long holdings via the 13F filings when handling US AUM $\geq \$ 100$ million. Institutional investment managers can be part of investment advisers, banks, insurance companies, broker-dealers, pension funds and corporations, etc. Therefore, non-financial institutions are also included in this form.
However, 13F disclosures are required only quarterly, filed in number of shares, miss the short positions and must be completed within 45 days after the financial quarter ends. Institutions are also allowed to file amendments (Form 13F-HR/A) if they mistakenly reported a position without any fees or alerts to follow, meaning that the most valuable bets are often kept hidden for a longer time, and \mc{deliberately disclosed with a delay},  as \cite{caohedge} shows for hedge funds.

Every market participant filing this report is identified by a Central Index Key (CIK). Investments pursued by different managers are identified by the \say{Other Manager} integer number, while the division between actual portfolios would be described in general by the SeriesID. Unfortunately, the latter identifier is not required in 13F filings and therefore we cannot access this most granular view. Assets that are compulsory to report are part of the SEC Section 13(f) securities list, meaning that there is a bias towards US funds and investments. The full list of securities covered includes $23,131$ securities as of Q4 2021 (fourth quarter of 2021). These are identified by the US CUSIP number (Committee on Uniform Securities Identification Procedures) and listed at \url{https://www.sec.gov/divisions/investment/13flists.htm}. Therein, one  can find equity securities that trade on an exchange, certain equity options and warrants, shares of closed-end investment companies, and some convertible debt securities. The shares of open-end investment companies are instead not required. 


US CUSIP codes are assigned with format \say{AAAAAABBC}, where \say{AAAAAA} represents the company issuing the financial security, \say{BB} is the issue number used to denote e.g. shares with/without voting rights, and \say{C} is the check digit. The informative part for our purposes is the \say{AAAAAA}, which we denote as CUSIP6. As an example, the AAPL stock has code \say{037833100}, where \say{037833} is Apple, Inc. Furthermore, \say{10} denotes Class A Shares and the final \say{0} is the check digit. It is a convention to assign \say{10} as first issue number and increasing numbers (\say{20},\say{30}... or \say{11},\say{12}...) for subsequent securities.

\subsection{Data pre-processing}

We consider holdings disclosed via Form 13F-HR from 2013-06-30 to 2021-09-30, where earlier quarters are not included due to low quality of reporting. We also do not correct positions for which an amendment was later filed, 
\dm{in order to investigate the trading signals directly available at the end of each quarter. However, these are not fully actionable signals due to the 45-days lag companies are allowed in their holdings' disclosure deadline.}
Our 13F data set shows long investments for $T=34$ points in time. For each quarter, the number of samples is larger than $10^6$, where each occurrence represents an agent holding a specific security of type \say{shares} at that point in time. The \say{Other Manager} key is discarded, while the related 
number of shares held is summed for each individual institution and security, for every period. Similarly, we aggregate stocks information to CUSIP6 level. The full data relate to $8,166$ unique CIKs, among which we see influential and diverse institutions such as banks (e.g. Bank of America, Wells Fargo \& Co), insurance companies (e.g. American International Group, AXA), pension funds (e.g. Canada Pension Plan Investment Board, Public sector pension investment board), hedge funds (e.g. UBS O'Connor, Renaissance Technologies), market makers (e.g. Cutler Group LP, Group One Trading)...

For each quarter ending at time $p$, we assume that we have all the related 13F filings immediately available, and build a holdings matrix $H_{p} = (F \times A)_p$ where rows are investment institutions $f \in F$ and columns are assets $a \in A$. Each cell is populated with the number of shares that $f$ is holding of asset $a$ at the end of quarter $p$, or otherwise $0$. Then, we compute differences in holdings between each two consecutive reporting periods and consider a new set of $T-1$ matrices $D_{p} = H_{p} - H_{p-1}, p=2,\ldots,T$. Between $30-40\%$ of securities are in every case found to have the related column fully populated with zeros. These are uncommon stocks chosen by only one or two funds that keep the position over longer horizons of time. These data amount to noise for our analyses and the related columns are thus dropped, completing the pre-processing.


\subsection{Close-to-close price returns}

To investigate whether disclosures on holdings can provide profitable investment signals, we download close-to-close adjusted daily returns for our stocks from the Center for Research in Security Prices (CRSP) dataset from Wharton Research Data Services (WRDS). The return of the  S\&P Composite Index is also selected to be able to compute market-excess returns. We map our CUSIP6 identifiers to tickers and obtain returns data for approximately $5,000$ securities per each reporting period.

\section{Methodology}
\label{sec:methodology}

The set of matrices $D_p$ describes differences in the amount of shares of stocks $a \in A$ held by institutions $f \in F$ between each consecutive periods $p-1$ and $p$.
\mc{Our goal is to leverage such increments for the construction of a predictive signal for the assets in question.} 
For each asset $a$, we proceed by calculating the volume of shares $B_{p,a}^{vol}, S_{p,a}^{vol} \geq 0$ that are bought or sold by market participants, respectively. Similarly, we compute the number of buying and selling trades,  $B_{p,a}^{tr}$, respectively  $S_{p,a}^{tr} \geq 0$ for each stock, by counting the number of entries with positive, respectively, negative sign. 

We define an \textit{imbalance} of allocations over asset $a$, between periods $p-1$ and $p$, by considering the normalised discrepancy in terms of volume 
\begin{equation}
    I_{p,a,N}^{vol} = \frac{B_{p,a}^{vol}-S_{p,a}^{vol}}{B_{p,a}^{vol}+S_{p,a}^{vol}} \in [-1;+1],
\label{imbalance-vol}
\end{equation}
\mc{
and define similarly the imbalance 
discrepancy in terms of trade counts} 
\begin{equation}
    I_{p,a,N}^{tr} = \frac{B_{p,a}^{tr}-S_{p,a}^{tr}}{B_{p,a}^{tr}+S_{p,a}^{tr}} \in [-1;+1].
\label{imbalance-tr}
\end{equation}
To increase the significance of our measure, we compute imbalances on securities that have at least $N$ institutions \say{active} on them. \say{Activity} of fund $f$ on stock $a$ is defined as having $D_{p}^{(f,a)} \neq 0$. In this way, we avoid e.g. the unreliable case of having irrelevant strongest signals $I_{p,a,1}^{vol,tr} = \pm 1$ associated to stocks being traded only by one institution.
\mc{Other normalisations could be considered for $ I_{p,a,N}^{vol}$, for example, by the total traded volume in the market over the same period.}   

As already mentioned, we assume that all data from Form 13F-HR are made public at the end of the last day of the financial quarter. At this point in time, we compute imbalances and employ them as signals to identify the set of assets to invest in, starting from the following trading day.
As a first \textit{vanilla strategy} that can leverage either $I^{vol}$ or $I^{tr}$, we simply go long each asset for which $\textsc{sign}(I_{p,a,N}^{vol,tr}) = +1$ and short the ones with $\textsc{sign}(I_{p,a,N}^{vol,tr}) = -1$. Moreover, we divide our imbalance signals between quantiles ranks $qr_i$ for $i \in \{1, 2, 3, 4, 5\}$ defined as
\begin{equation*}
    qr_i = \text{top } \frac{100}{i} \text{\% imbalances largest in magnitude}, 
\end{equation*}
where we neglect imbalances equal to zero. This allows us to focus on the strongest segments of our signals available and reinforces the $N$ threshold.

\begin{figure}[h]
    \centering
    \includegraphics[width=0.6\textwidth]{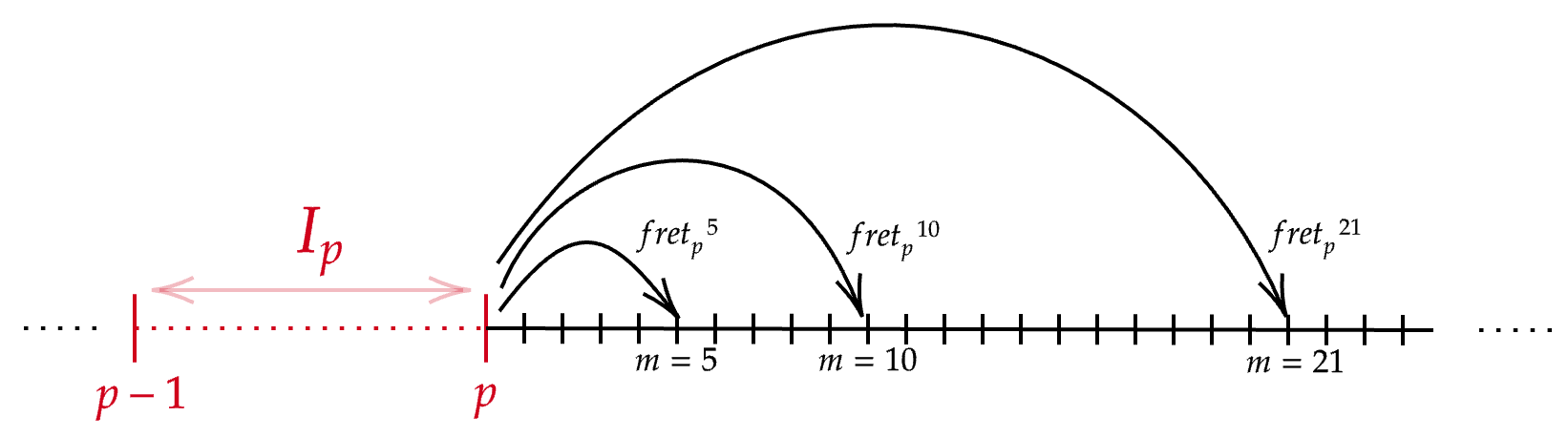}
    \caption{\dm{Illustration of the timeline on which we extract imbalances and compute future returns for each quarter $p$.}}
    \label{fig:diagram}
\end{figure}

\dm{For each one of our \say{strategies} defined by the choices of $I^{vol,tr}$ and $qr_i$, we consider future markout horizons of $m \in \{5, 10, 21, 42, 63\}$ trading days ahead of the end of each quarter $p$ (considering that there are $21$ trading days on average in a month). The raw future return $fret_{p, a}^{m,RAW}$ of an investment on asset $a$ is computed summing close-to-close adjusted daily returns over the horizon of interest, and then we define the future market-excess return $fret_{p, a}^{m,MER}$ as 
\begin{equation}
    fret_{p, a}^{m,MER} = fret_{p, a}^{m,RAW} - fret_{p, SPY}^{m,RAW},
\end{equation}
where SPY is the S\&P Composite Index ETF. Market excess returns allow us to remain orthogonal to the market component and hedge our investments. The related \textit{profit-and-loss} ($PnL$) from our strategy amounts to
\begin{equation}
    PnL_{p, a}^{m,MER} = fret_{p, a}^{m,MER} \times \textsc{sign}(I_{p,a}).
\end{equation}
For clarity, we represent in the diagram in Fig. \ref{fig:diagram} our quantities and timelines of interest.
}

\dm{At each period $p$, we allocate capital uniformly (i.e. $\$\frac{1}{|A_p|}$) to each one of the stocks $a \in A_p$ chosen by our strategy and to SPY.} This choice allows us to have comparable long-short portfolios in terms of capital invested. A different weighting scheme can also be implemented, e.g. by trading stocks proportionally to the magnitude of their imbalance.
However, this type of modification is redundant to the quantile approach and can lower the expected profit. Thus, we decide to proceed with the uniform weighting scheme, for simplicity.

We iterate the above investment procedure at the end of each financial quarter, i.e. when we assume that we can see the newly released 13F filings. Next, we compute summary statistics to judge the actual information provided by imbalances and their effectiveness at different markouts. \dm{We compute the $PnL$ for each period $p$ as
\begin{equation}
    PnL_p = \sum_{a} PnL_{p, a}^{m,MER}.
\end{equation}}
The full performance of our strategy is given by adding each 
result in time, i.e. $PnL_{tot} = \sum_{p=p_i}^{p_f} PnL_p$ where $p_{i,f}$ are the first and last quarter with available imbalances' data. 
We keep track of the \textit{profit-per-trade} ($PPT$) by computing
\begin{equation}
    PPT = \textsc{AVG}_{p} \Big ( \frac{ PnL_p}{|A_p|} \Big ) 
\end{equation}
for each configuration. This metric is useful to qualitatively compare the profits after execution costs for each portfolio. 

We also compute the \textit{Sharpe Ratio} $S$ as
\begin{equation}
    S = \frac{\textsc{avg}_{p=p_i}^{p_f} (PnL_p)}{\textsc{std}_{p=p_i}^{p_f}(PnL_p)} \times \sqrt{4}, 
\end{equation}
to value the profit achieved by unit of risk taken. Since S is usually reported annualised, we process it accordingly by considering that we are investing once per quarter in each year.
The work of \cite{deflated_SR}, \cite{LEDOIT_signif_Sharpe} and \cite{mihai_sharpe} further highlight the importance of checking the significance of Sharpe Ratios when back-testing a sample of hypothetical strategies. We implement the methodology developed in \cite{deflated_SR} to compute a  confidence level $C_k$ of the performance of each one of our strategies $k \in K$. This is defined as
\begin{equation}
    C_k = Z \Big[ \frac{
    (S_k-S_0)\sqrt{L-1}
    }{
    \sqrt{1-\gamma_3 S_k + (\gamma_4 -1)S_k^2/4}
    } \Big], 
\end{equation}
where $Z$ is the cumulative function of the standard Normal distribution, $S_k$ the Sharpe Ratio of the specific strategy we are testing, $L$ its sample length, $\gamma_3$ the skewness of the related returns distribution and $\gamma_4$ its kurtosis. $|K|$ is the total number of strategies developed and tested. Finally, 
\begin{equation}
    S_0 = \textsc{std}(\{S_k\})\Big((1-\gamma)Z^{-1}[1-\frac{1}{|K|}] + \gamma Z^{-1}[1-\frac{e^{-1}}{|K|}]\Big), 
\end{equation}
is a term that deflates our Sharpe Ratio to account for selection bias and overfitting due to multiple testing. The parameter $\gamma \sim 0.5772$ is the Euler-Mascheroni constant.
\section{Preliminary analyses of imbalances}
\label{sec:preliminary}

\subsection{Considerations on threshold $N$}

\begin{figure}
     \centering
     \begin{subfigure}[b]{0.49\textwidth}
         \centering
         \includegraphics[width=\textwidth]{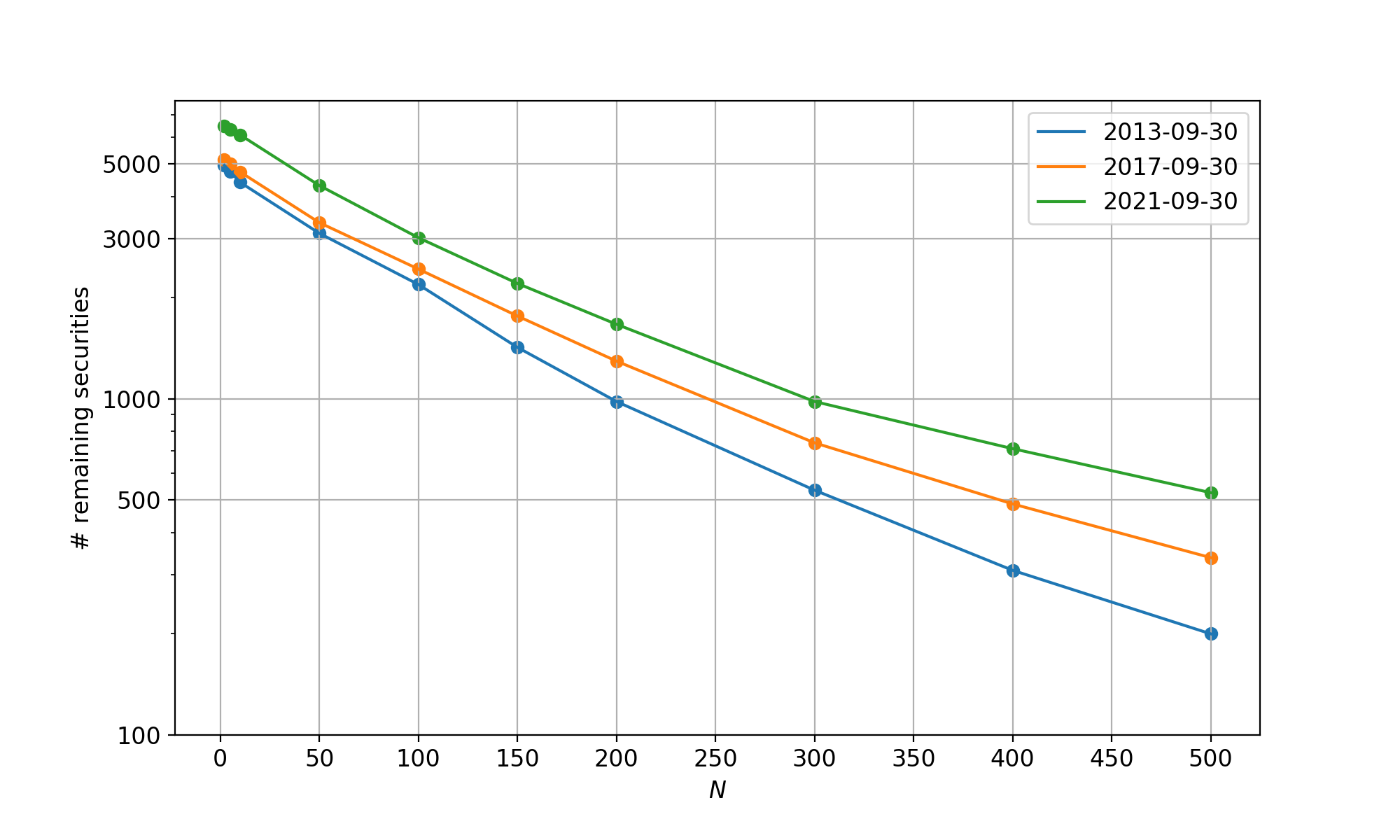}
         \caption{}
         \label{fig:threshold_number_funds}
     \end{subfigure}
     \hfill
     \begin{subfigure}[b]{0.49\textwidth}
         \centering
         \includegraphics[width=\textwidth]{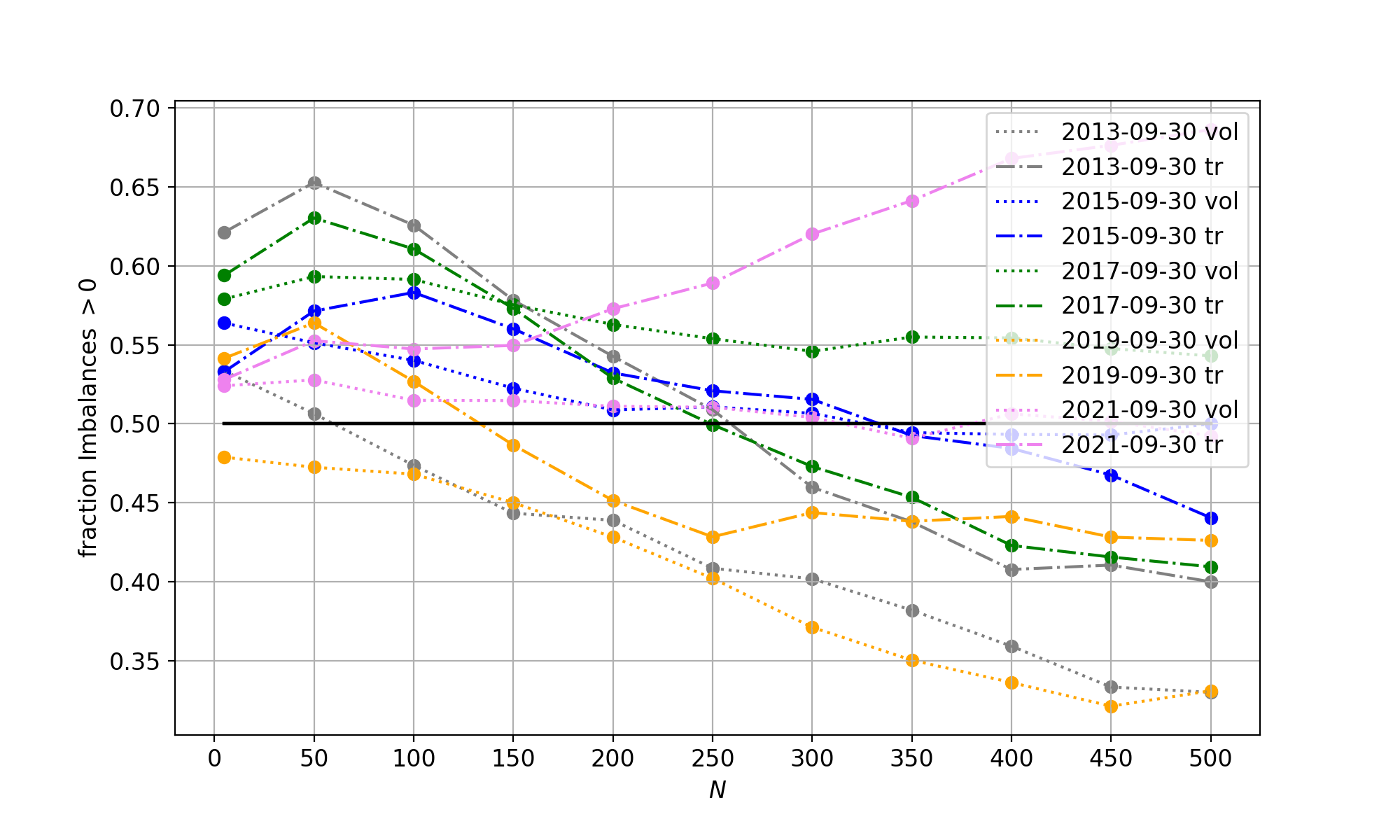}
         \caption{}
         \label{fig:frac_pos_number_funds}
     \end{subfigure}
        \caption{\dm{We vary the threshold $N$ on the x-axis and (a) look at how many securities (i.e. number of imbalances) are left to possibly trade, (b) compute the fraction of volume and trades imbalances with positive sign.}}
        \label{fig:vary-N}
\end{figure}

Before investigating whether 13F imbalances carry a profitable signal, we check the implications of choosing a different threshold $N$. For each period $p$, we vary $N \in [0,500]$ with step size $=50$ and count the remaining number of securities, i.e. how many assets have at least $N$ funds active on them at that point in time. We also compute the related imbalances, following Eqs. \ref{imbalance-vol} and \ref{imbalance-tr}. 
Figure \ref{fig:threshold_number_funds} shows the number of remaining securities for different $N$ in Q3 of 2013, 2017 and 2021. This is a linear-log plot and the witnessed linear relationship implies that we can approximate the number of surviving securities via an exponential decay dependent on the  threshold $N$. The shift between the three temporal lines expresses the increase in active institutions over the years, as intuitively expected from the expansion of markets, but does not influence the shape of the relationship.

\begin{figure}[h]
    \centering
    \includegraphics[width=0.7\textwidth]{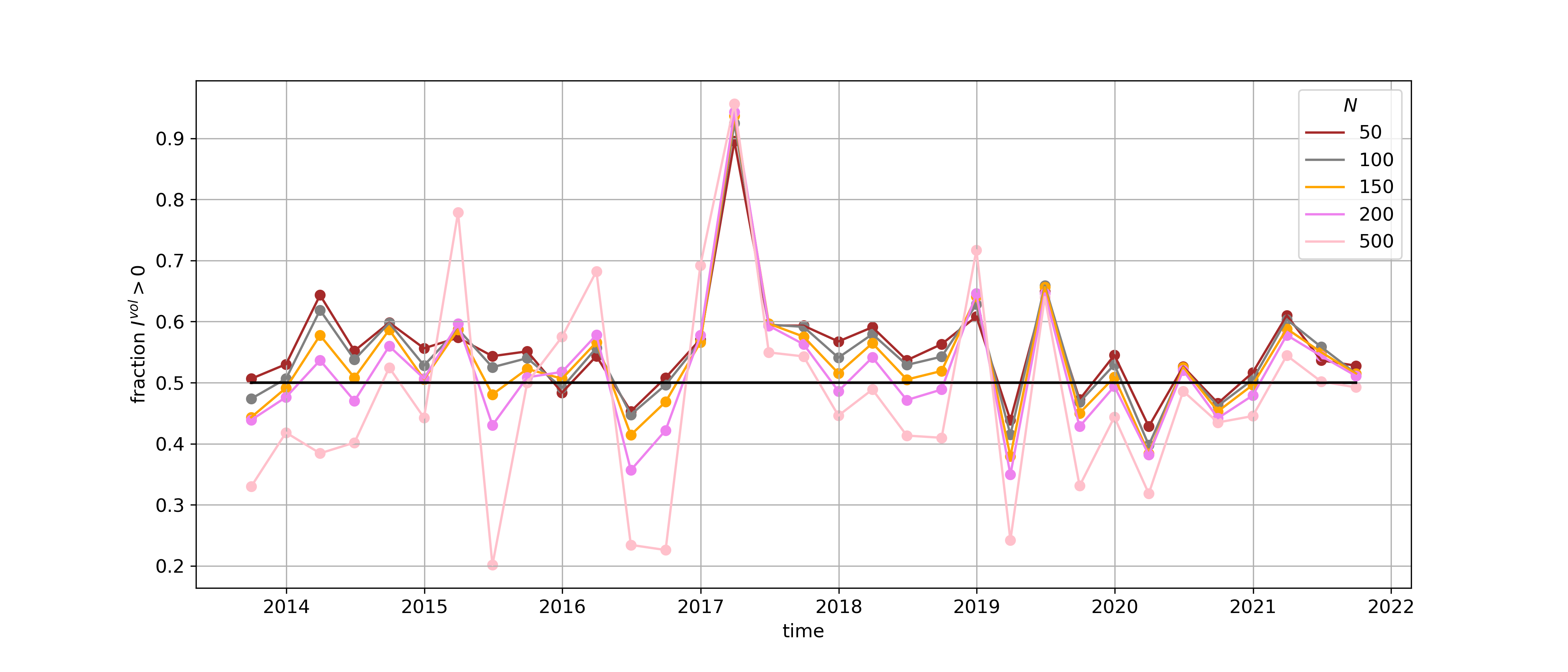}
    \caption{Evolution in time of the fraction of volume imbalances with positive sign  for different choice of threshold $N$.}
    \label{fig:frac_pos_vol}
\end{figure}

\begin{figure}[h]
    \centering
    \includegraphics[width=0.7\textwidth]{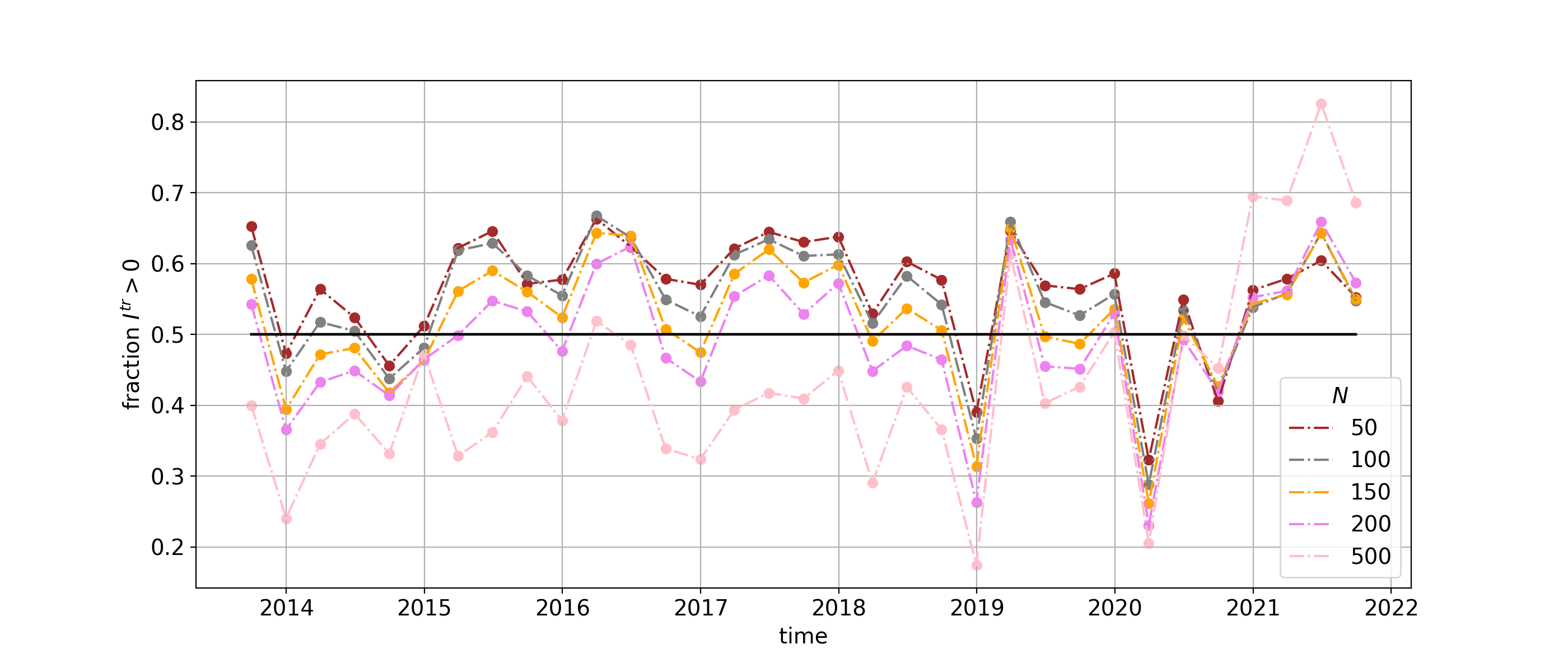}
    \caption{Evolution in time of the fraction of trades imbalances with positive sign for different choice of threshold $N$.}
    \label{fig:frac_pos_trade}
\end{figure}

In Fig. \ref{fig:frac_pos_number_funds}, we decompose the above relationship between remaining securities with positive or negative imbalance and compute the related ratio. We also look at the evolution of this ratio in time in Figs. \ref{fig:frac_pos_vol} and \ref{fig:frac_pos_trade}, for imbalances computed in terms of volume or trades, respectively. We observe that there is a predominance of buying (selling) behaviour with decreasing (increasing) $N$. However, there is also a stable tendency of reversion towards the balanced proportion. This is especially true for $I^{vol}$, while $I^{tr}$ shows a more persistent bias in time. This difference is due to the two natures of the quantities. While the former considers the actual amount of shares traded in the market, but can include special strong long niche bets from highly knowledgeable investors, the latter relates specifically to the number of buying or selling trades and will strongly react to market conditions. We can thus 
infer that investors tend to start or increase their common positions at different points in time, but they will decrease or exit them more synchronously. This result is again in agreement with intuition.
Finally, the oscillations of Fig. \ref{fig:frac_pos_vol} can be related to liquidity injections or withdrawal.

For the purpose of our analysis, we will consider and compare imbalances generated by $N \in \{50, 150, 500\}$. The case $N=150$ will be the least unbalanced one \mc{in terms of sign proportions}, but also $N=50$ does not divert too strongly from equilibrium. These two  options allow us to consider more than $1,000$ or $3,000$ securities, respectively. We are also interested in investigating the results of trading those securities on which investors are most active on, and to this end, we further consider $N=500$. Due to the biases witnessed, we also build a data set  where we \mc{cross-sectionally} demean the signals from imbalances, in order to investigate the impact on the performance of the strategy.

\subsection{\imp{Price impact of imbalances on contemporaneous returns}}
\label{sec:market-impact}

Before proceeding to the analysis of imbalances as a signal to forecast the future returns of assets, we first aim to gain an understanding of the impact of imbalances over contemporaneous returns. The goal is to extract insights into the extent to which fund managers, which are entering into a position over a given quarter, impact the price return of the asset over the same quarter. 
We perform a linear regression of raw returns (and market-excess returns) onto the related imbalances $I^{vol, tr}_N$ with $N \in \{50, 150, 500\}$, at each quarter $p$. Quarterly returns are calculated summing the adjusted close-to-close daily returns covering the three months within each quarter, but these are first winsorized to control for the lowest and highest $10\%$ of values, which could include noisy outliers. 
Finally, we compute the $R^2$ of each regression resulting from the different combination of variables and quarter $p$ considered.

In Table \ref{table-impact}, we report the average $R^2$ over the whole set of periods $p$, i.e. $\forall p$, for each one of the above combinations. To investigate periodical patterns, we also report the average $R^2$ aggregated across each quarter (Q$1$, Q$2$, Q$3$ and Q$4$), over the years available in the study. Imbalances computed over trade counts generally attain higher average $R^2$ compared to the volume ones, and their magnitudes are particularly enhanced for Q$2$ with $N=50$, and Q$3$ with $N=500$. One possible explanation of the latter result is that major investors might act very similarly in entering/exiting positions during Q$3$ following the yearly Russell Index Rebalancing. 
All $R^2$ values are fairly low, which is somewhat expected since we are dealing with noisy financial data, and are considering long periods of $3$ months. Moreover, we have already mentioned that fund managers tend to hide their \say{best bets} and file amendments to the 13F filings later in time. Thus, the related market impact is also concealed.

\begin{figure}[h]
    \centering
    \includegraphics[width=0.7\textwidth]{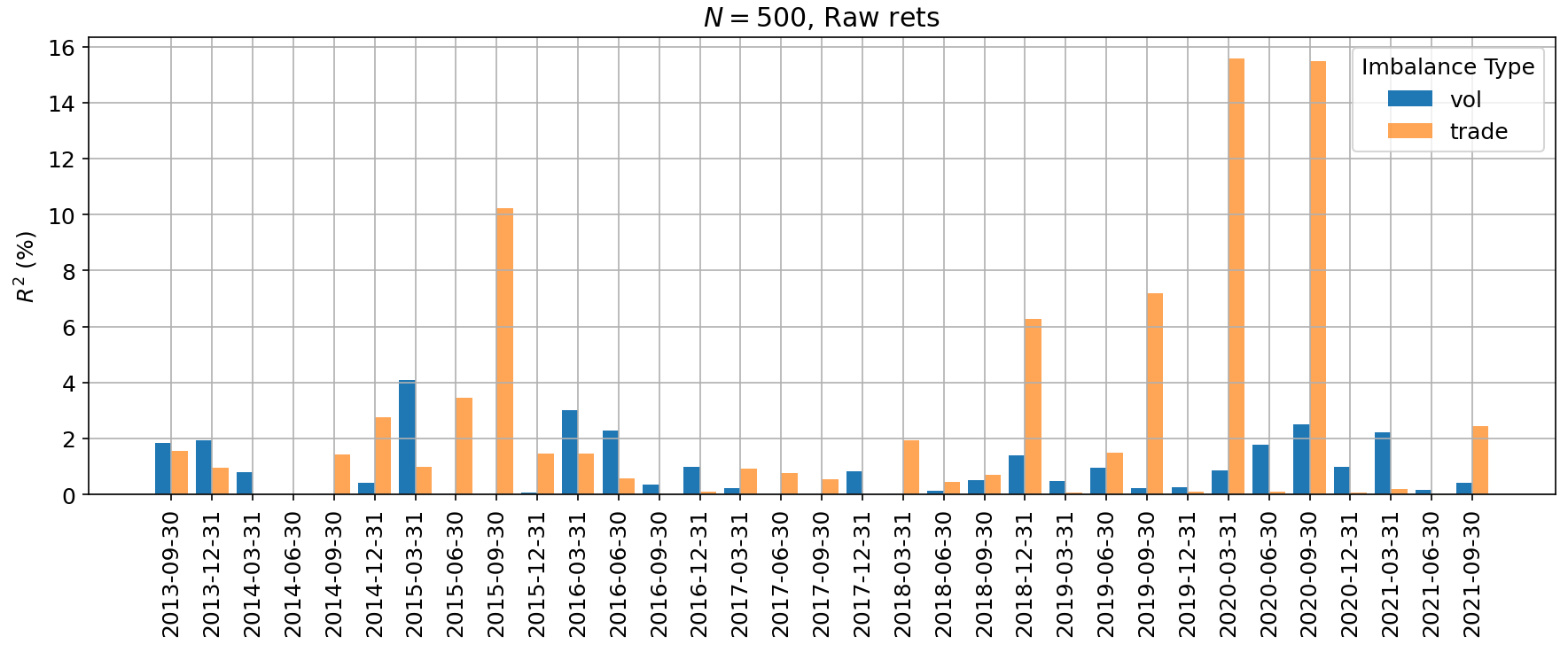}
    \caption{For each quarter $p$, we report the $R^2$ from the regression of contemporaneous raw returns on $I^{vol, tr}_{N=500}$.}
    \label{impact}
\end{figure}

The full view on the temporal evolution of $R^2$'s is shown in Figs. \ref{fig:impact-raw-rets} and \ref{fig:impact-mers} of Appendix \ref{sec:A0}, for raw returns and MERs respectively. For clarity, Fig. \ref{impact} depicts below the mentioned trend for the case of raw returns and $N=500$. Surprisingly, imbalances computed in terms of trade counts show extreme spikes in explanatory power with $R^2$ reaching $10-15\%$. The related quarters could be interesting to examine in future work, but for now we proceed to the investigation of imbalances as investment signals forecasting future returns.

\begin{longtable}{ |p{1.5cm}|p{1cm}|p{1cm}||p{1.8cm}||p{1.8cm}|p{1.8cm}|p{1.8cm}|p{1.8cm}|}
\hline
\textbf{Returns} & $\mathbf{N}$ & $\mathbf{I^{vol, tr}}$ &
\textbf{$\mathbf{R^2\%}$, $\mathbf{\forall p}$} & 
\textbf{$\mathbf{R^2\%}$, Q1} & 
\textbf{$\mathbf{R^2\%}$, Q2} & 
\textbf{$\mathbf{R^2\%}$, Q3} & 
\textbf{$\mathbf{R^2\%}$, Q4} \\
\hline
\hline
Raw rets & $500$ & $I^{vol}$ & $0.90$ & $1.46$ & $0.67$ & $0.65$ & $0.86$ \\
\hline
Raw rets & $500$ & $I^{tr}$ & $2.40$ & $2.64$ & $0.87$ & $4.40$ & $1.46$ \\
\hline
Raw rets & $150$ & $I^{vol}$ & $0.47$ & $0.31$ & $0.58$ & $0.53$ & $0.44$ \\
\hline
Raw rets & $150$ & $I^{tr}$ & $1.08$ & $1.22$ & $0.86$ & $1.76$ & $0.40$ \\
\hline
Raw rets & $50$ & $I^{vol}$ & $1.05$ & $1.22$ & $1.47$ & $0.90$ & $0.61$ \\
\hline
Raw rets & $50$ & $I^{tr}$ & $1.98$ & $2.01$ & $2.71$ & $2.37$ & $0.80$ \\
\hline
\hline
MERs & $500$ & $I^{vol}$ & $0.86$ & $1.16$ & $0.83$ & $0.69$ & $0.77$ \\
\hline
MERs & $500$ & $I^{tr}$ & $2.59$ & $2.79$ & $1.78$ & $3.91$ & $1.71$ \\
\hline
MERs & $150$ & $I^{vol}$ & $0.45$ & $0.21$ & $0.63$ & $0.58$ & $0.35$ \\
\hline
MERs & $150$ & $I^{tr}$ & $1.09$ & $1.23$ & $0.87$ & $1.57$ & $0.61$ \\
\hline
MERs & $50$ & $I^{vol}$ & $1.04$ & $1.14$ & $1.62$ & $0.91$ & $0.52$ \\
\hline
MERs & $50$ & $I^{tr}$ & $1.89$ & $1.91$ & $2.67$ & $2.19$ & $0.77$ \\
\hline
\caption{For each combination of return type and imbalances $I^{vol}_{N}$ and  $I^{tr}_{N}$, we report the average $R^2$ in percentage of the related regressions over the whole set of periods $p$. To investigate periodical patterns, we also show the average $R^2$ calculated for each quarter (i.e. Q$1$, Q$2$, Q$3$, Q$4$) over the years}
\label{table-impact}
\end{longtable}

\section{Strategy construction and profitability}
\label{sec:results}

\subsection{Vanilla strategy}

\begin{figure}[h]
    \centering
    \includegraphics[width=0.85\textwidth]{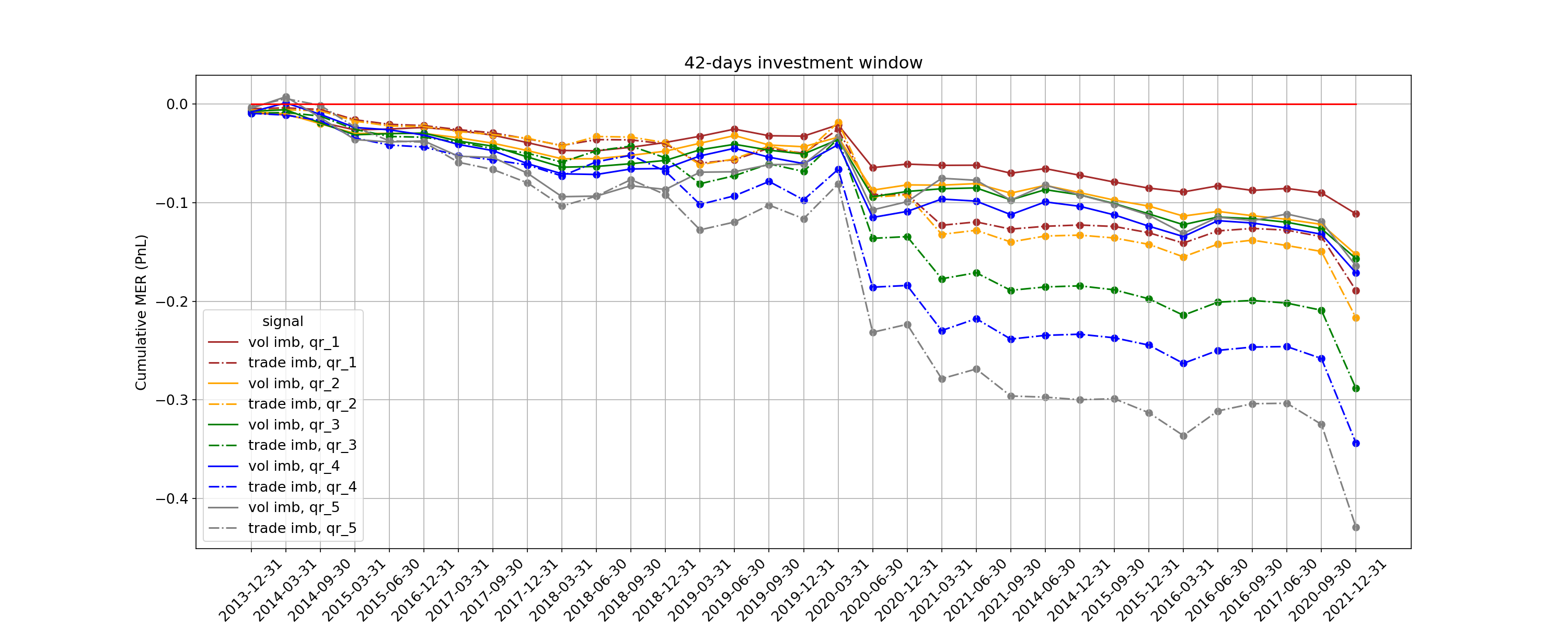}
    \caption{Cumulative PnL from MERs on a horizon of $2$ months when trading the sign of imbalances with $N=150$.}
    \label{fig:vanilla-negative}
\end{figure}

\begin{figure}[h]
\centering
\begin{subfigure}{.49\textwidth}
  \centering
  \includegraphics[width=.99\linewidth]{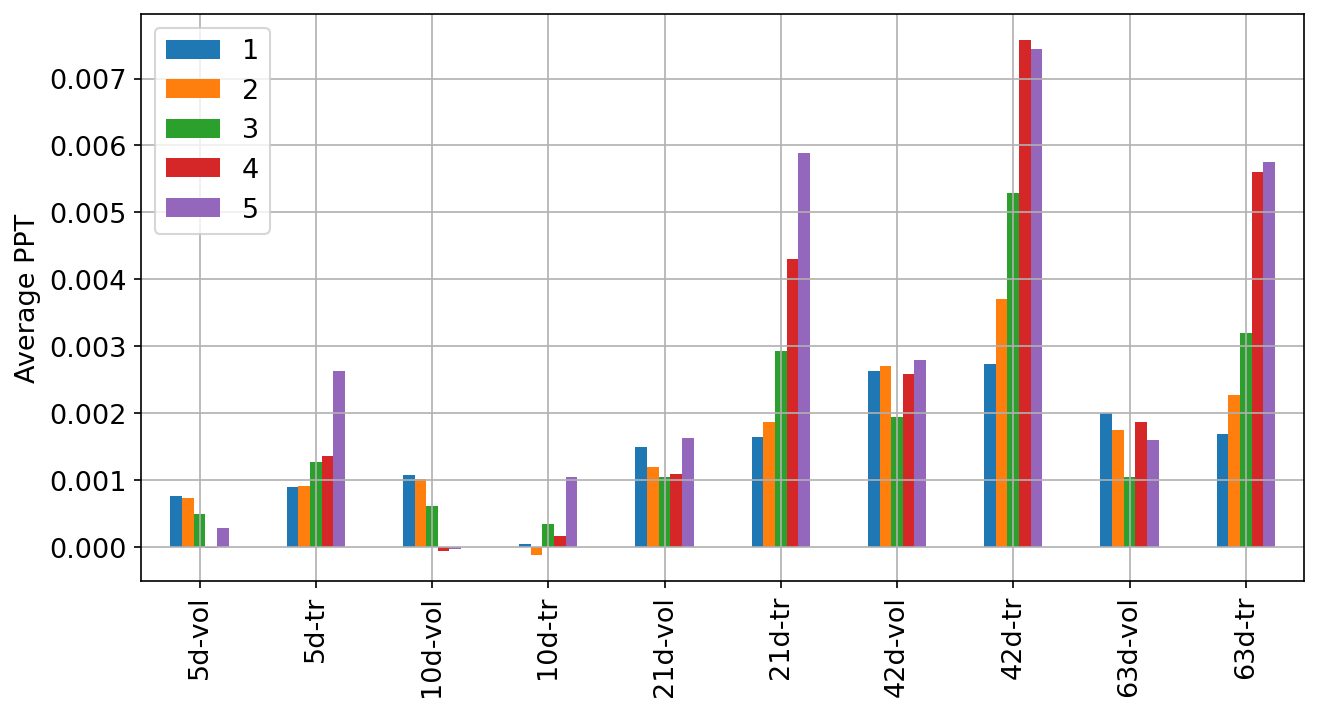}  
  \caption{$N=500$, average PPT}
\end{subfigure}
\begin{subfigure}{.49\textwidth}
  \centering
  \includegraphics[width=.99\linewidth]{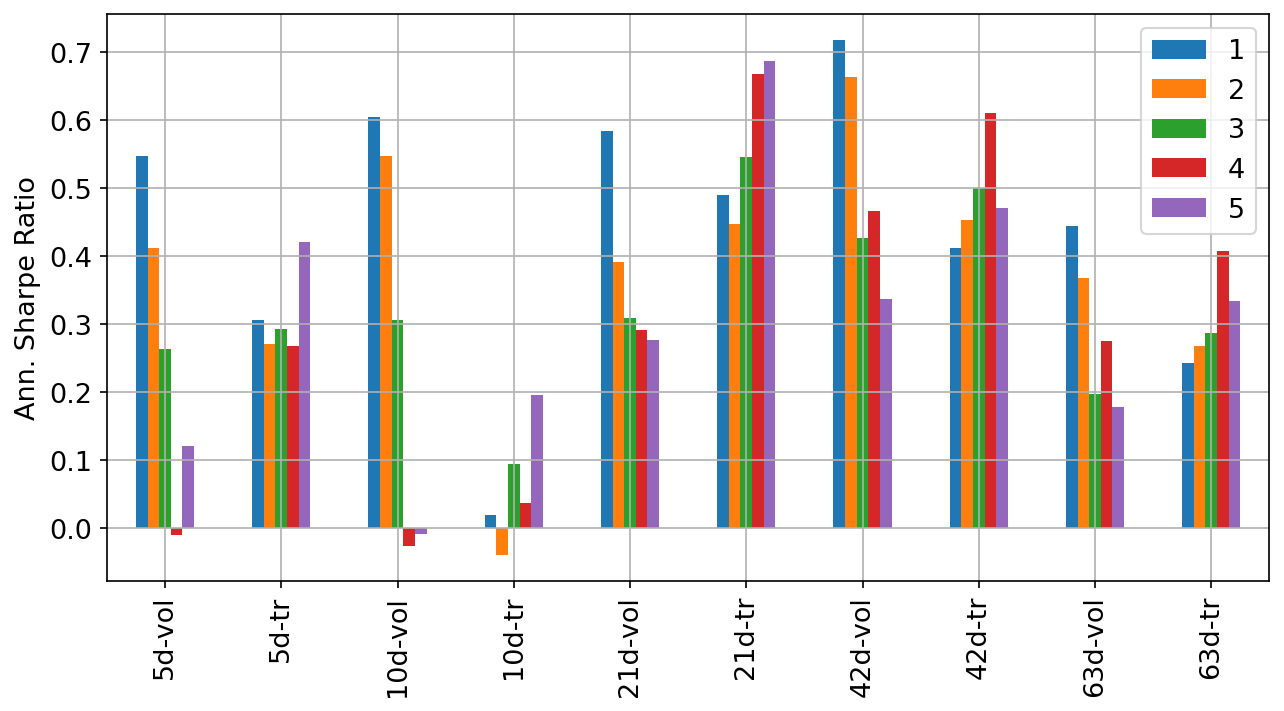}  
  \caption{$N=500$, Sharpe Ratio}
\end{subfigure}\newline

\begin{subfigure}{.49\textwidth}
  \centering
  \includegraphics[width=.99\linewidth]{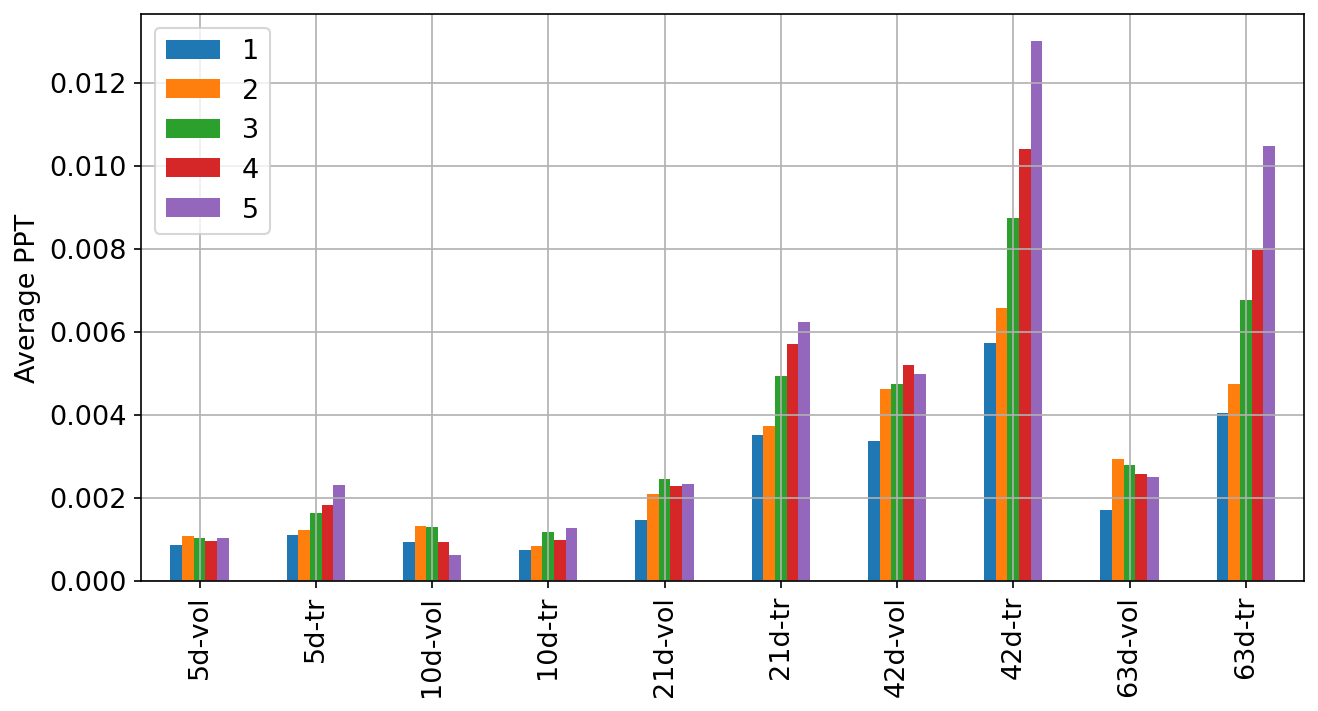}  
  \caption{$N=150$, average PPT}
\end{subfigure}
\begin{subfigure}{.49\textwidth}
  \centering
  \includegraphics[width=.99\linewidth]{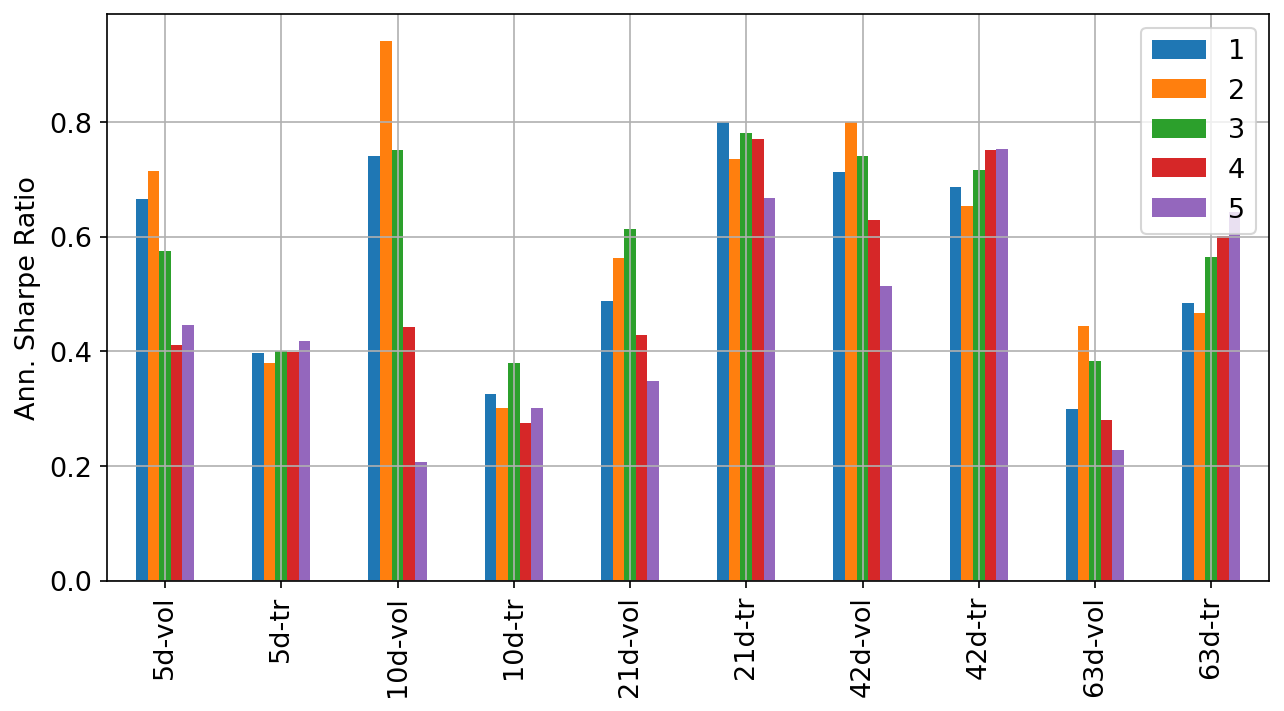}  
  \caption{$N=150$, Sharpe Ratio}
\end{subfigure}\newline

\begin{subfigure}{.49\textwidth}
  \centering
  \includegraphics[width=.99\linewidth]{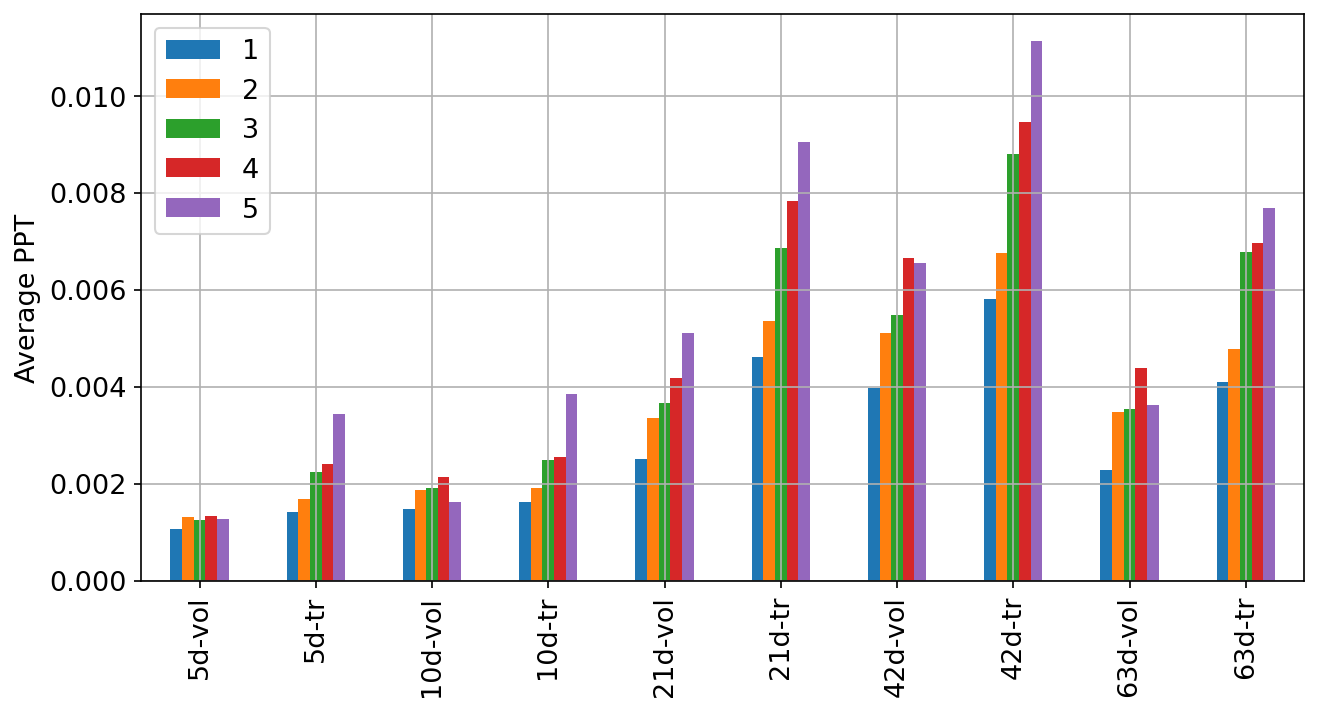}  
  \caption{$N=50$, average PPT}
\end{subfigure}
\begin{subfigure}{.49\textwidth}
  \centering
  \includegraphics[width=.99\linewidth]{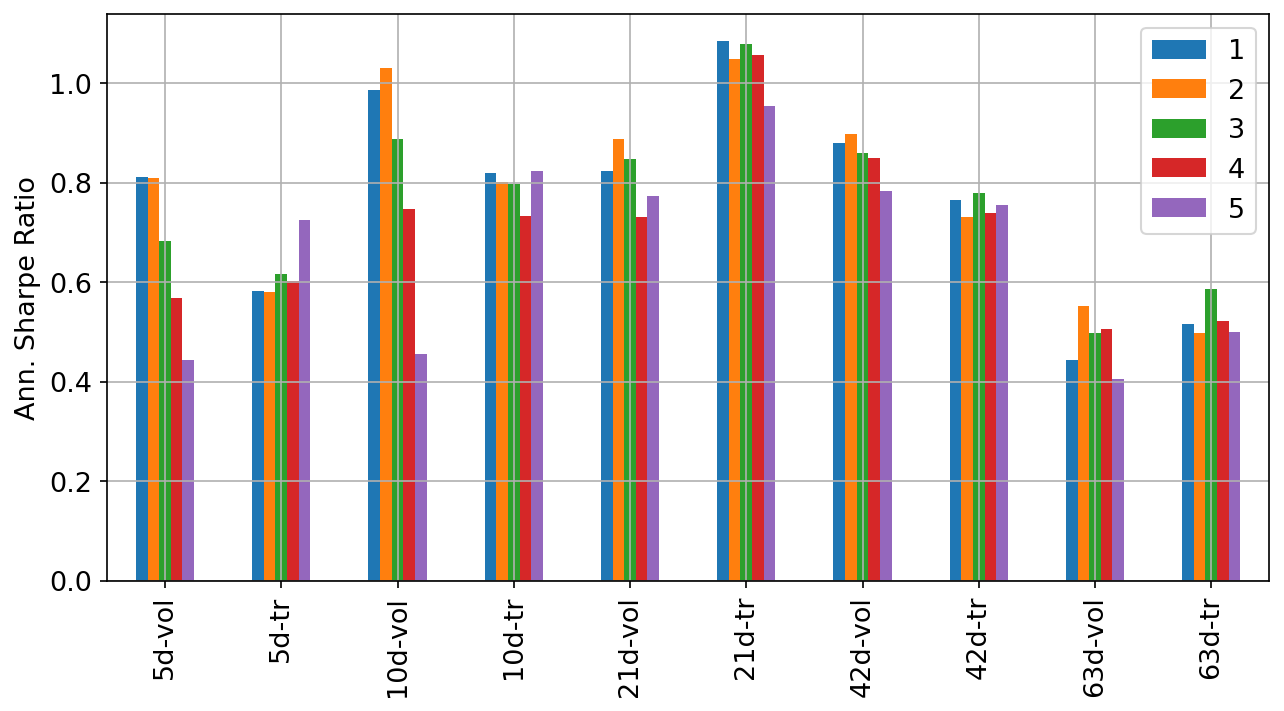}  
  \caption{$N=50$, Sharpe Ratio}
\end{subfigure}

\caption[]{Average $PPT$ and Sharpe Ratio statistics on our contrarian vanilla strategy for $I^{vol,tr}$. The different colors relate to the different quantile ranks as highlighted in the legend.}
\label{bars-vanilla}
\end{figure}

\begin{figure}[h]
\centering
\begin{subfigure}{.49\textwidth}
  \centering
  \includegraphics[width=.99\linewidth]{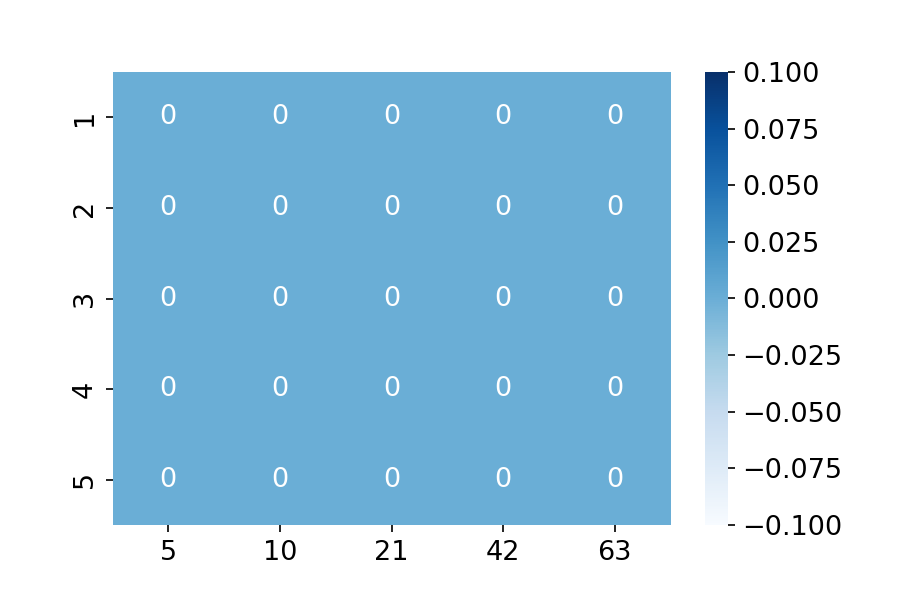}  
  \caption{$N=500$, vanilla strategy on volume imbalances}
\end{subfigure}
\begin{subfigure}{.49\textwidth}
  \centering
  \includegraphics[width=.99\linewidth]{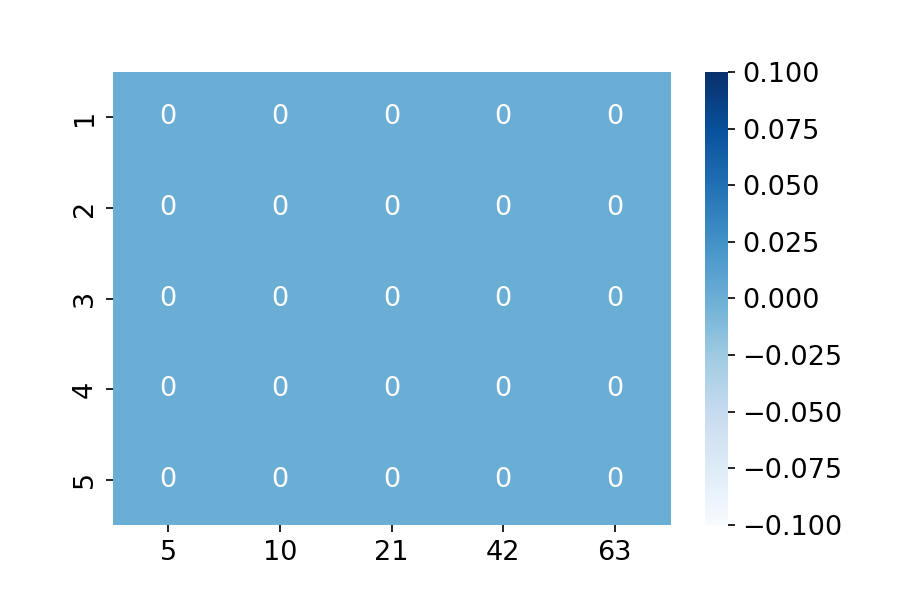}  
  \caption{$N=500$, vanilla strategy on trade imbalances}
\end{subfigure}\newline

\begin{subfigure}{.49\textwidth}
  \centering
  \includegraphics[width=.99\linewidth]{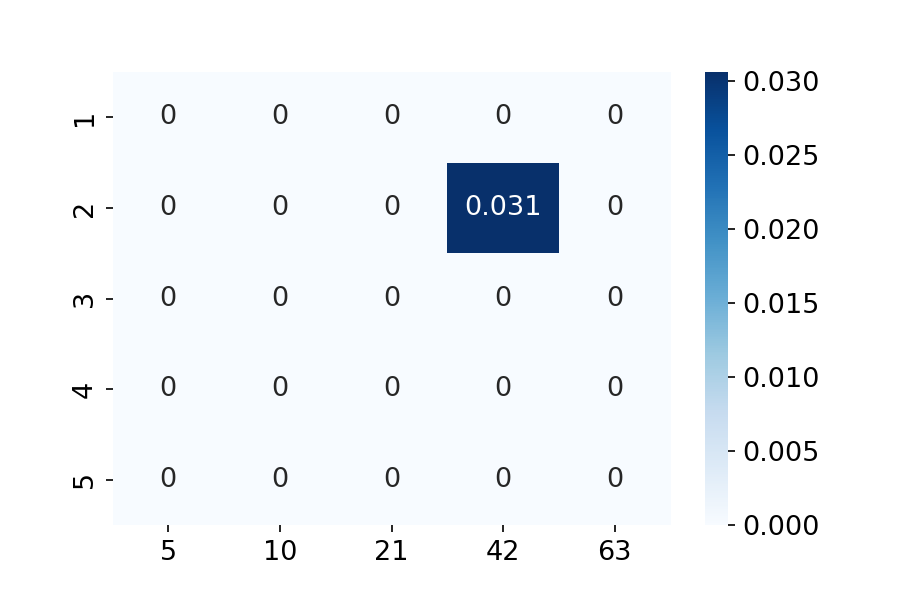}  
  \caption{$N=150$, vanilla strategy on volume imbalances}
\end{subfigure}
\begin{subfigure}{.49\textwidth}
  \centering
  \includegraphics[width=.99\linewidth]{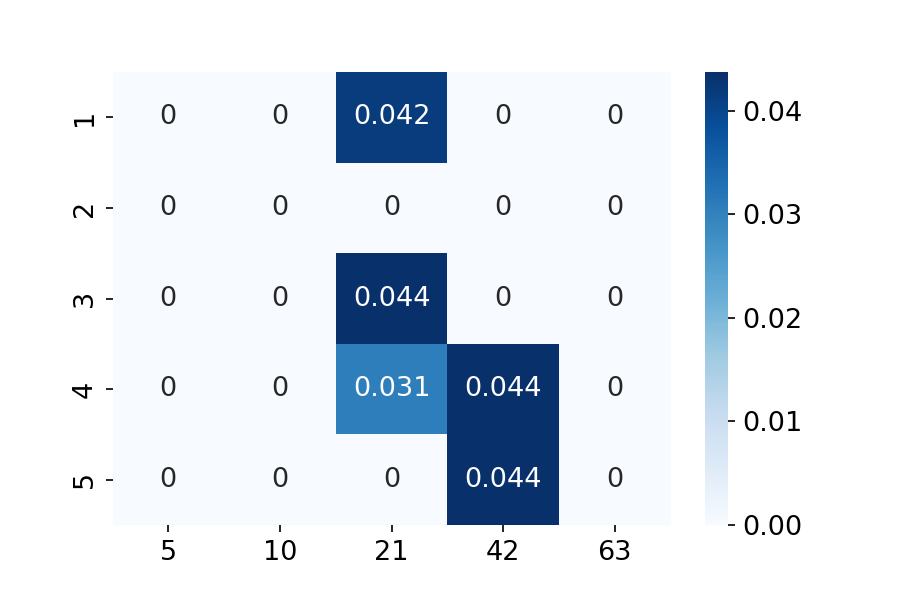}  
  \caption{$N=150$, vanilla strategy on trade imbalances}
\end{subfigure}\newline

\begin{subfigure}{.49\textwidth}
  \centering
  \includegraphics[width=.99\linewidth]{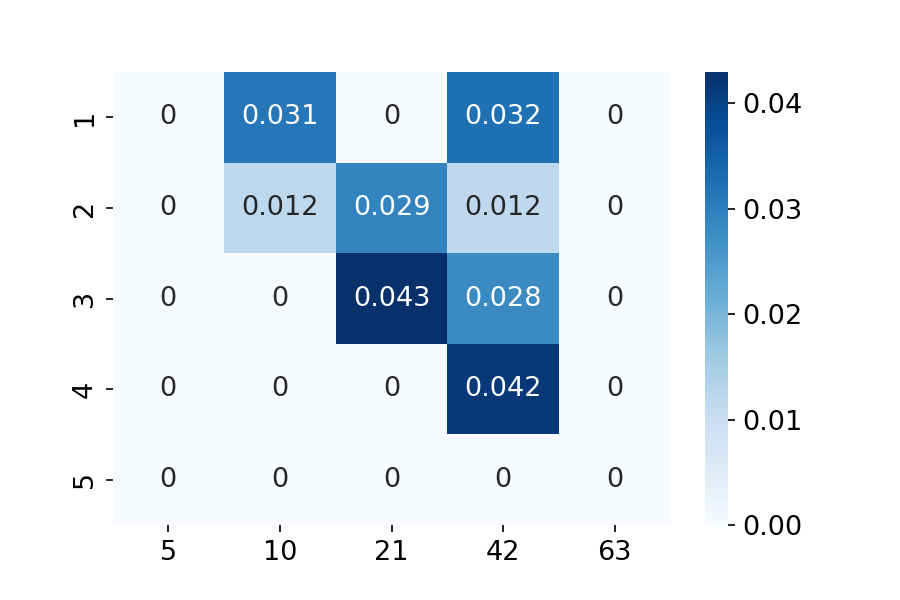}  
  \caption{$N=50$, vanilla strategy on volume imbalances}
\end{subfigure}
\begin{subfigure}{.49\textwidth}
  \centering
  \includegraphics[width=.99\linewidth]{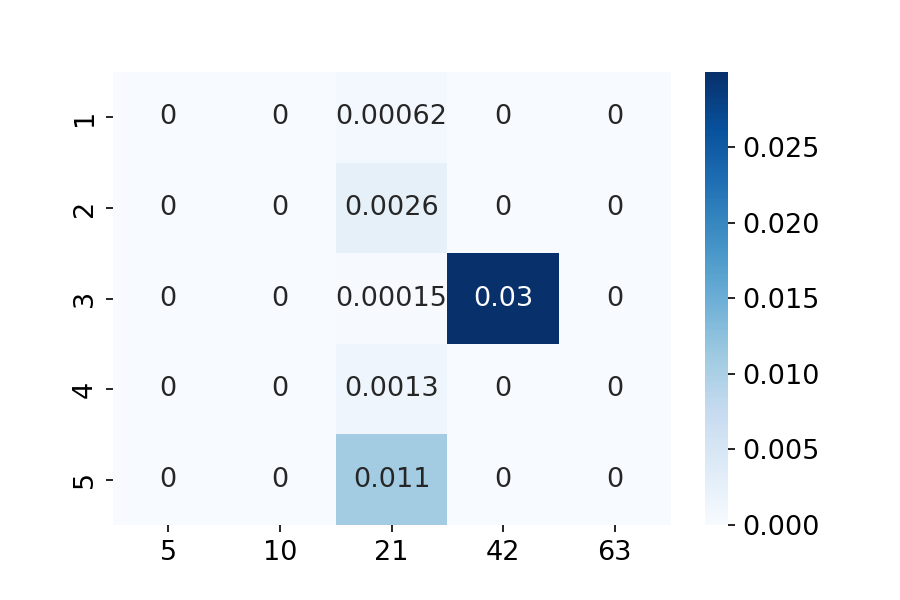}  
  \caption{$N=50$, vanilla strategy on trade imbalances}
\end{subfigure}

\caption[]{Results when testing the significance levels of our Sharpe Ratios. Strategies rejected by the test are reported with precise zero-value for ease of visualisation. In each heatmap, the x-axis is the horizon of the strategy and the y-axis is the quantile rank used for trading.}
\label{heat-vanilla}
\end{figure}

Imbalances in volumes and trades are computed for $N \in \{50, 150, 500\}$, following the methodology described in Section \ref{sec:methodology} without de-meaning at first. We invest on horizons of $m \in \{5, 10, 21, 42, 63\}$ trading days and compute the cumulative PnL from future MERs. The full set of results are reported in Figs. \ref{cum50}, \ref{cum150}, \ref{cum500} in Appendix \ref{sec:A1}, while here we show the performance on the $42$ trading days horizon in Fig. \ref{fig:vanilla-negative}. The striking feature, which is common to all the simulations and for which Fig. \ref{fig:vanilla-negative} acts as a representative, is that trading on the sign of imbalances generally leads to negative PnL for market excess returns. Therefore, we proceed to compute the $PPT$ and $PnL$ of each strategy but by trading $-1 \times \textsc{sign}(I^{vol, tr})$ as the signal, i.e. betting against the imbalances. The results are shown in Fig. \ref{bars-vanilla}, and in Fig. \ref{heat-vanilla} we report the significance levels of the Sharpe Ratios at less than $0.05$. For ease of visualisation, strategies that did not pass the significance test are reported with precise zero-value.
We observe that no strategy has significant Sharpe Ratio when $N=500$ and thus, we run all experiments also with de-meaning the signals from imbalances. However, no significance arises at $N=500$ and the performances and significance at $N=50,150$ are negatively affected. We conclude that de-meaning is not suitable on 13F filings imbalances.

From a pure investment purpose, we compare Sharpe Ratios and average PPT of the strategies passing the significance test. It is indeed meaningful to look at the profit-per-trade to have a qualitative view of performances when we consider costs of executions from rebalancing. Our vanilla strategy achieves a highly desirable Sharpe Ratio $S>1$ if we trade against trade imbalances on a horizon of $21$ trading days, for $N=50$ and quantiles $qr_4, qr_5$. Otherwise, we can achieve good $PPT$ and $S>0.8$ if we trade against volume imbalances on a longer horizon of $42$ days again for $N=50$.
This result is in line with our intuition. Trade imbalances are more suitable to show crowding and over-heating of trades and indeed are our best performing signal. Regarding the timeline, 13F filings are fully made public within 45 calendar days after each financial quarter ends, making copycat investors start trading the positions and indeed further fading our contrarian strategy performance.

\clearpage


\subsection{Contrarian to Market Momentum and to Market Mean-Reversion strategy}

To assess  whether we can further enhance the profit achieved by our strategies, we condition the imbalances that we trade according to the past performance of the related security. We now hold positions for horizons of $10, 21, 42$-trading days but further considering whether the MER of each security was positive or negative over the past $10, 21, 42$ days. We define our strategies being:
\begin{itemize}
    \item \textit{contrarian to market momentum (CMM)} if we trade contrarian to imbalances that have positive (negative) sign while the related securities showed past positive (negative) performance,
    \item \textit{contrarian to market mean-reversion (CMR)} if we trade contrarian to imbalances that have positive (negative) sign while the related securities showed past negative (positive) performance.
\end{itemize}

\begin{figure}[h]
\centering
\begin{subfigure}{.49\textwidth}
  \centering
  \includegraphics[width=.99\linewidth]{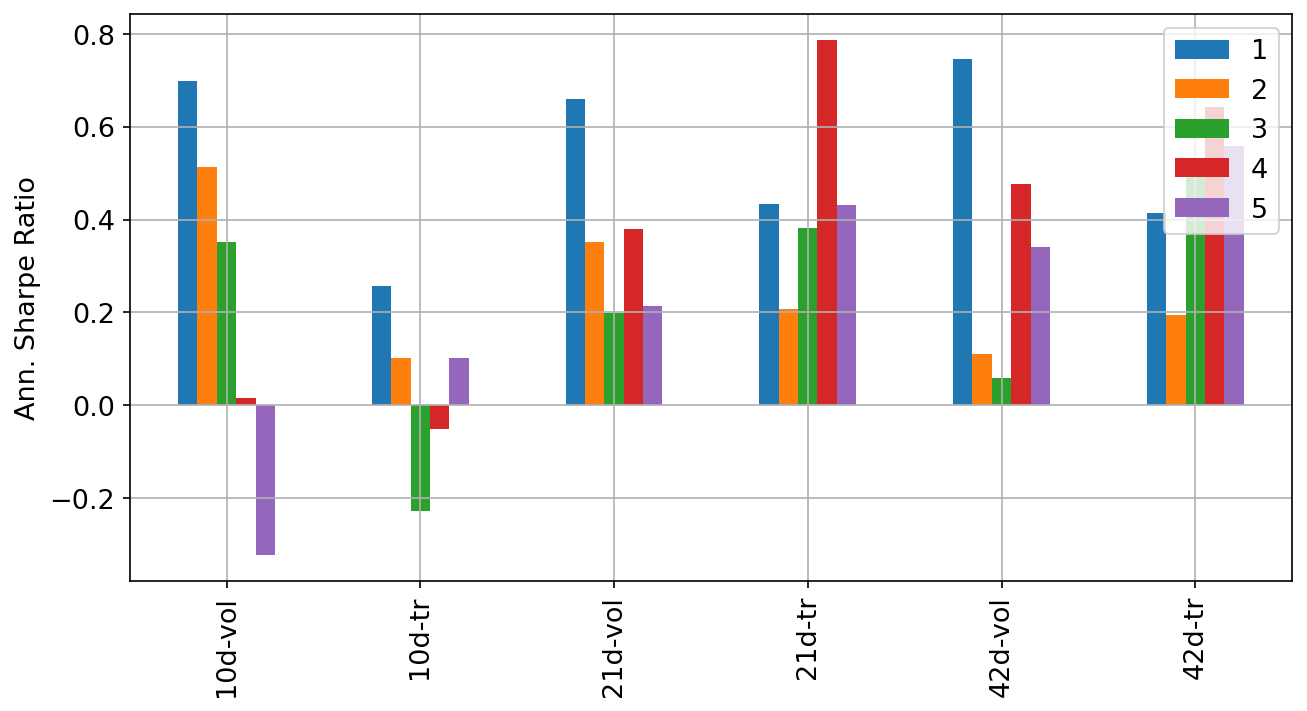}  
  \caption{$N=500$, CMM}
\end{subfigure}
\begin{subfigure}{.49\textwidth}
  \centering
  \includegraphics[width=.99\linewidth]{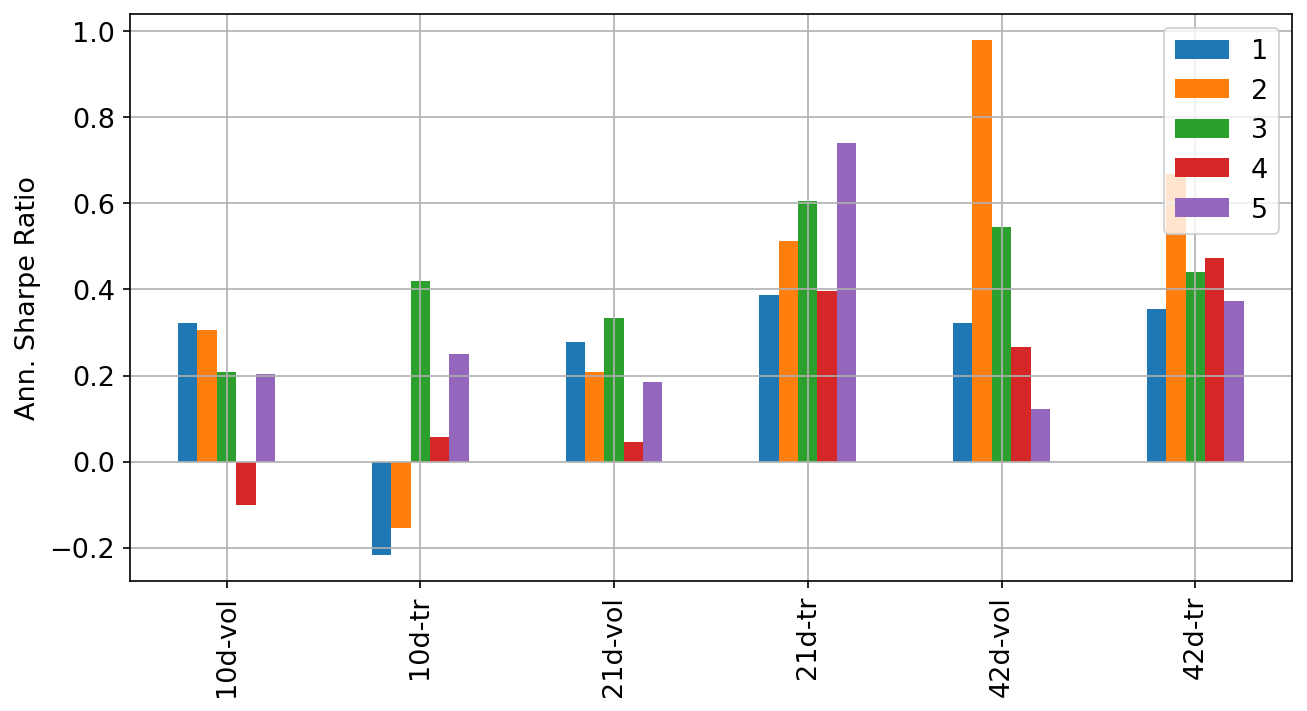}  
  \caption{$N=500$, CMR}
\end{subfigure}\newline

\begin{subfigure}{.49\textwidth}
  \centering
  \includegraphics[width=.99\linewidth]{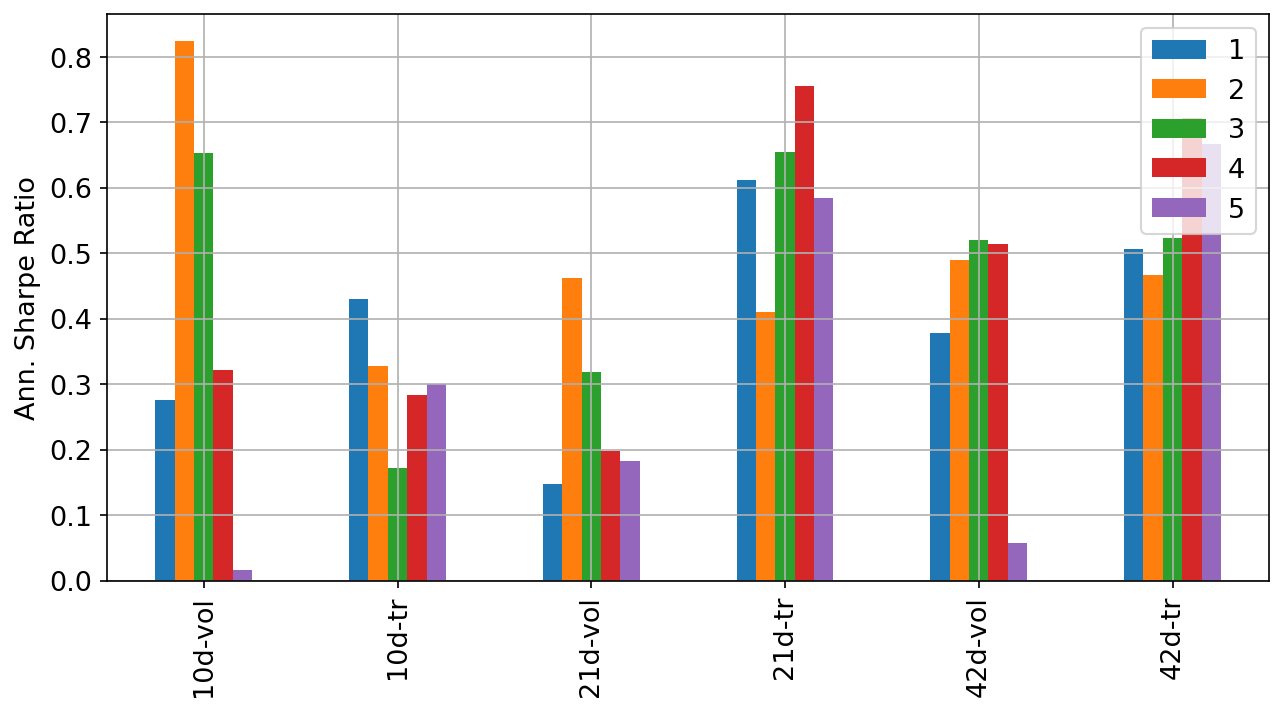}  
  \caption{$N=150$, CMM}
\end{subfigure}
\begin{subfigure}{.49\textwidth}
  \centering
  \includegraphics[width=.99\linewidth]{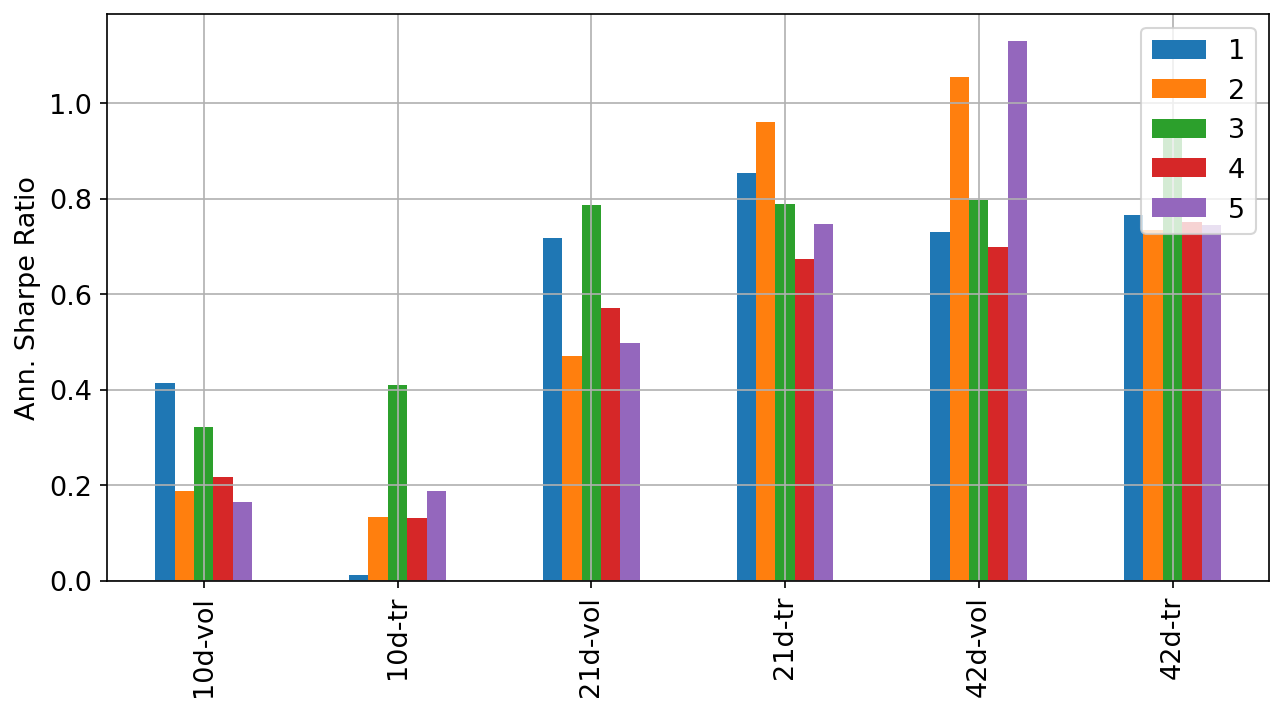}  
  \caption{$N=150$, CMR}
\end{subfigure}\newline

\begin{subfigure}{.49\textwidth}
  \centering
  \includegraphics[width=.99\linewidth]{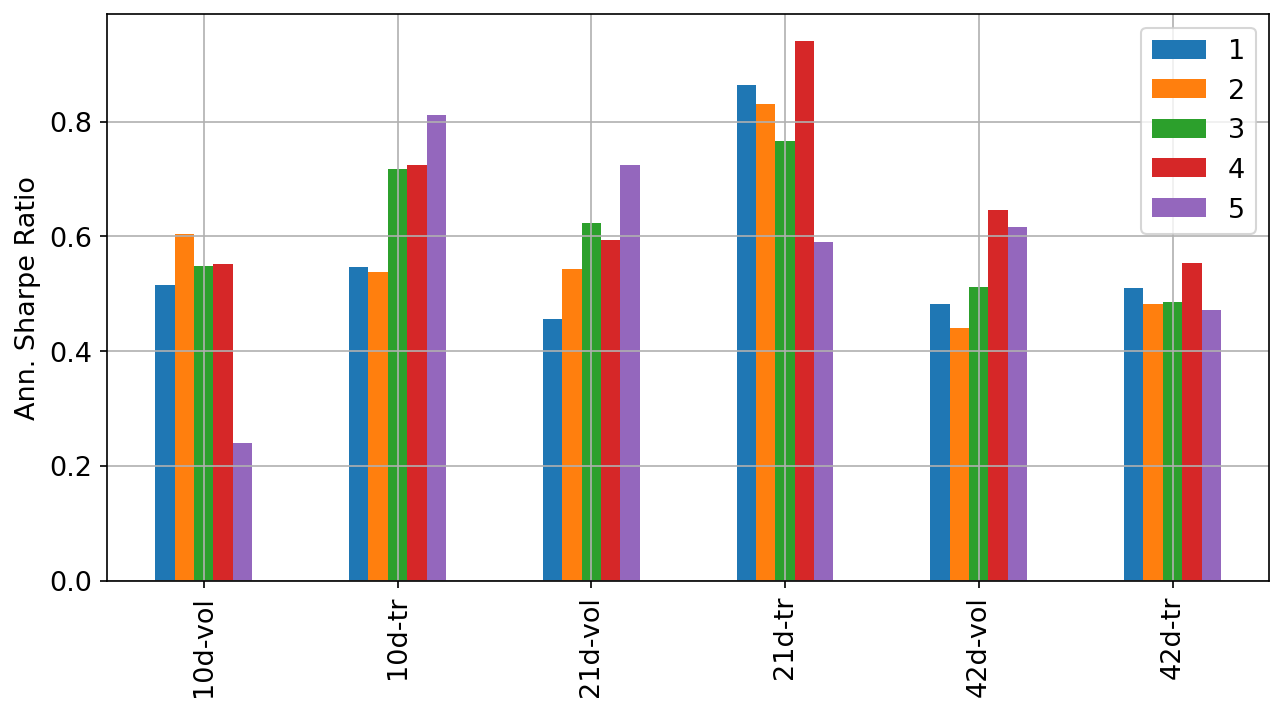}  
  \caption{$N=50$, CMM}
\end{subfigure}
\begin{subfigure}{.49\textwidth}
  \centering
  \includegraphics[width=.99\linewidth]{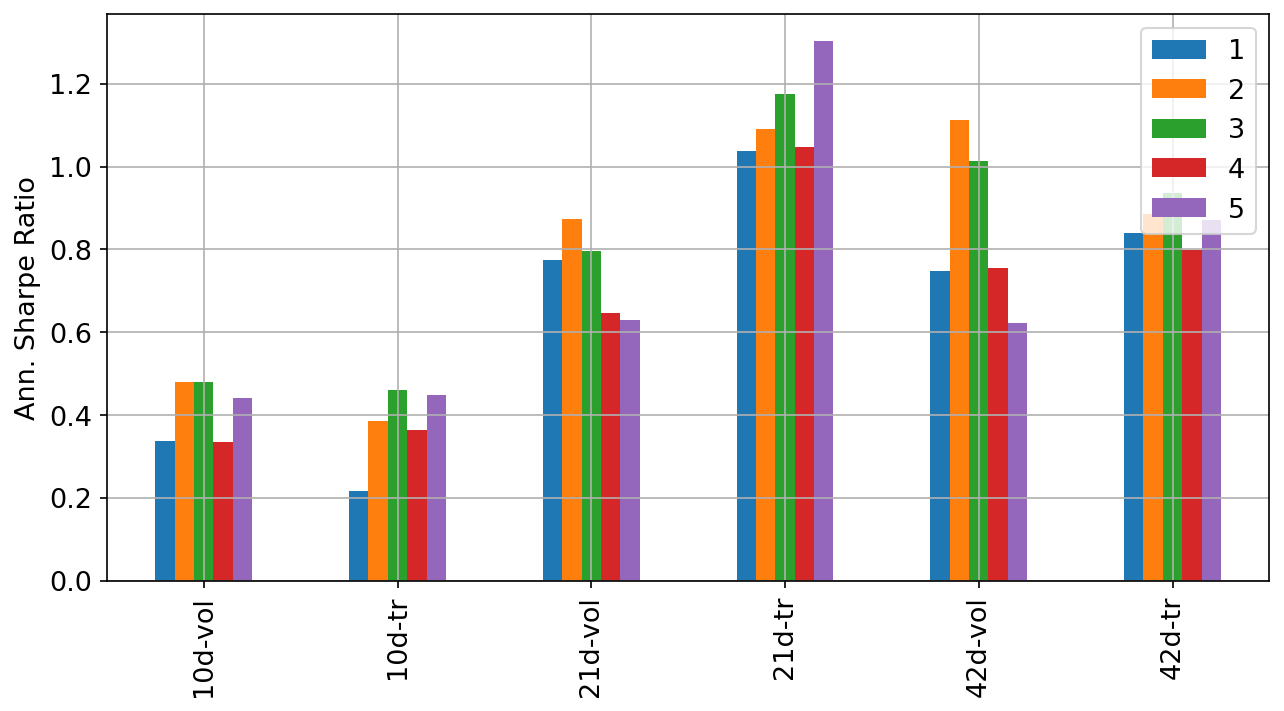}  
  \caption{$N=50$, CMR}
\end{subfigure}

\caption[]{Sharpe Ratios achieved for our Contrarian to Market Momentum imbalances (CMM) and Contrarian to Market Mean-Reversion imbalances (CMR) strategies.}
\label{bars-mom-rev}
\end{figure}

\begin{figure}[h]
\centering
\begin{subfigure}{.49\textwidth}
  \centering
  \includegraphics[width=.99\linewidth]{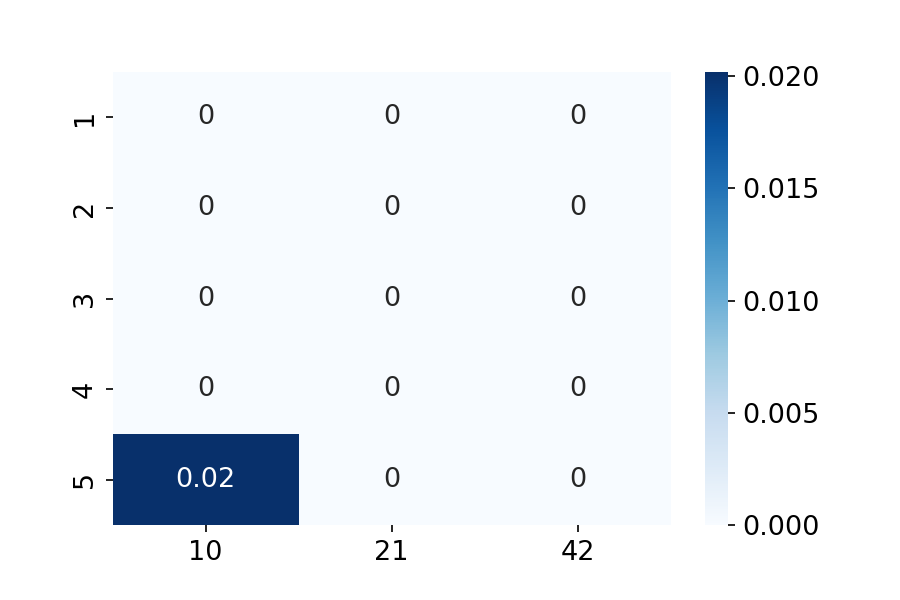}  
  \caption{$N=500$, CMM strategy on volume imbalances}
\end{subfigure}
\begin{subfigure}{.49\textwidth}
  \centering
  \includegraphics[width=.99\linewidth]{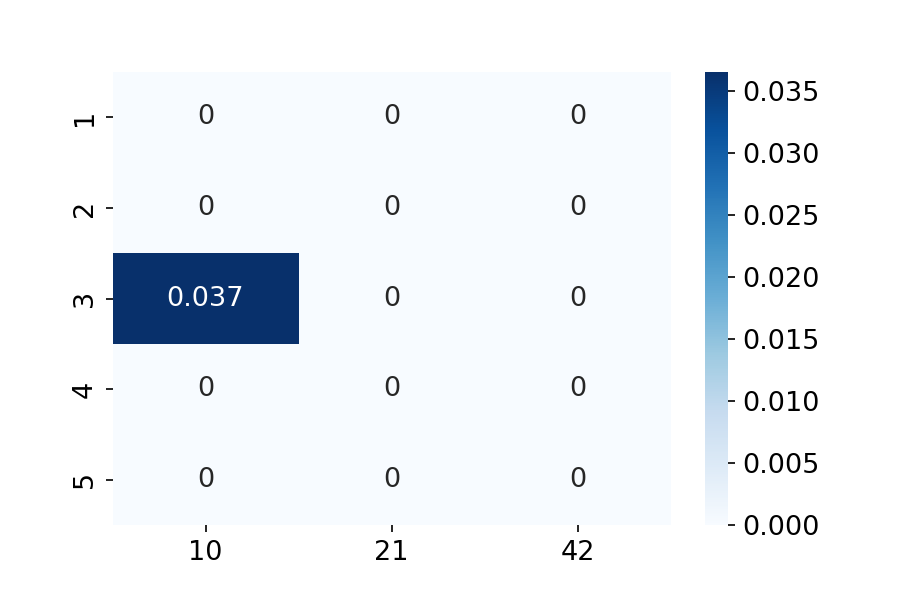}  
  \caption{$N=500$, CMM strategy on trade imbalances}
\end{subfigure}\newline

\begin{subfigure}{.49\textwidth}
  \centering
  \includegraphics[width=.99\linewidth]{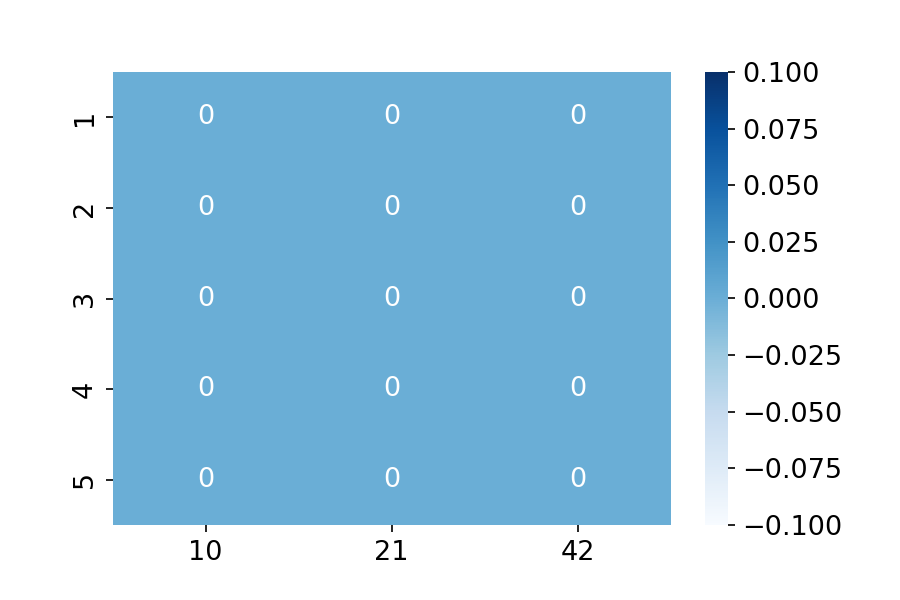}  
  \caption{$N=150$, CMM strategy on volume imbalances}
\end{subfigure}
\begin{subfigure}{.49\textwidth}
  \centering
  \includegraphics[width=.99\linewidth]{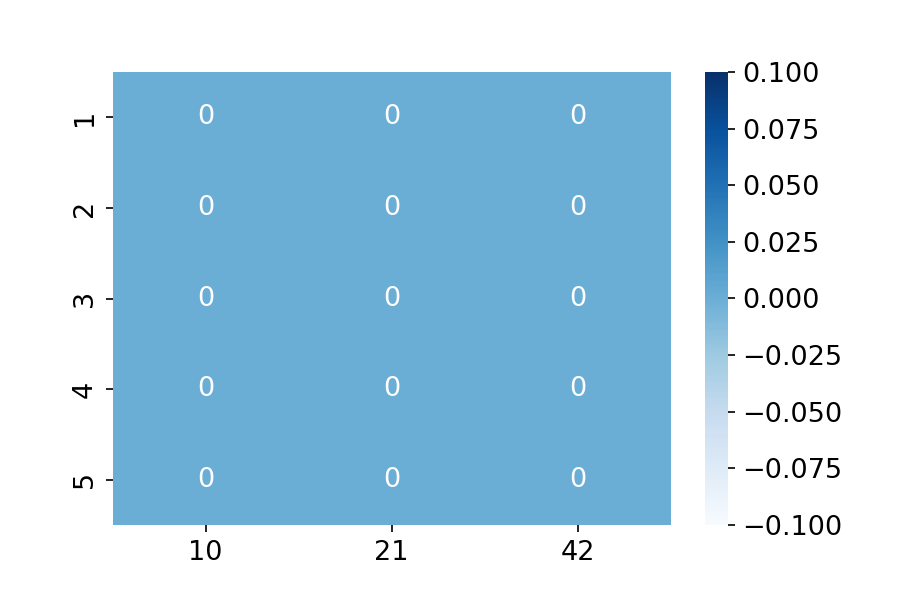}  
  \caption{$N=150$, CMM strategy on trade imbalances}
\end{subfigure}\newline

\begin{subfigure}{.49\textwidth}
  \centering
  \includegraphics[width=.99\linewidth]{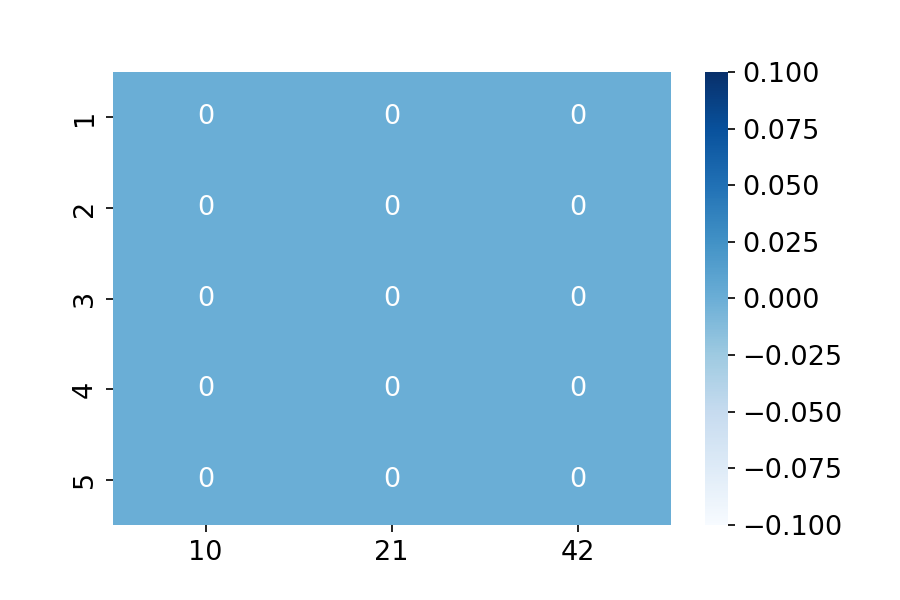}  
  \caption{$N=50$, CMM strategy on volume imbalances}
\end{subfigure}
\begin{subfigure}{.49\textwidth}
  \centering
  \includegraphics[width=.99\linewidth]{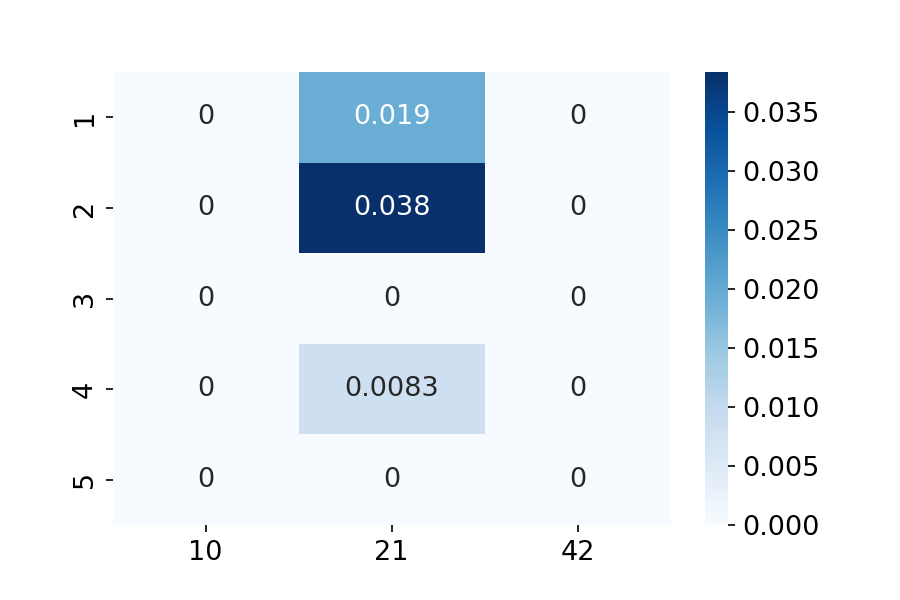}  
  \caption{$N=50$, CMM strategy on trade imbalances}
\end{subfigure}

\caption[]{Significance level tests of our CMM Sharpe Ratios. Strategies rejected by the test are reported with precise zero-value for ease of visualisation. In each heatmap, the x-axis is the horizon of the strategy, and the y-axis is the quantile rank used for trading.}
\label{heat-mom}
\end{figure}

\begin{figure}[h]
\centering
\begin{subfigure}{.49\textwidth}
  \centering
  \includegraphics[width=.99\linewidth]{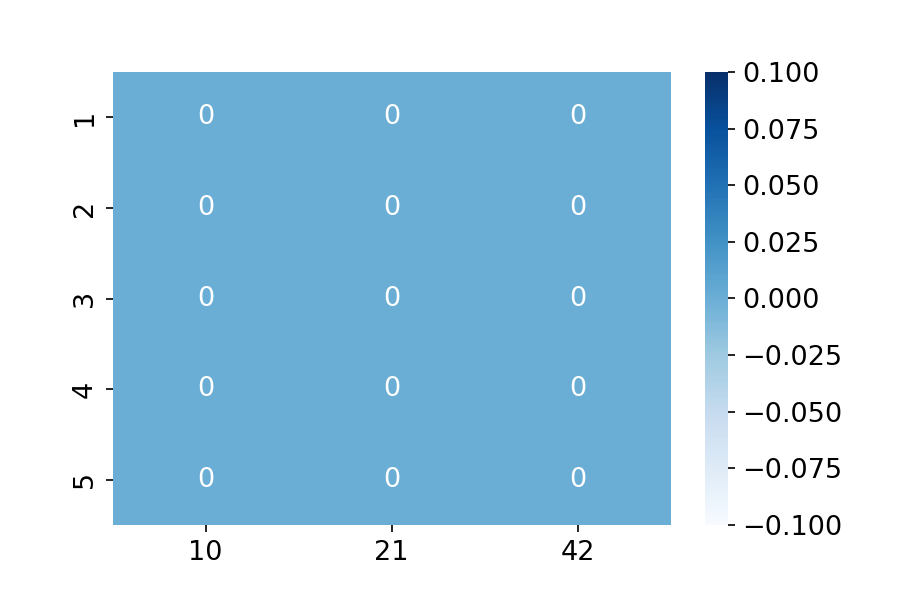}  
  \caption{$N=500$, CMR strategy on volume imbalances}
\end{subfigure}
\begin{subfigure}{.49\textwidth}
  \centering
  \includegraphics[width=.99\linewidth]{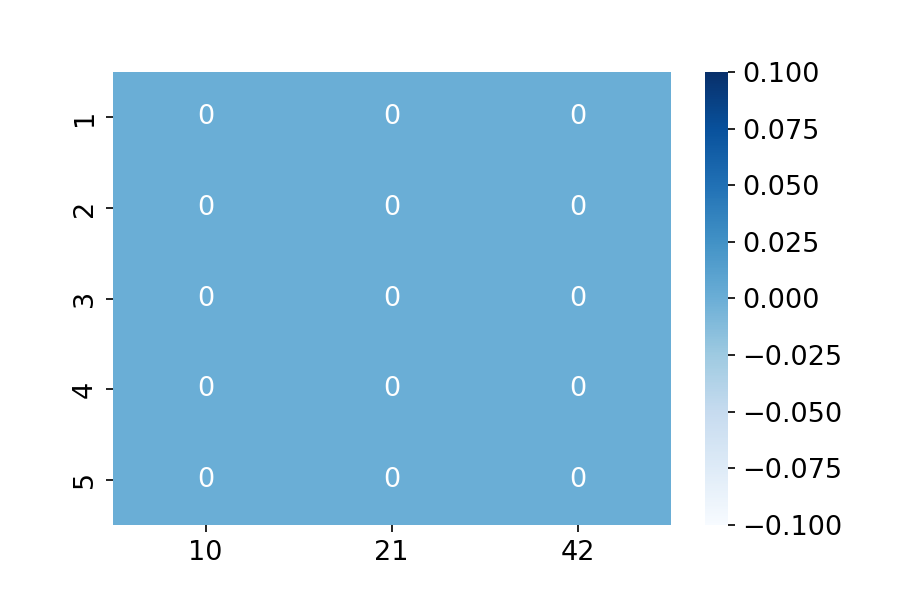}  
  \caption{$N=500$, CMR strategy on trade imbalances}
\end{subfigure}\newline

\begin{subfigure}{.49\textwidth}
  \centering
  \includegraphics[width=.99\linewidth]{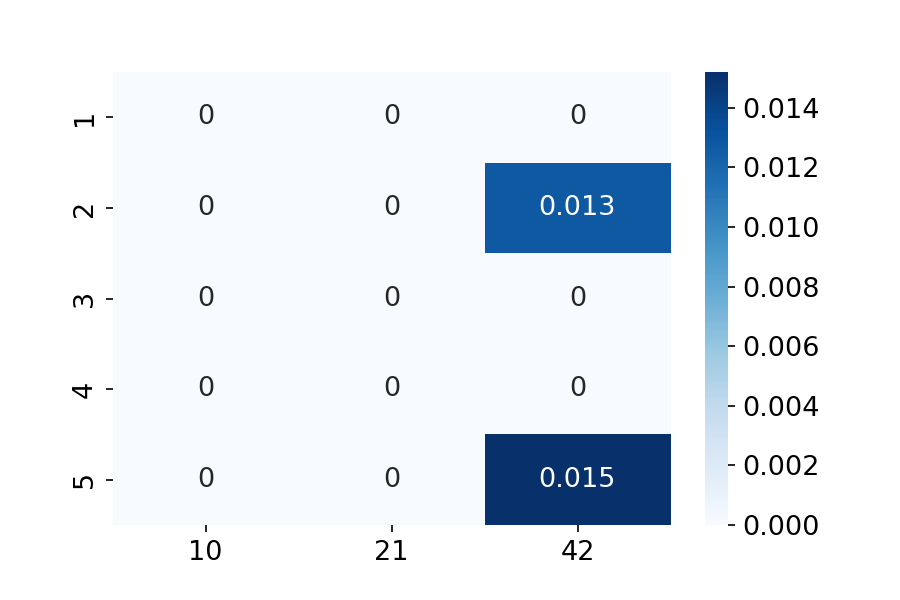}  
  \caption{$N=150$, CMR strategy on volume imbalances}
\end{subfigure}
\begin{subfigure}{.49\textwidth}
  \centering
  \includegraphics[width=.99\linewidth]{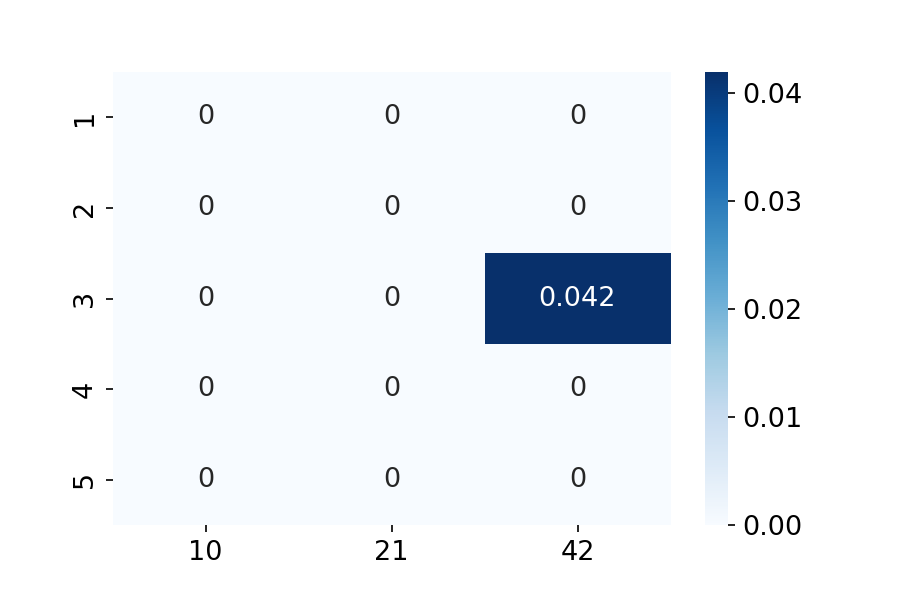}  
  \caption{$N=150$, CMR strategy on trade imbalances}
\end{subfigure}\newline

\begin{subfigure}{.49\textwidth}
  \centering
  \includegraphics[width=.99\linewidth]{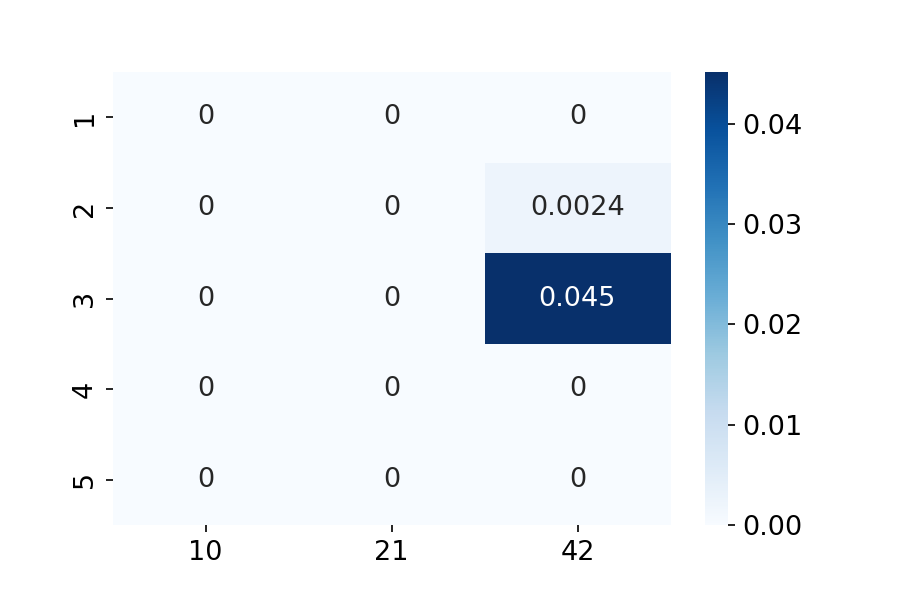}  
  \caption{$N=50$, CMR strategy on volume imbalances}
\end{subfigure}
\begin{subfigure}{.49\textwidth}
  \centering
  \includegraphics[width=.99\linewidth]{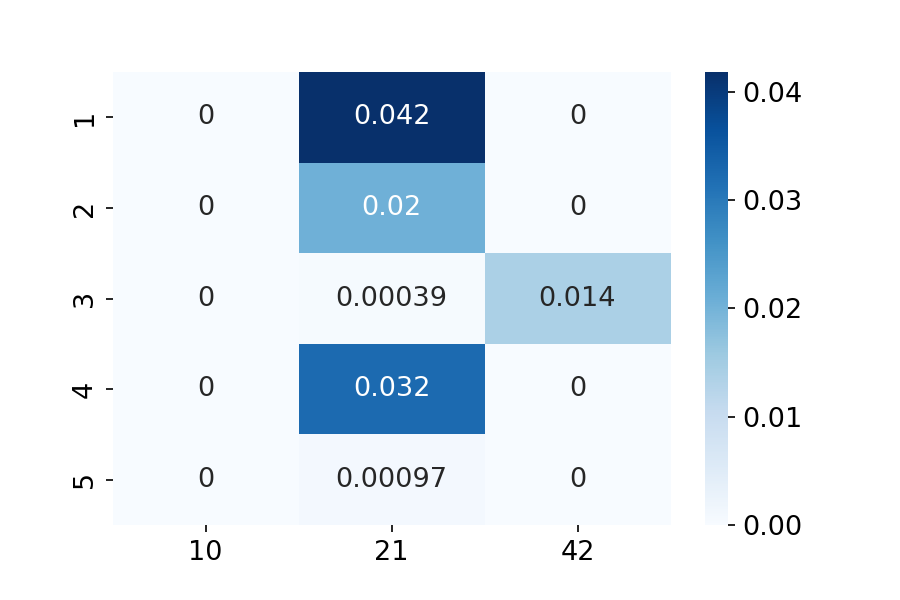}  
  \caption{$N=50$, CMR strategy on trade imbalances}
\end{subfigure}

\caption[]{Significance level tests  of our CMR Sharpe Ratios. Strategies rejected by the test are reported with precise zero-value for ease of visualisation. In each heatmap, the x-axis is the horizon of the strategy,  and the y-axis is the quantile rank used for trading.}
\label{heat-rev}
\end{figure}

Figure \ref{bars-mom-rev} shows the Sharpe Ratios following $I^{vol,tr}$ for the $5$ quantile ranks, while Figs. \ref{heat-mom} and \ref{heat-rev} report the related significance levels for CMM and CMR performances respectively. Interestingly, we mainly see that high and significant Sharpe Ratios are achieved when trading contrarian to trade imbalances on a horizon of $21$ days for $N=50$ for both CMM and CMR strategies. 
We would have expected a stronger positive relationship between imbalances and past market returns that would have further hint to the over-heating of a trade. However, this result suggests possible independence of the performance of imbalances to past prices, but further analysis is needed for more reliable conclusions in this direction.


\subsection{Sector-specific investing}

\begin{figure}[h]
\centering
\begin{subfigure}{\textwidth}
\centering
\includegraphics[width=.75\linewidth]{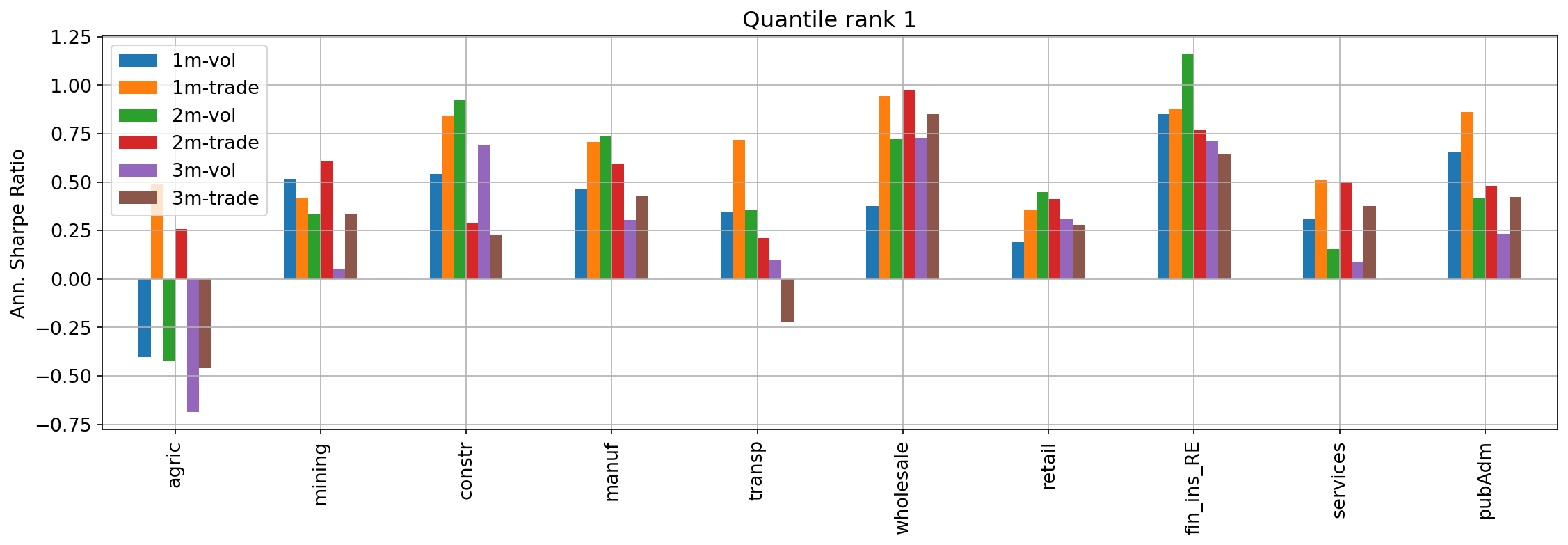}  
\end{subfigure}\newline
\begin{subfigure}{\textwidth}
\centering
\includegraphics[width=.75\linewidth]{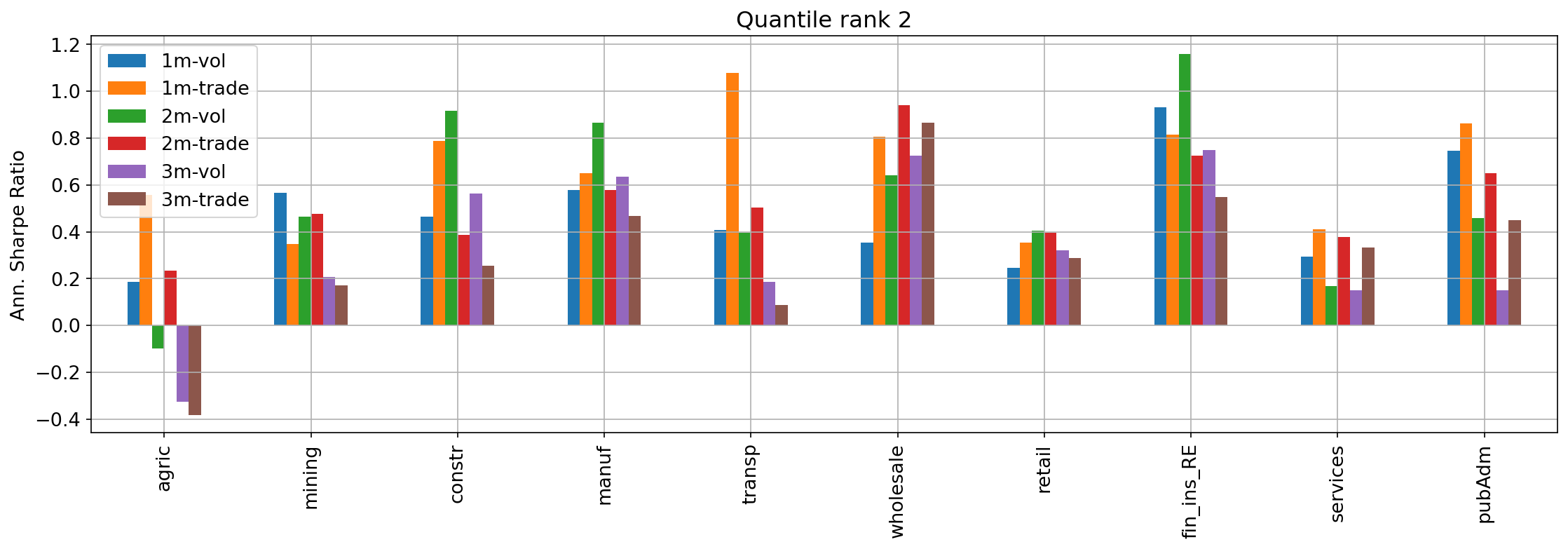}  
\end{subfigure}\newline
\begin{subfigure}{\textwidth}
\centering
\includegraphics[width=.75\linewidth]{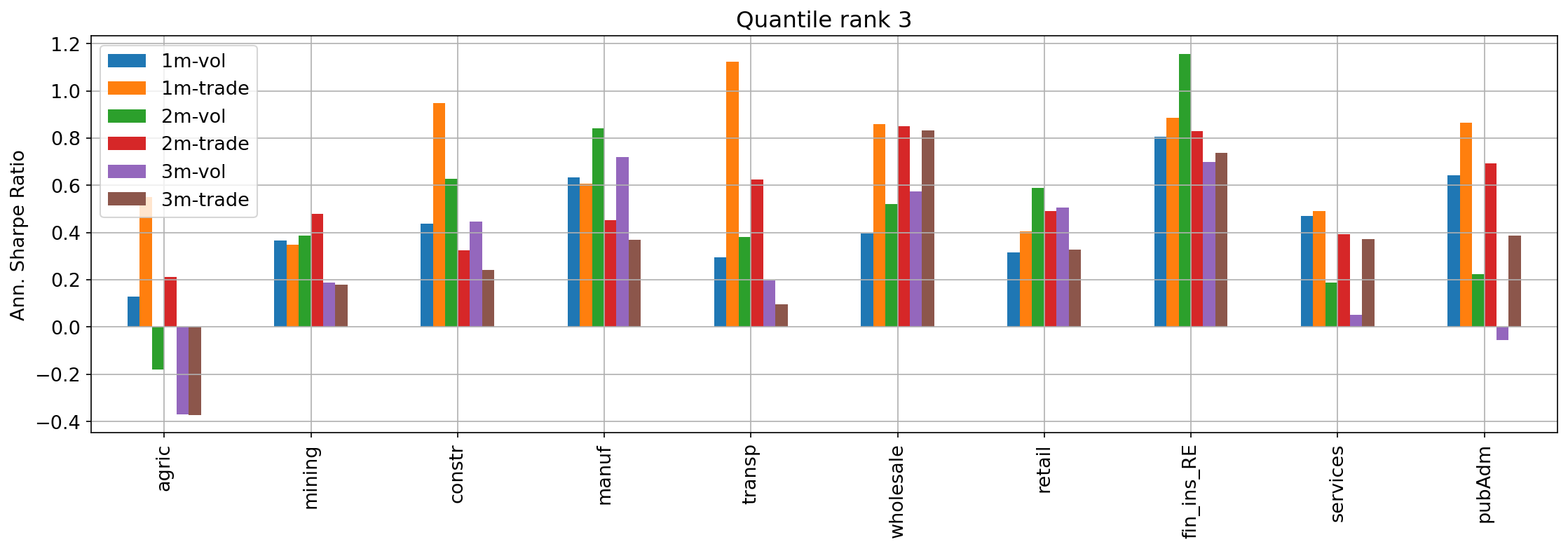}  
\end{subfigure}\newline
\begin{subfigure}{\textwidth}
\centering
\includegraphics[width=.75\linewidth]{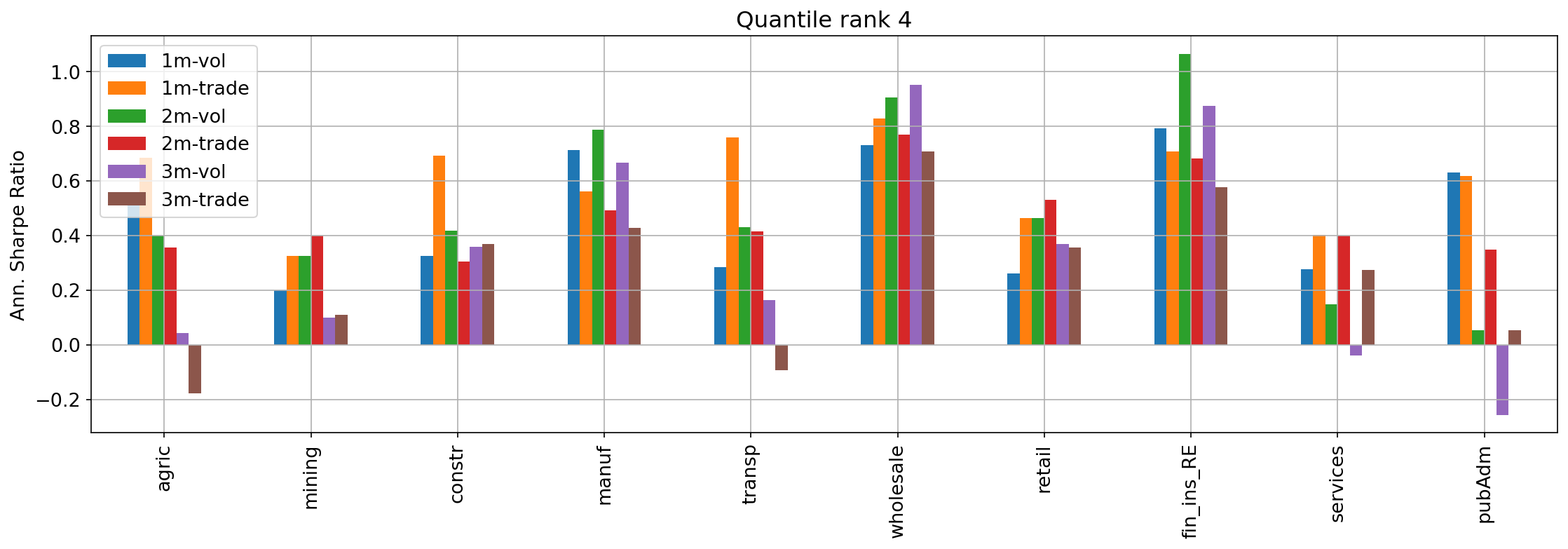}  
\end{subfigure}\newline
\begin{subfigure}{\textwidth}
\centering
\includegraphics[width=.75\linewidth]{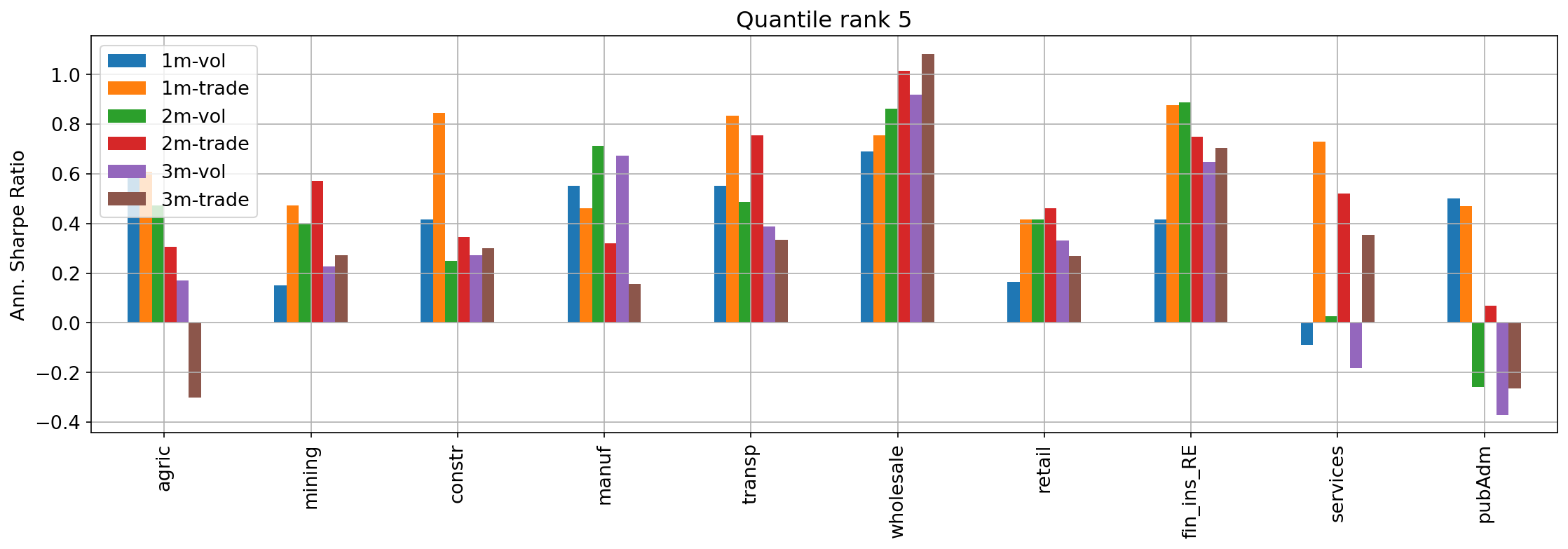}  
\end{subfigure}

\caption[]{Sharpe Ratios achieved by our contrarian vanilla strategy when restricting the trading to the different sectors. We use a threshold $N=50$ and test horizons of $m \in \{21, 42, 63\}$ trading days.}
\label{sectors}
\end{figure}

We perform  one final experiment and project imbalances of stocks onto sector memberships, using data from Wharton Research Data Services providing SIC (Standard Industrial Classification) Major Group for a set of securities. This is a code that classifies stocks into $10$ sectors and its details can be found on the SIC website itself at \url{https://siccode.com/page/structure-of-sic-codes}.
Table \ref{table-sectors} provides a summary of the number of securities for which we have an imbalance and a SIC code, and assigns to each category a short label that we will use in our plots. We then run our contrarian vanilla strategy on each group for horizons of $21,42,63$-trading days (i.e. $1,2,3$ months). We require only $N=50$, since a few sectors have already a limited number of securities to possibly trade. The resultant Sharpe Ratios are shown in Fig. \ref{sectors}, where each subfigure relates to a different quantile. We also compute the related significance levels, but conclude that we can accept only medium performances, while the most profitable ones are rejected by the test.

We conclude that the major information that Form 13F carries is a suggestion to go contrarian to what imbalances reveal. Further conditioning is not able to uncover any stronger signals that could further enhance profits.  However, we think that there is one further view to investigate that we will address in the future. This is the connection between average number (\textit{popularity}) of active funds on the securities of a sector and the related growth and expectation of further inflows. We plot a few initial trends in Fig. \ref{popularity} of Appendix \ref{sec:A2} to intuitively show our thoughts. Once we manage to increase the proportion of securities for which we have available  imbalances, returns and SIC data, then we could monitor variations in the popularity of stocks within sectors to identify signals of changes in inflows, and leverage them to achieve stronger profits.

\begin{longtable}{ |p{7cm}|p{2cm}|p{3cm}|}
\hline
\textbf{SIC CATEGORY} & \textbf{LABEL} & \textbf{No. SECURITIES} \\
\hline
\hline
Agriculture, Forestry, Fishing & agric & $5/16$ \\
\hline
Mining & mining & $98/352$ \\
\hline
Construction & constr & $20/56$ \\
\hline
Manufacturing & manuf & $521/1227$ \\
\hline
Transportation \& Public Utilities & transp & $161/441$ \\
\hline
Wholesale Trade & wholesale & $70/149$ \\
\hline
Retail Trade & retail & $71/194$ \\
\hline
Finance Insurance, Real Estate & fin-ins-RE & $796/5136$ \\
\hline
Services & services & $153/597$ \\
\hline
Public Administration & pubAdm & $75/2897$ \\
\hline
\caption{SIC sectors on which we investigate the performance of our contrarian vanilla strategy. We report the number of securities for which we have available a SIC code for each sector (denominator in the right column), along with the number of them for which we have also imbalances and returns data (numerator)}
\label{table-sectors}
\end{longtable}

\clearpage






\section{Conclusions}
\label{sec:conclusions}

This study investigates the hidden information that is disclosed within the SEC Form 13F-HR. We leverage on the concept of \textit{imbalance} in buying vs. selling behaviour and show that a significant opportunity for profit arises if an external investor is willing to trade contrarian to these 13F filings imbalances.
At first glance, the reliability of our signals might be thought to increase with the number of active funds $N$ on a security. However, the risk is to invest in very liquid stocks without real opportunities for profit as shown by the lack of significance of our Sharpe Ratios for $N=500$.
On the other hand, we find that a  threshold of $N=50$ provides a good trade-off between reliability of the imbalance signals and variety of the remaining stocks to possibly trade.
Our simplest contrarian vanilla strategy achieves a desirable Sharpe Ratio $S \sim 1$ that is significant at the $0.05$ level when trading quantiles $qr_4, qr_5$ of $I^{tr}$. These two options also show higher average $PPT$.
Our imbalances capture the amount of information already consumed in the market and can be inflated by herding despite the very eclectic set of institutions considered. The strongest imbalances point to overheated trades, for which we expect a reversion of prices against the sign of imbalances, which is also confirmed by our results.
The highest profitability is achieved when following trade imbalances, since they better depict crowding in the market. We conclude that the optimal holding horizon is around $21$ trading days. This agrees with the fact that plain copycat investors become active around $45$ calendar days after the financial quarter ends.

For future investigations, an interesting direction to explore concerns the effect of correcting reports with their amendments (Form 13F-HR/A) prior to running the analyses, and compare the results. We expect the signals to be affected in a tangible way, but it is not trivial to establish this in advance. \imp{However, it is very likely that we will gain stronger insights into the extent to which fund managers, which are entering into a position over a given quarter, impact the price return of the asset over the same quarter.}

Yet another direction worth pursuing concerns the investigation of network effects in terms of the cross-impact of the imbalance associated to a given asset, in either the contemporaneous or future return of another asset, in the spirit of the recent work in \cite{price_impact_imbalance}.
On a longer timescale, we also plan to pursue the described investment analyses on N-PORT holdings reports, once more history is available. At the beginning of October 2016, the SEC introduced the form N-PORT to modernise portfolio reporting of both assets and liabilities of Registered Investment Companies and exchange-traded funds (ETFs) organised as unit investment trusts. This form became compulsory in May 2019 and provides a further level of market transparency and holdings details, with potentially valuable information to analyse.



\section*{Acknowledgements}
Deborah Miori's research was supported by the \textit{EPSRC CDT in Mathematics of Random Systems} (EPSRC Grant  EP/S023925/1).

\bibliographystyle{unsrt}  
\bibliography{zz_main}  

\appendix
\section{\imp{Price impact of imbalances over quarters}}
\label{sec:A0}

\begin{figure}[h]
    \centering
    \includegraphics[width=0.7\textwidth]{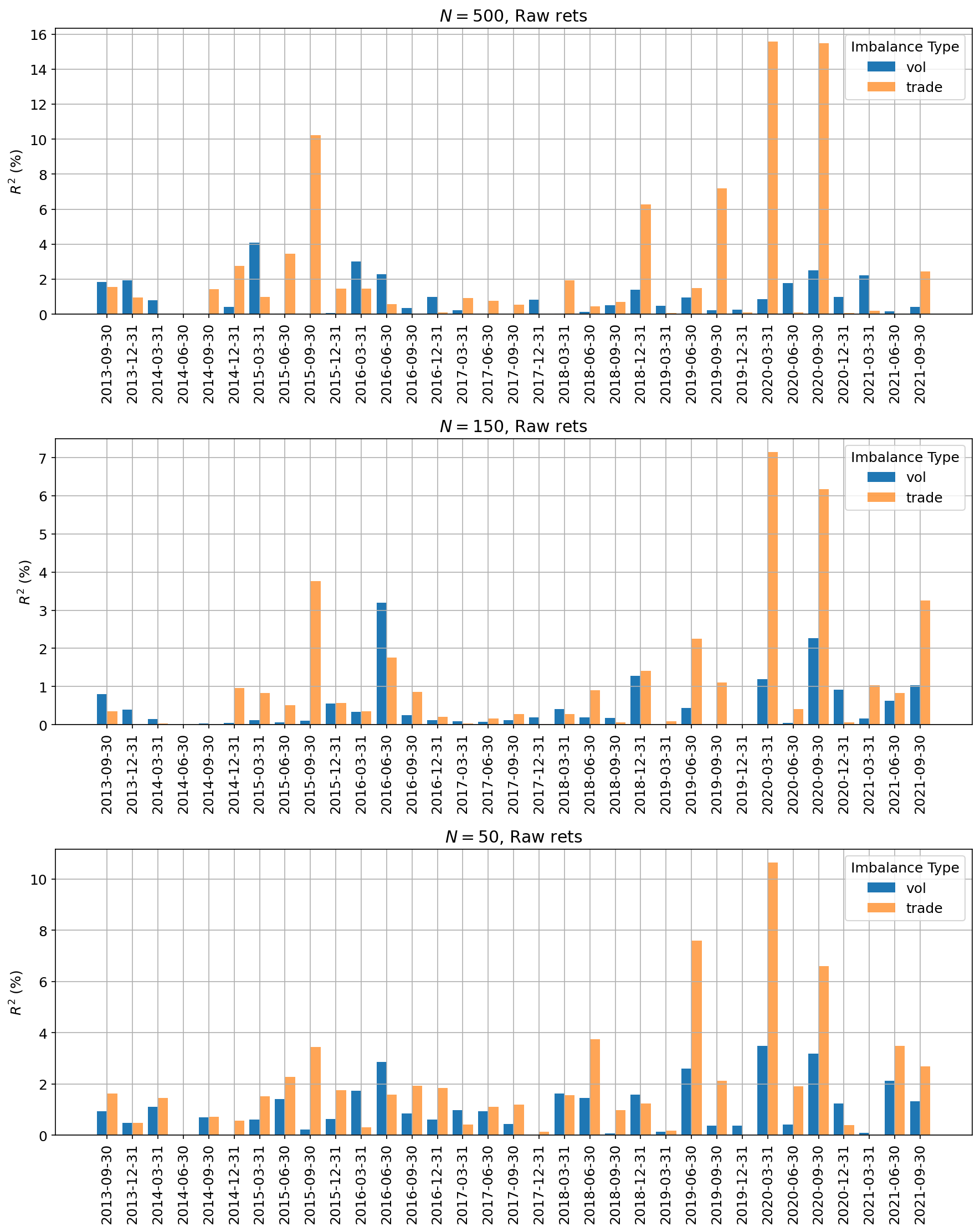}
    \caption{For each quarter $p$, we report the $R^2$ from the regression of contemporaneous raw returns on $I^{vol, tr}$ for $N \in \{50, 150, 500\}$.}
    \label{fig:impact-raw-rets}
\end{figure}

\begin{figure}[h]
    \centering
    \includegraphics[width=0.7\textwidth]{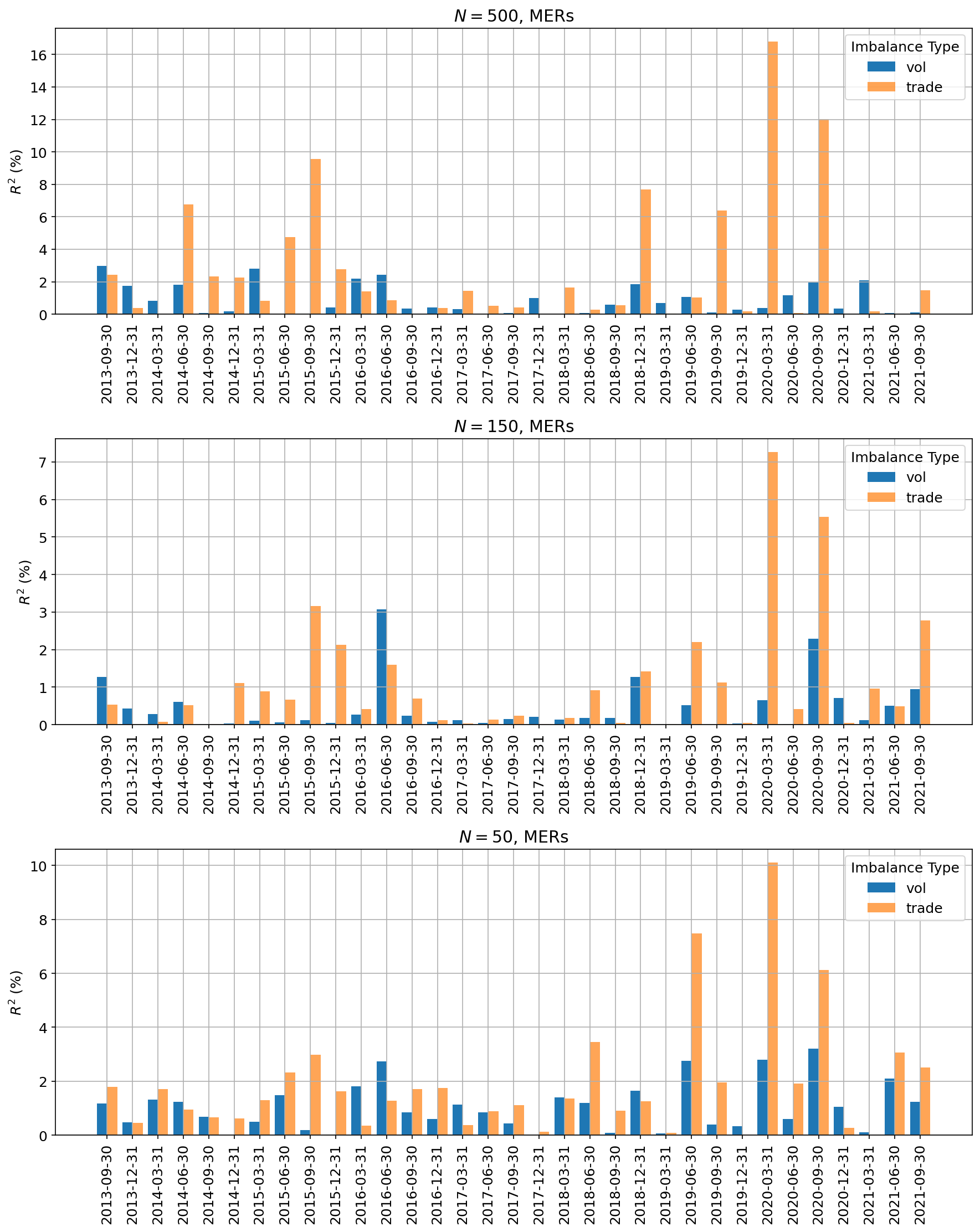}
    \caption{For each quarter $p$, we report the $R^2$ from the regression of contemporaneous MERs on $I^{vol, tr}$ for $N \in \{50, 150, 500\}$.}
    \label{fig:impact-mers}
\end{figure}

We regress contemporaneous raw returns (raw rets) and market-excess returns (MERs) on imbalances to weight the impact of fund managers' new allocations on the price returns of the related assets over each quarter $p$. We consider both $I^{vol}$ and $I^{tr}$, and $N \in \{50, 150, 500\}$. For each point in time, we compute the $R^2$ of the related regression and plot it as percentage in Figs. \ref{fig:impact-raw-rets} and \ref{fig:impact-mers}, for raw rets and MERs respectively. 

As highlighted in Section \ref{sec:market-impact}, imbalances computed in terms of trade counts show higher explanatory power for asset returns, especially in Q$3$s. Moreover, it is interesting to see here how this measure has spikes in $R^2$ reaching $10-15\%$, despite our data being extremely noisy and addressing very long periods of $3$ months at each iteration.
    
\clearpage

\clearpage

\section{Cumulative PnL for the vanilla strategy}
\label{sec:A1}

\begin{figure}
\centering
\begin{subfigure}[b]{0.6\textwidth}
   \includegraphics[width=1\linewidth]{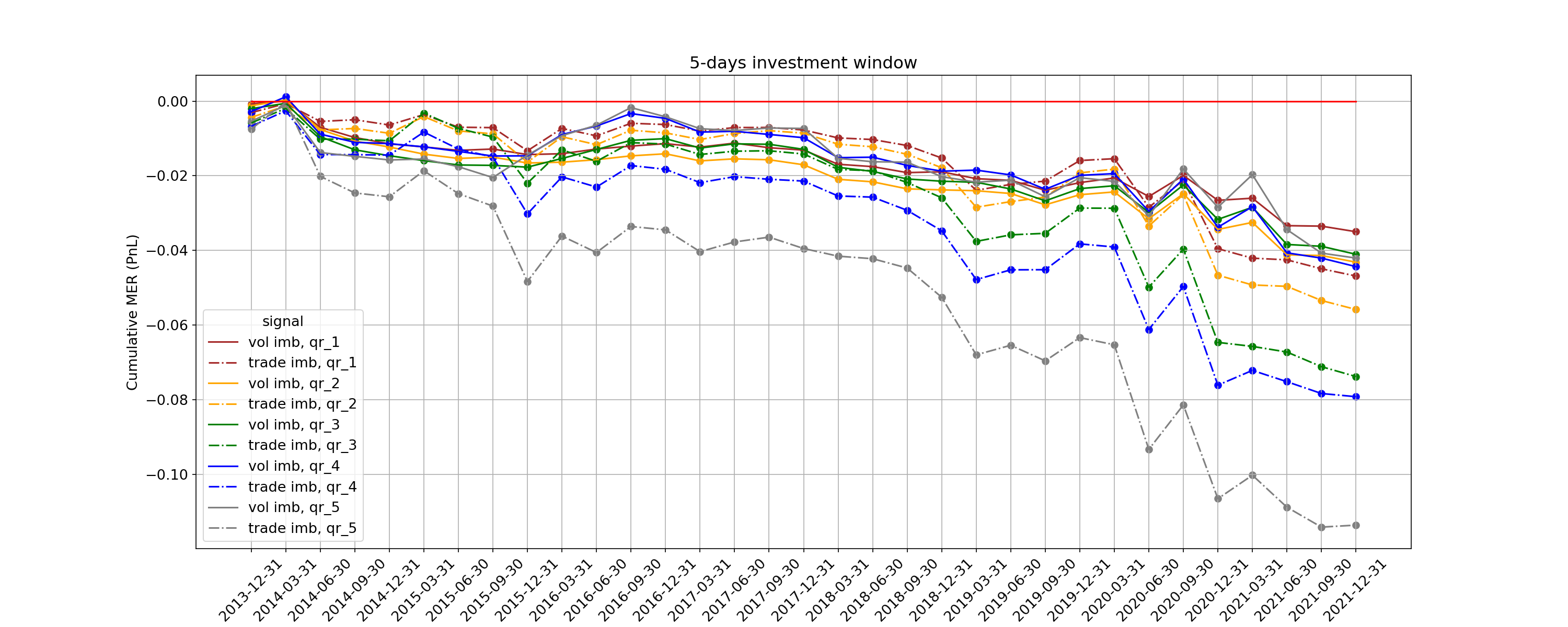}
\end{subfigure}
\vspace{0.3cm}
\begin{subfigure}[b]{0.6\textwidth}
   \includegraphics[width=1\linewidth]{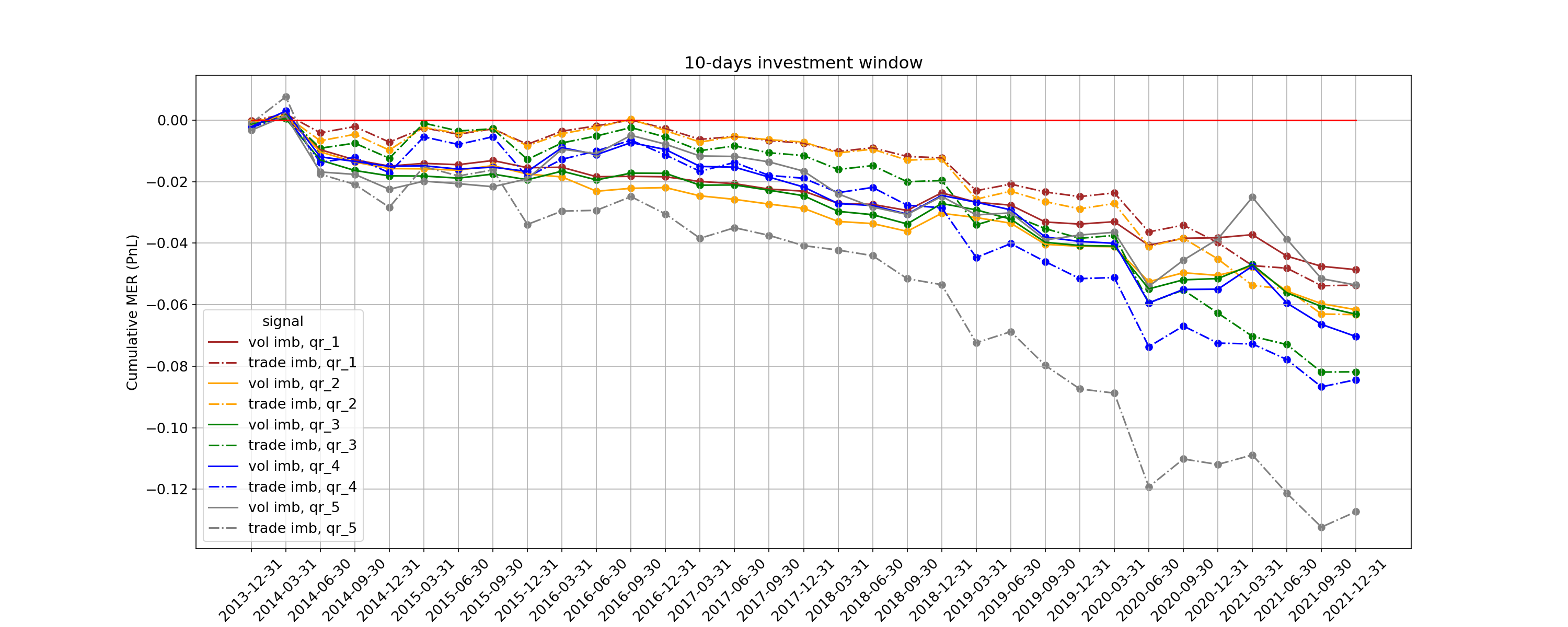}
\end{subfigure}
\vspace{0.3cm}
\begin{subfigure}[b]{0.6\textwidth}
   \includegraphics[width=1\linewidth]{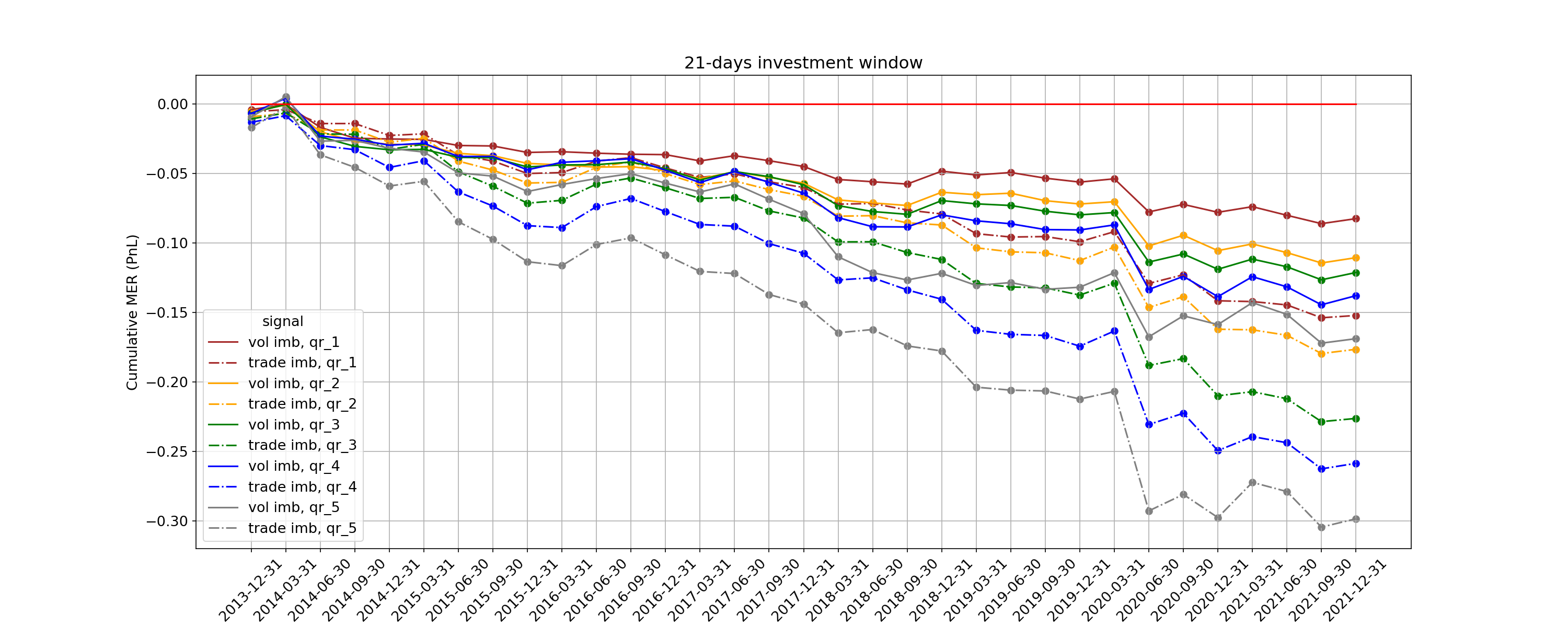}
\end{subfigure}
\vspace{0.3cm}
\begin{subfigure}[b]{0.6\textwidth}
   \includegraphics[width=1\linewidth]{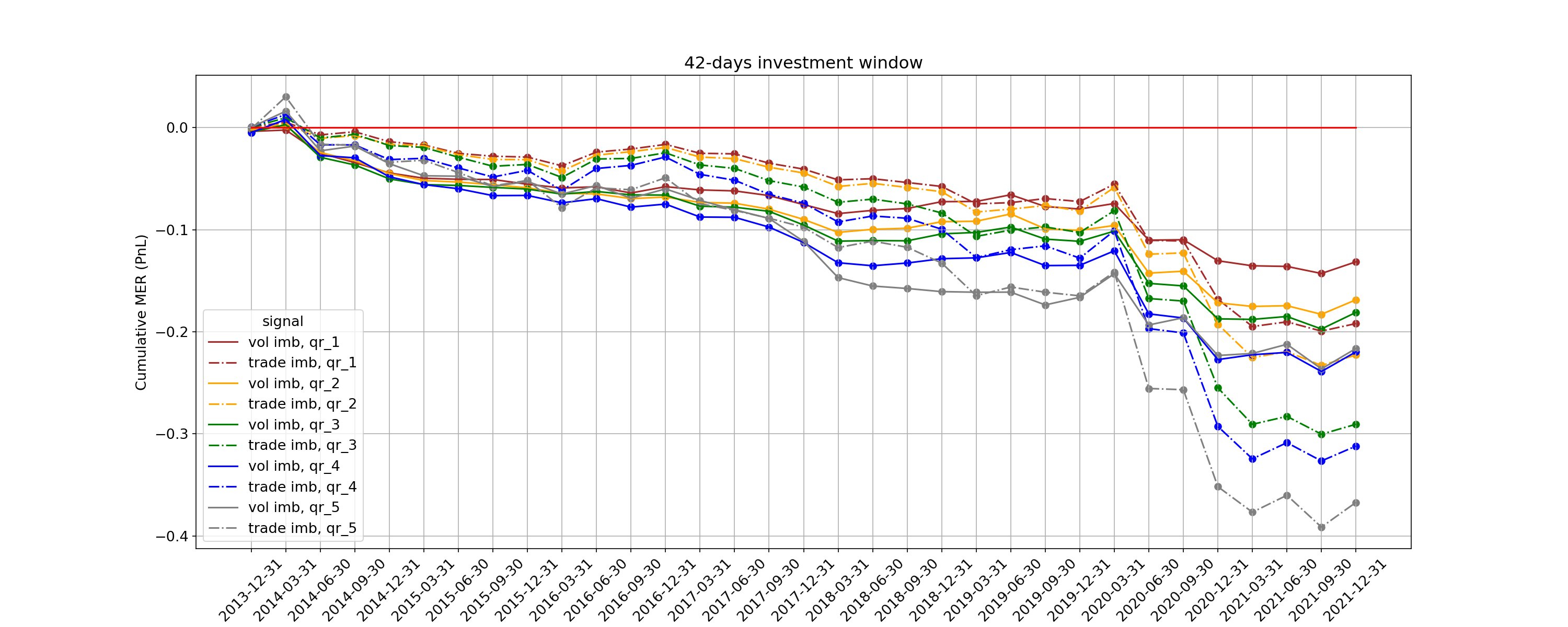}
\end{subfigure}
\vspace{0.3cm}
\begin{subfigure}[b]{0.6\textwidth}
   \includegraphics[width=1\linewidth]{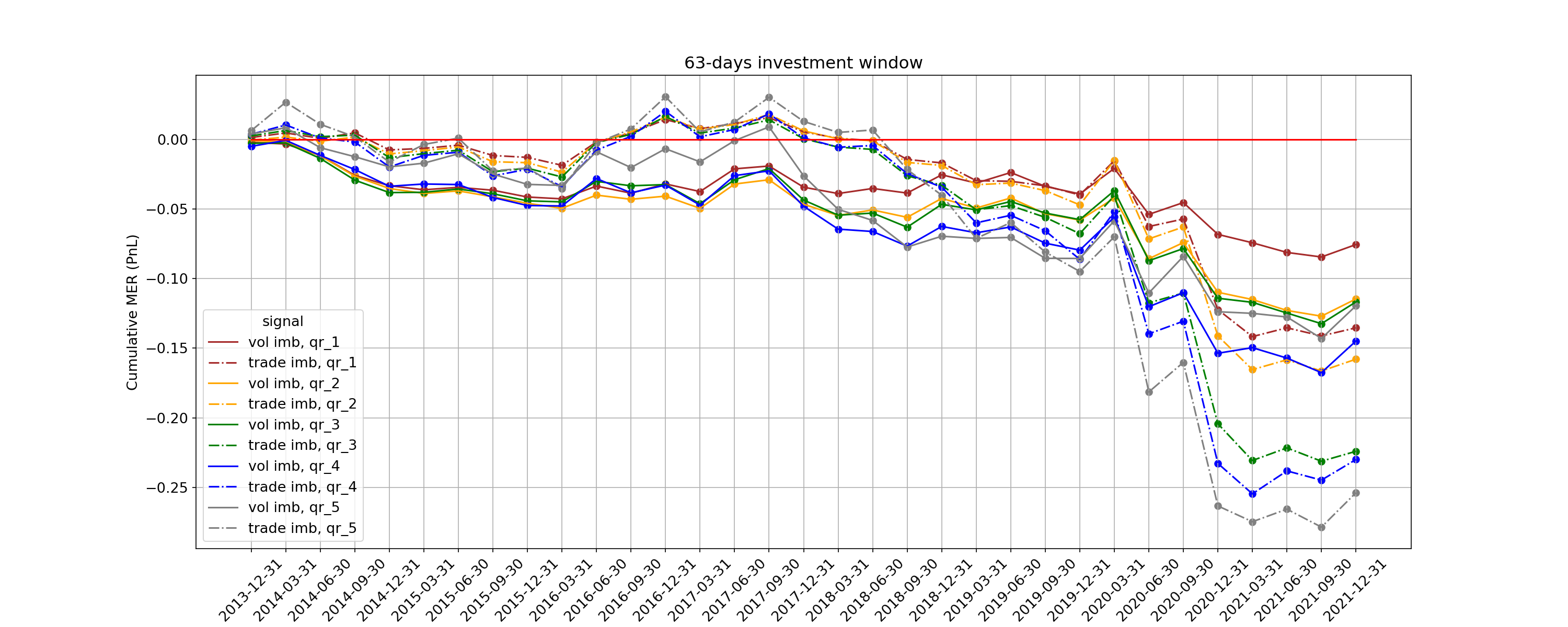}
\end{subfigure}
\caption[]{Time series of cumulative PnL when trading imbalances applying the threshold $N=50$.}
\label{cum50}
\end{figure}

\begin{figure}
\centering
\begin{subfigure}[b]{0.6\textwidth}
   \includegraphics[width=1\linewidth]{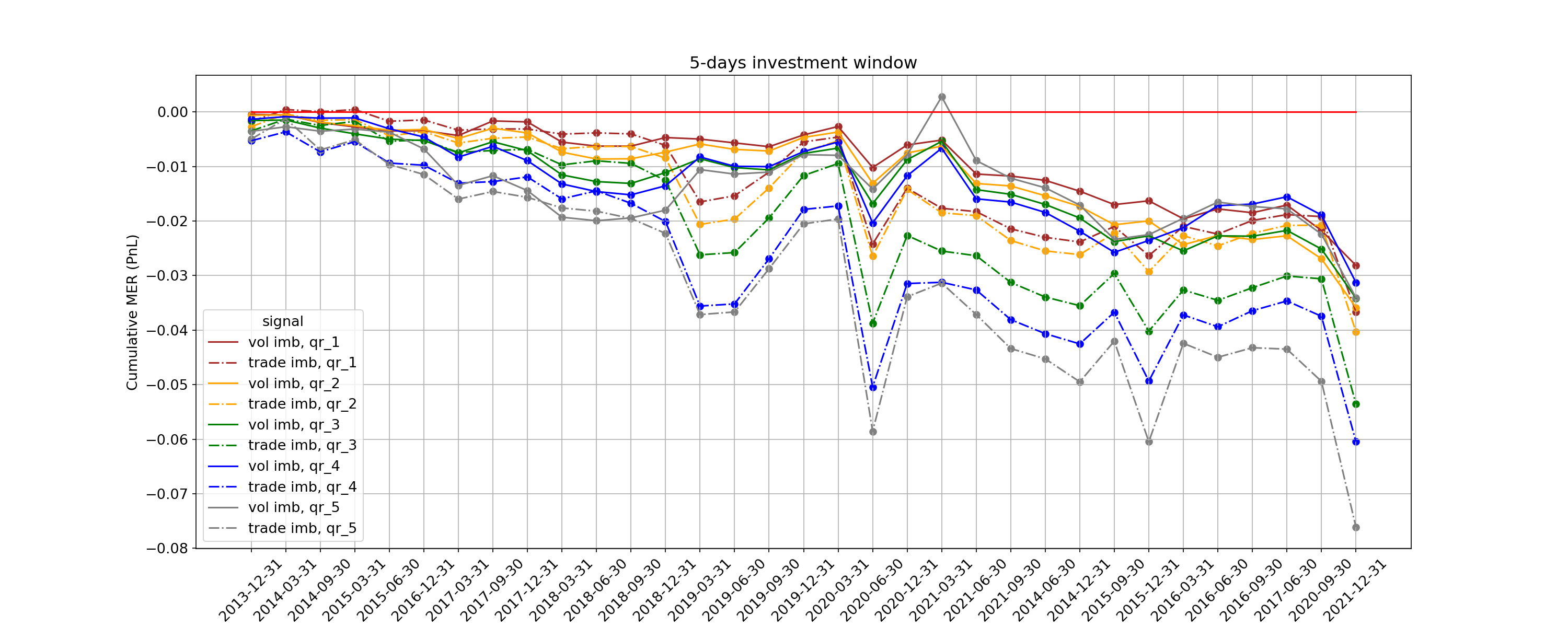}
\end{subfigure}
\vspace{0.3cm}
\begin{subfigure}[b]{0.6\textwidth}
   \includegraphics[width=1\linewidth]{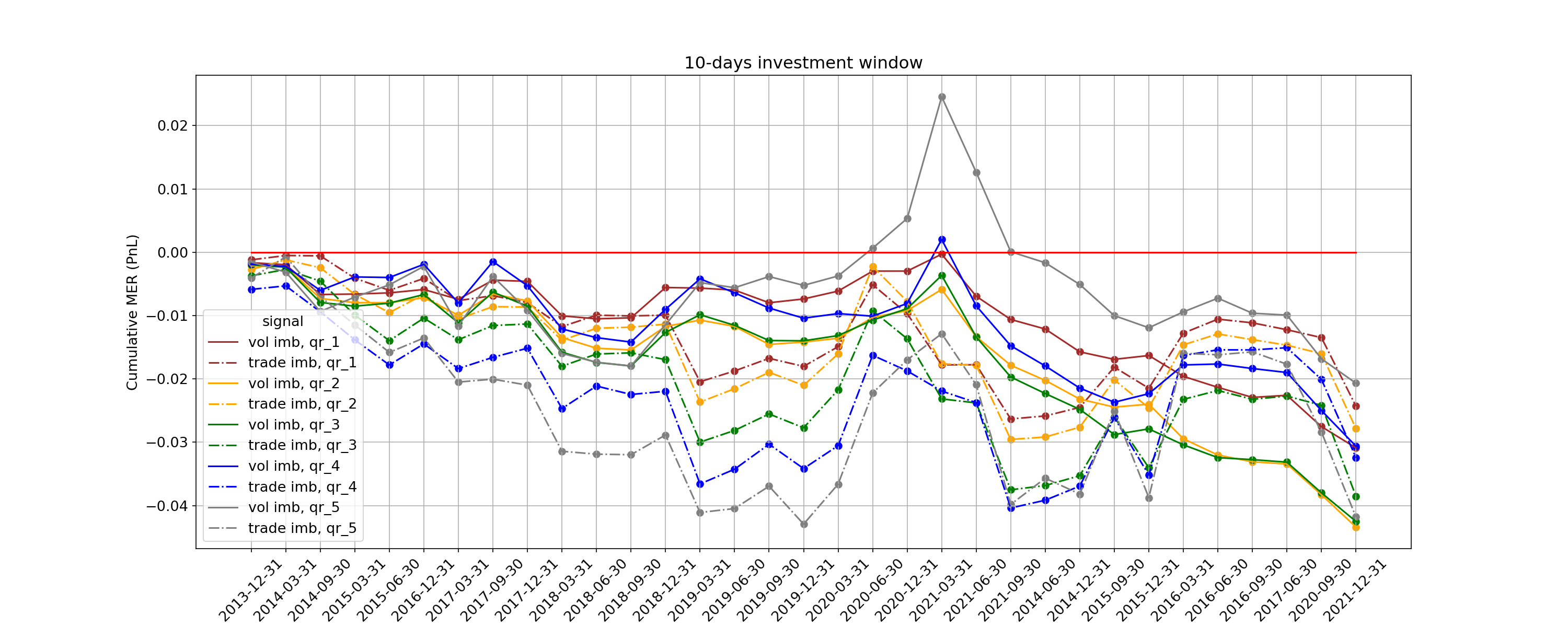}
\end{subfigure}
\vspace{0.3cm}
\begin{subfigure}[b]{0.6\textwidth}
   \includegraphics[width=1\linewidth]{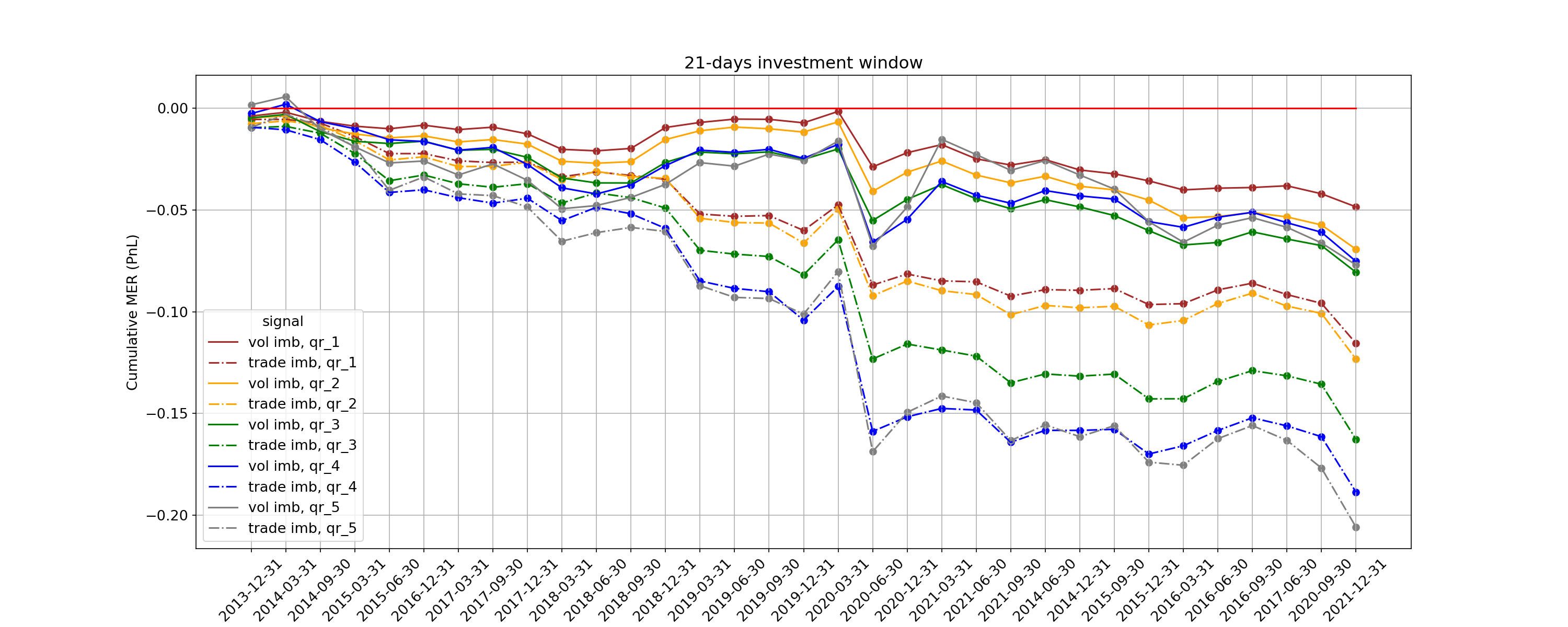}
\end{subfigure}
\vspace{0.3cm}
\begin{subfigure}[b]{0.6\textwidth}
   \includegraphics[width=1\linewidth]{cum/cum42-150.png}
\end{subfigure}
\vspace{0.3cm}
\begin{subfigure}[b]{0.6\textwidth}
   \includegraphics[width=1\linewidth]{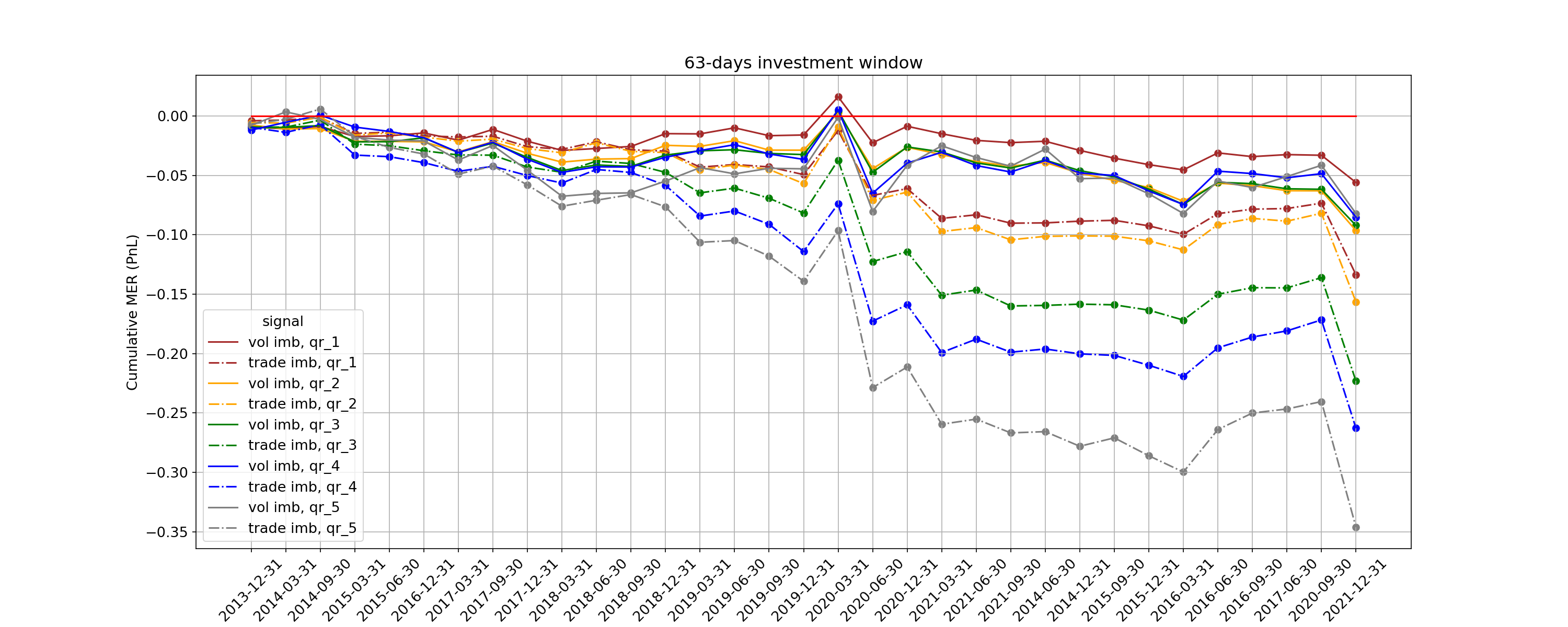}
\end{subfigure}
\caption[]{Time series of cumulative PnL when trading imbalances applying the threshold $N=150$.}
\label{cum150}
\end{figure}

\begin{figure}
\centering
\begin{subfigure}[b]{0.6\textwidth}
   \includegraphics[width=1\linewidth]{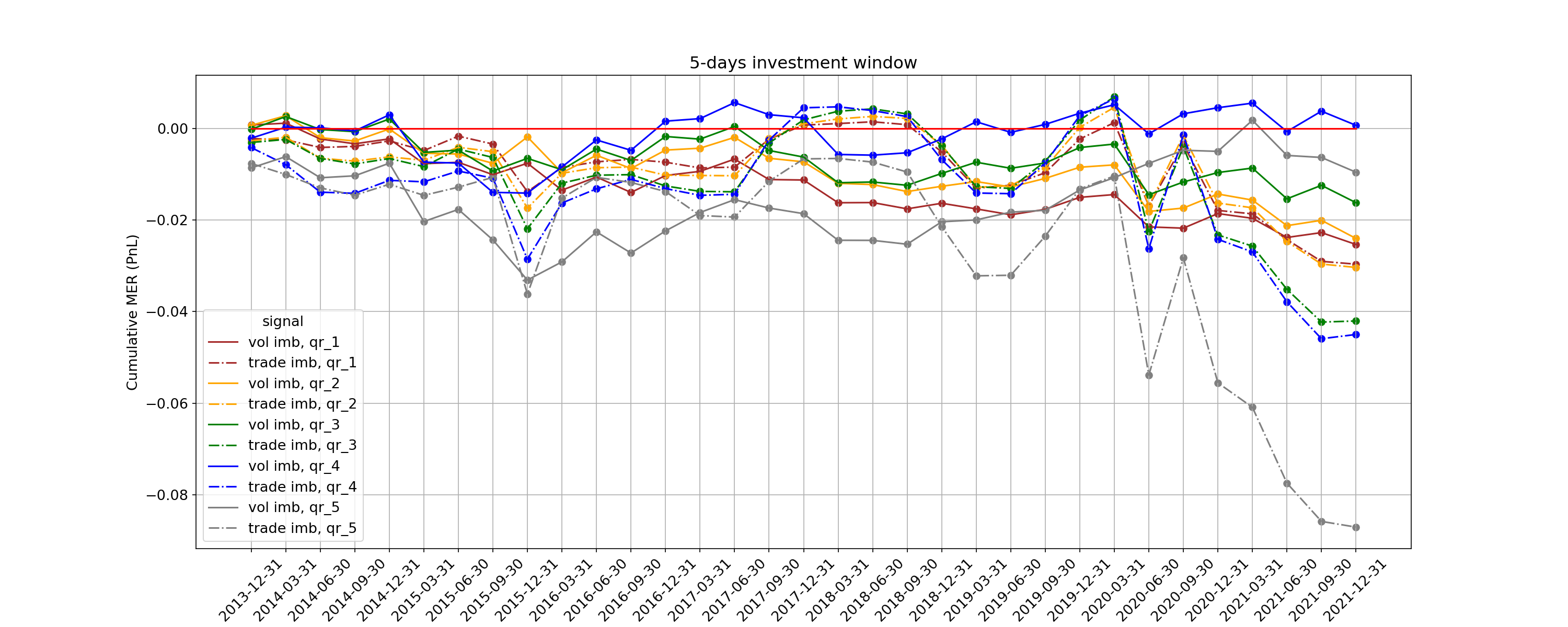}
\end{subfigure}
\vspace{0.3cm}
\begin{subfigure}[b]{0.6\textwidth}
   \includegraphics[width=1\linewidth]{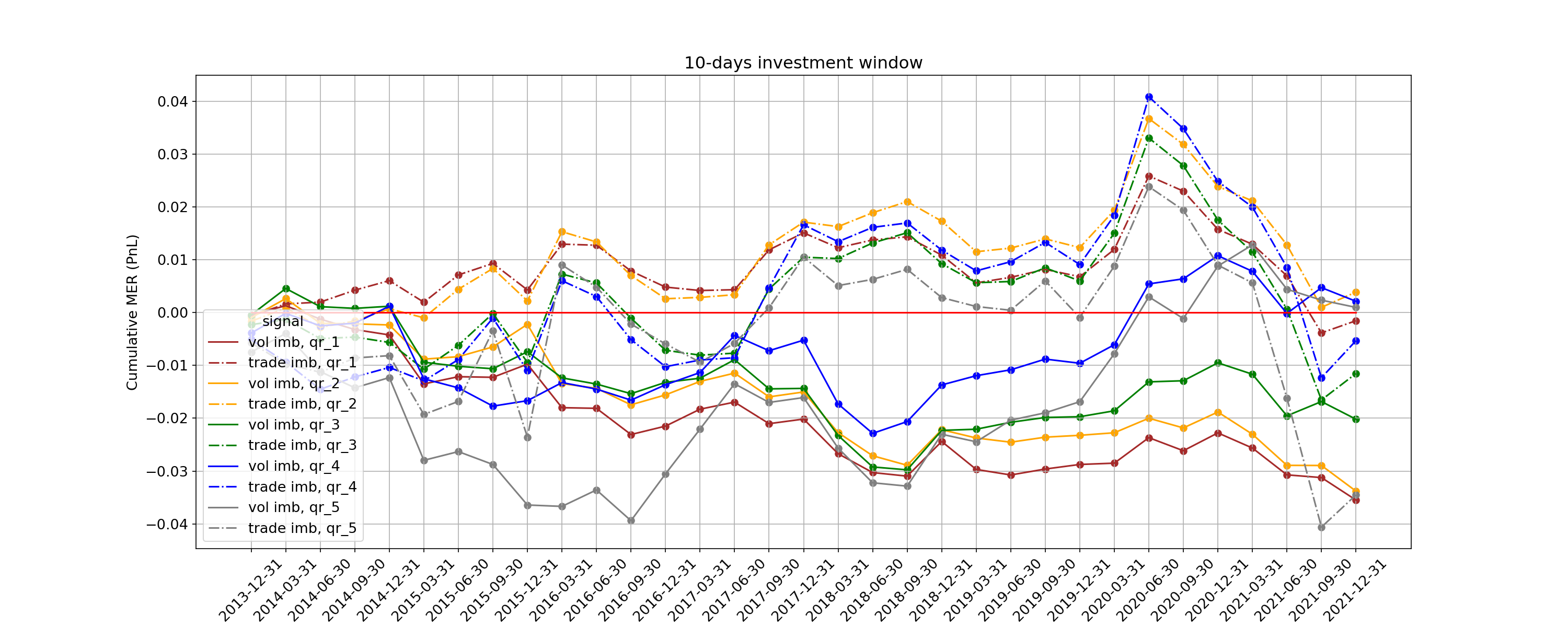}
\end{subfigure}
\vspace{0.3cm}
\begin{subfigure}[b]{0.6\textwidth}
   \includegraphics[width=1\linewidth]{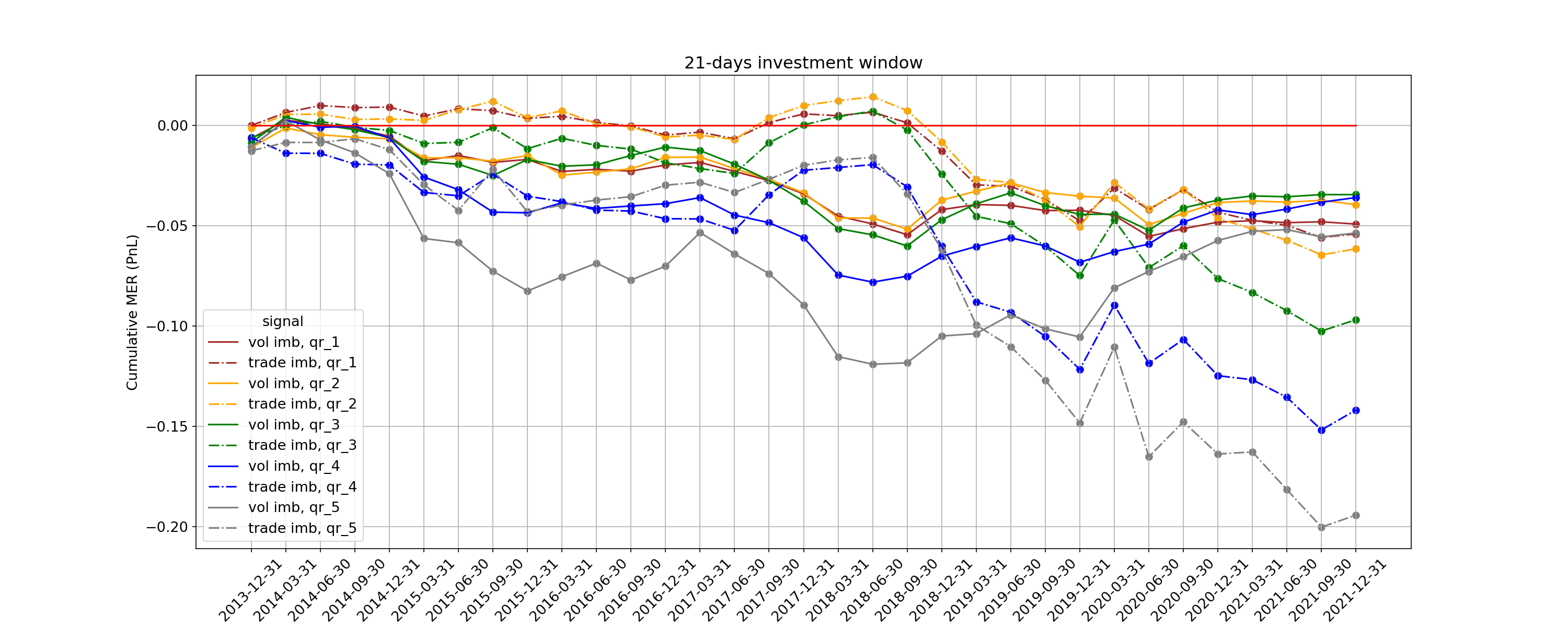}
\end{subfigure}
\vspace{0.3cm}
\begin{subfigure}[b]{0.6\textwidth}
   \includegraphics[width=1\linewidth]{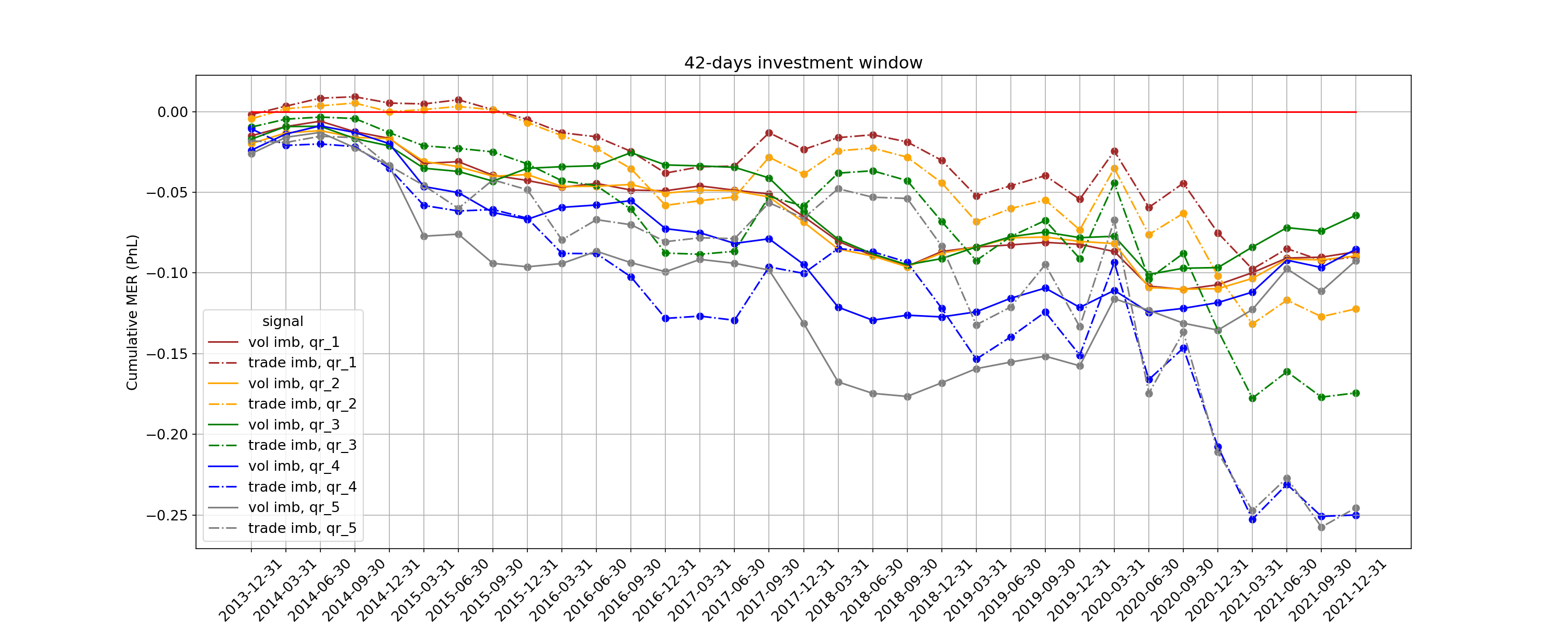}
\end{subfigure}
\vspace{0.3cm}
\begin{subfigure}[b]{0.6\textwidth}
   \includegraphics[width=1\linewidth]{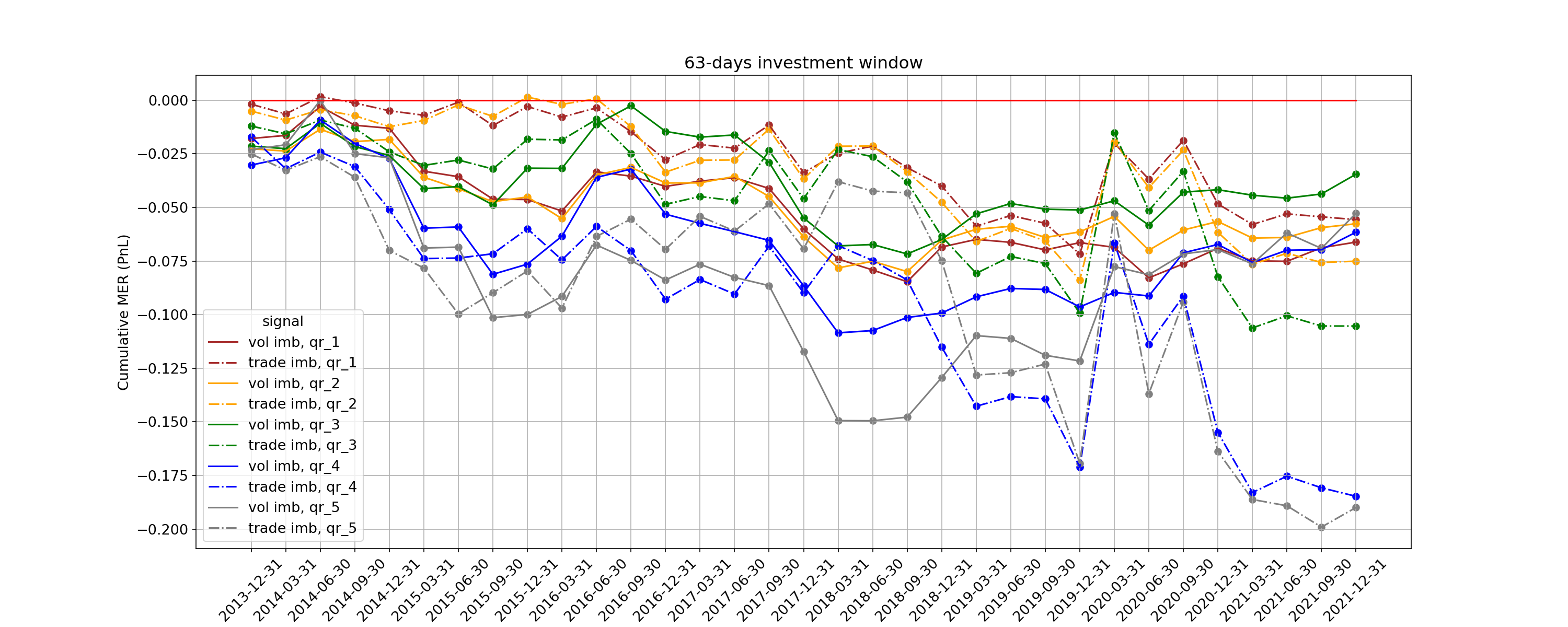}
\end{subfigure}
\caption[]{Time series of cumulative PnL when trading imbalances applying threshold the  $N=500$.}
\label{cum500}
\end{figure}

We compute time series of the cumulative PnL gained while trading the plain imbalance signal, i.e. buying (selling) stocks if the related imbalance is positive (negative). Figures \ref{cum50}, \ref{cum150} and \ref{cum500} show the results for minimum number of active funds $N \in \{50, 150, 500\}$ required on each security. In each case, we plot the performance for quantile ranks $qr_i$ with $i \in \{1,2,3,4,5\}$ and horizons of $m \in \{5, 10, 21, 42, 63\}$ trading days. We also divide between trading $I^{vol}$ versus $I^{tr}$. The results agree that a general opportunity for profit arises if the investor is willing to trade \textit{contrarian} to imbalances from 13F filings.

\clearpage

\section{Evolution of the popularity of stocks within SIC sectors}
\label{sec:A2}

\begin{figure}[h]
\centering
\begin{subfigure}{.45\textwidth}
  \centering
  \includegraphics[width=.99\linewidth]{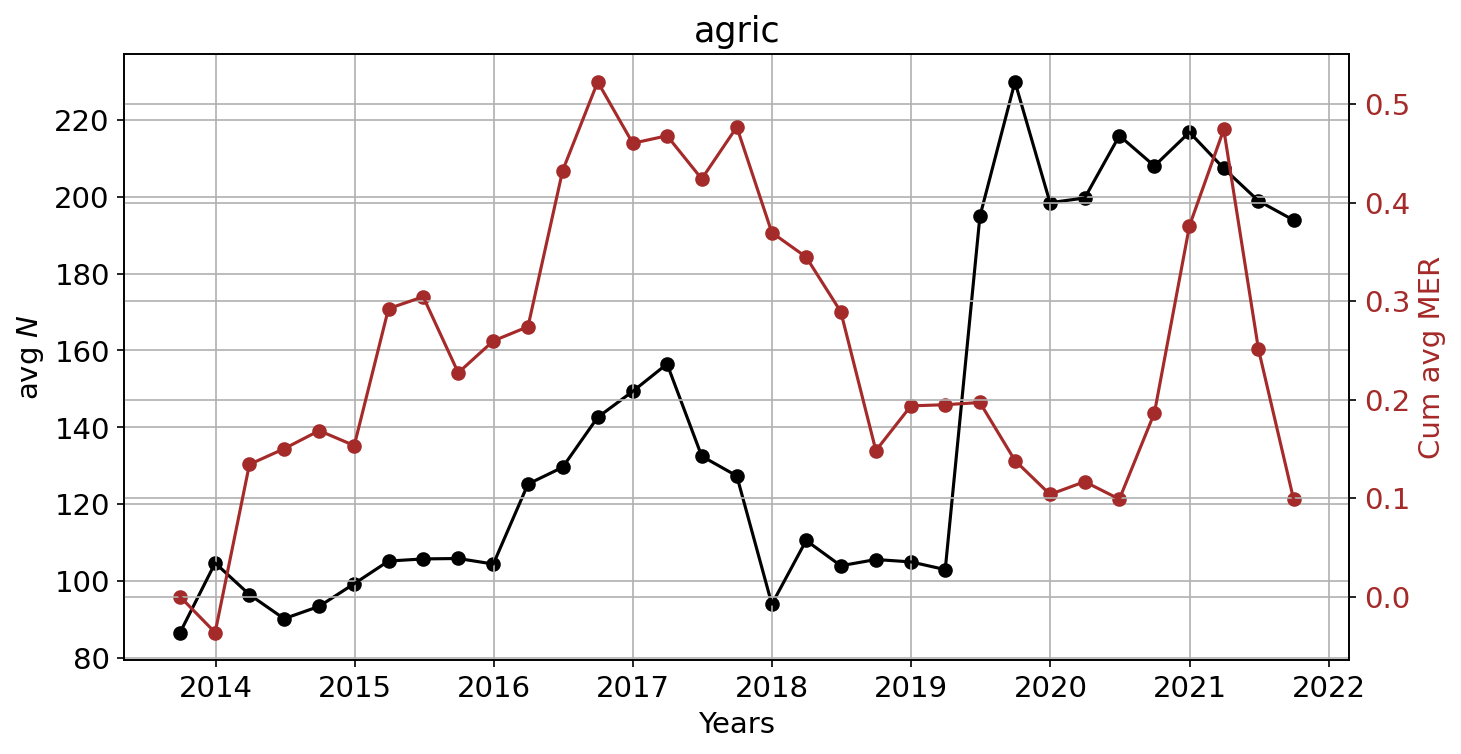}  
\end{subfigure}
\begin{subfigure}{.45\textwidth}
  \centering
  \includegraphics[width=.99\linewidth]{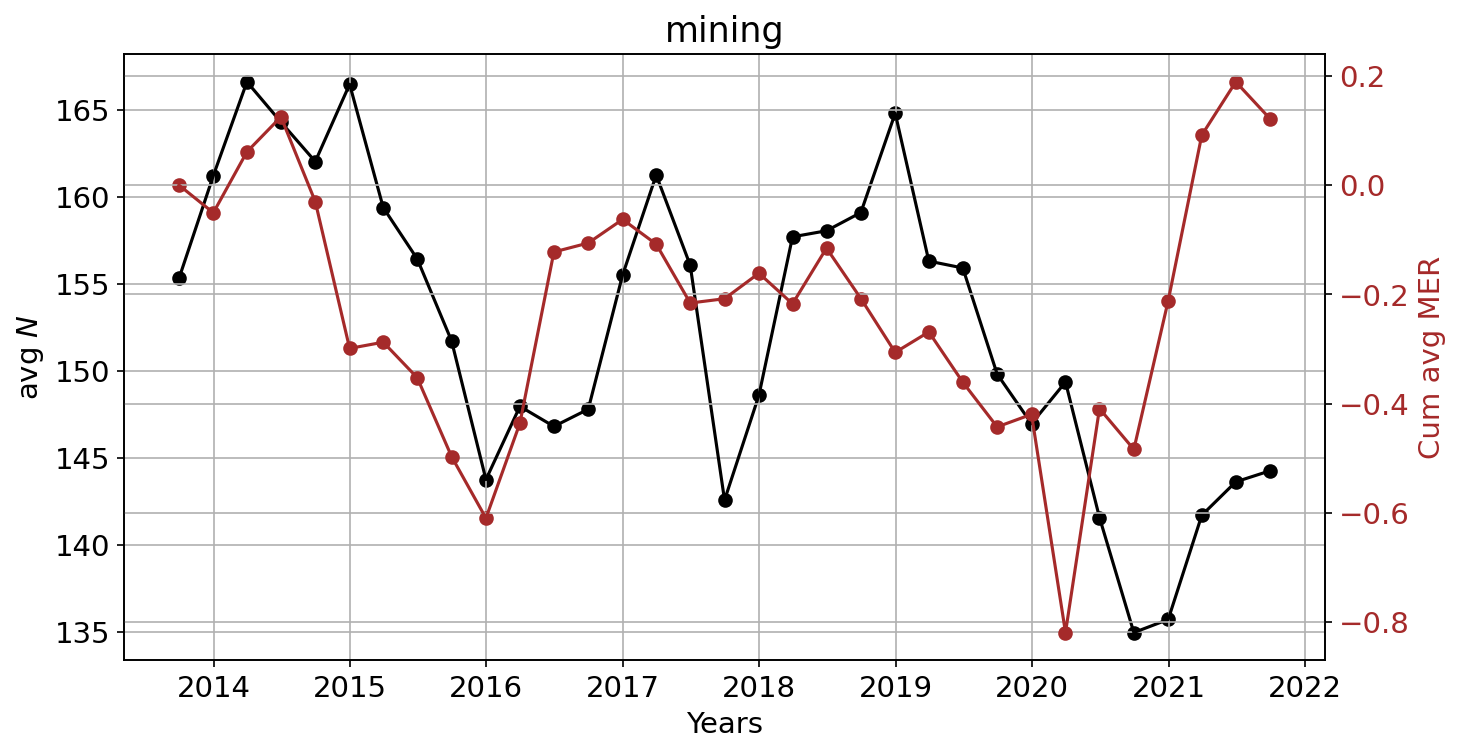}  
\end{subfigure}\newline

\begin{subfigure}{.45\textwidth}
  \centering
  \includegraphics[width=.99\linewidth]{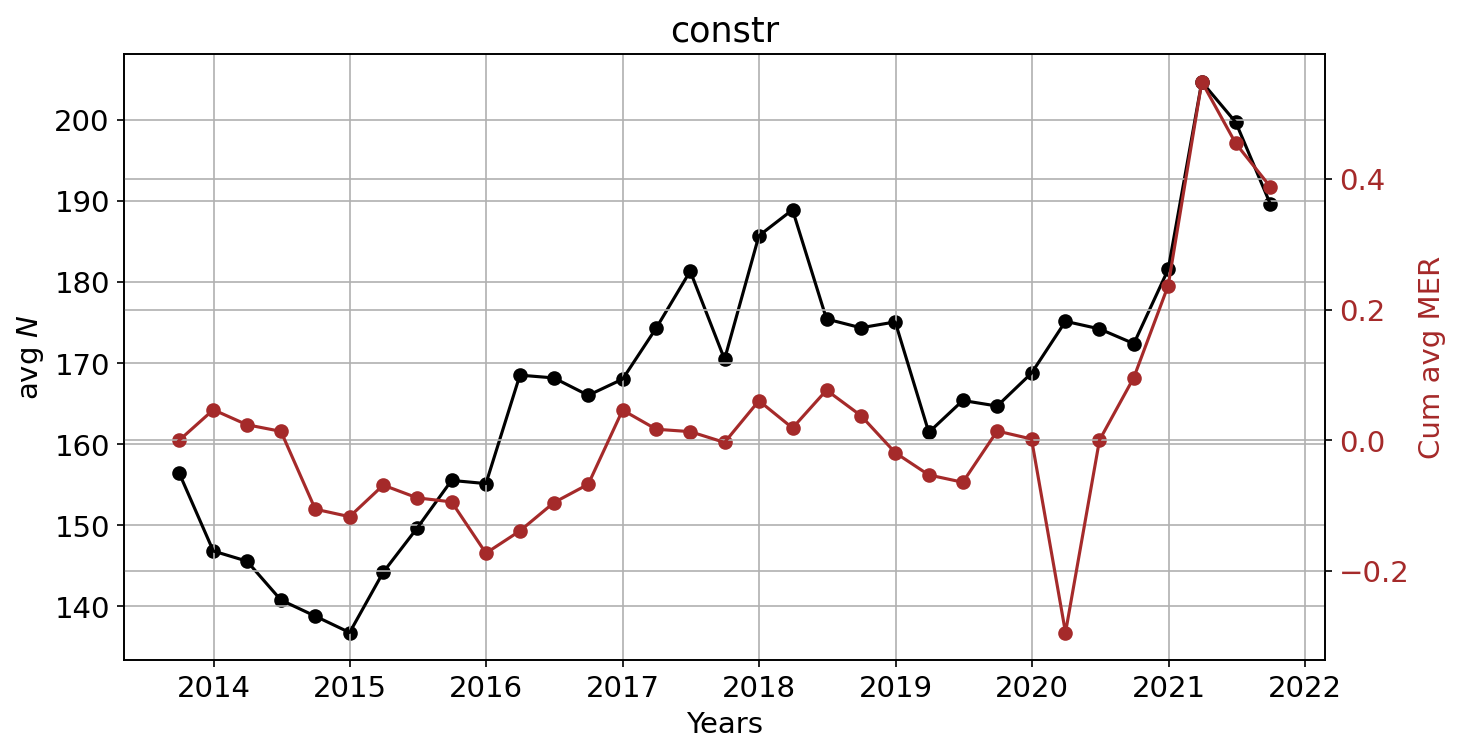}  
\end{subfigure}
\begin{subfigure}{.45\textwidth}
  \centering
  \includegraphics[width=.99\linewidth]{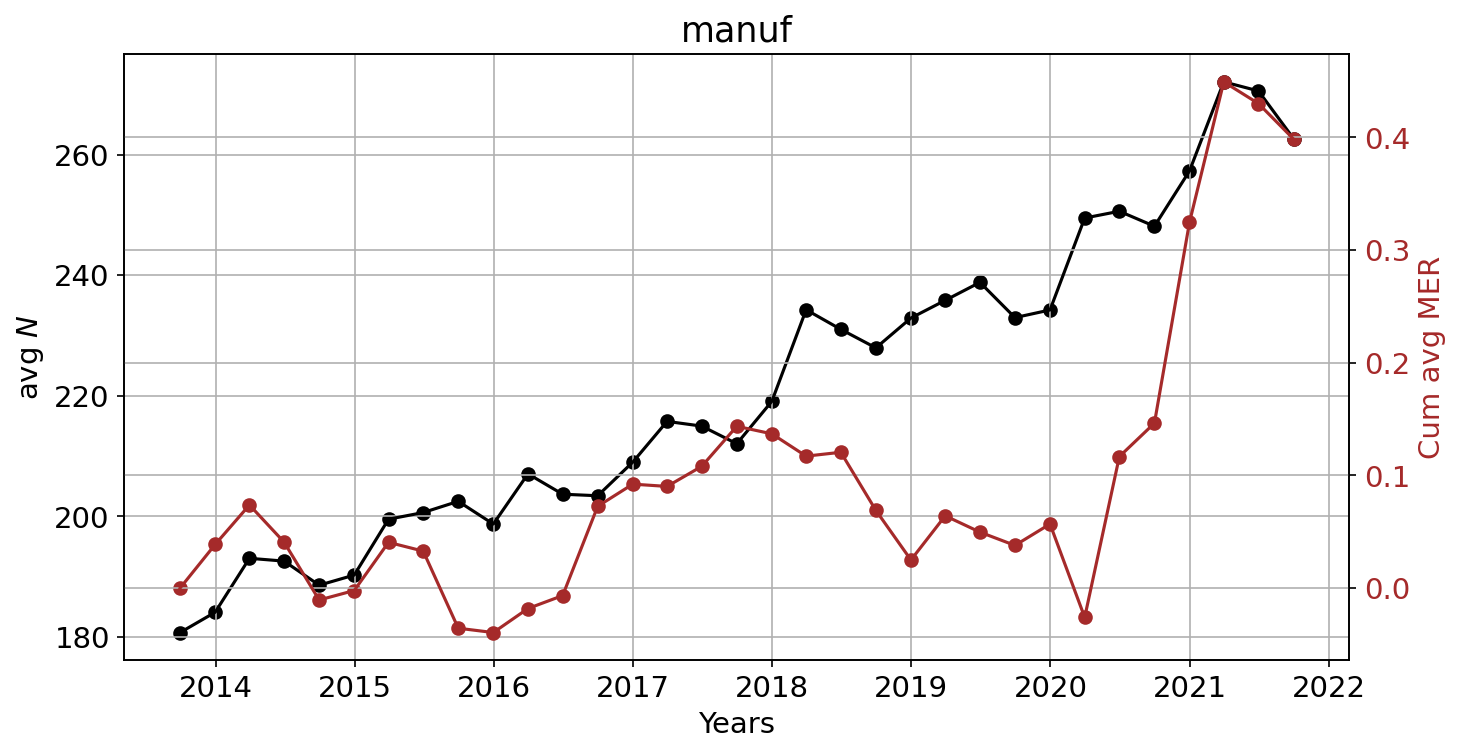}  
\end{subfigure}\newline

\begin{subfigure}{.45\textwidth}
  \centering
  \includegraphics[width=.99\linewidth]{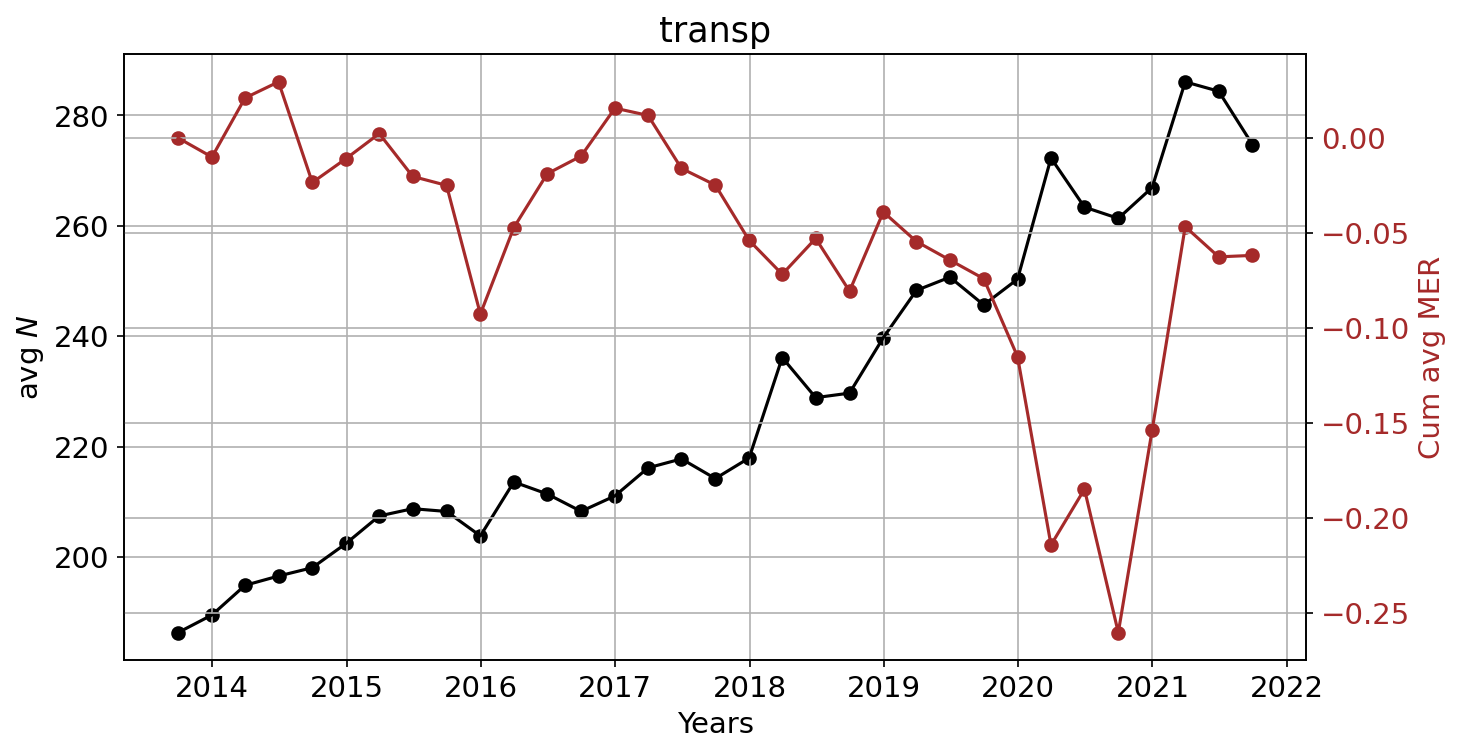}  
\end{subfigure}
\begin{subfigure}{.45\textwidth}
  \centering
  \includegraphics[width=.99\linewidth]{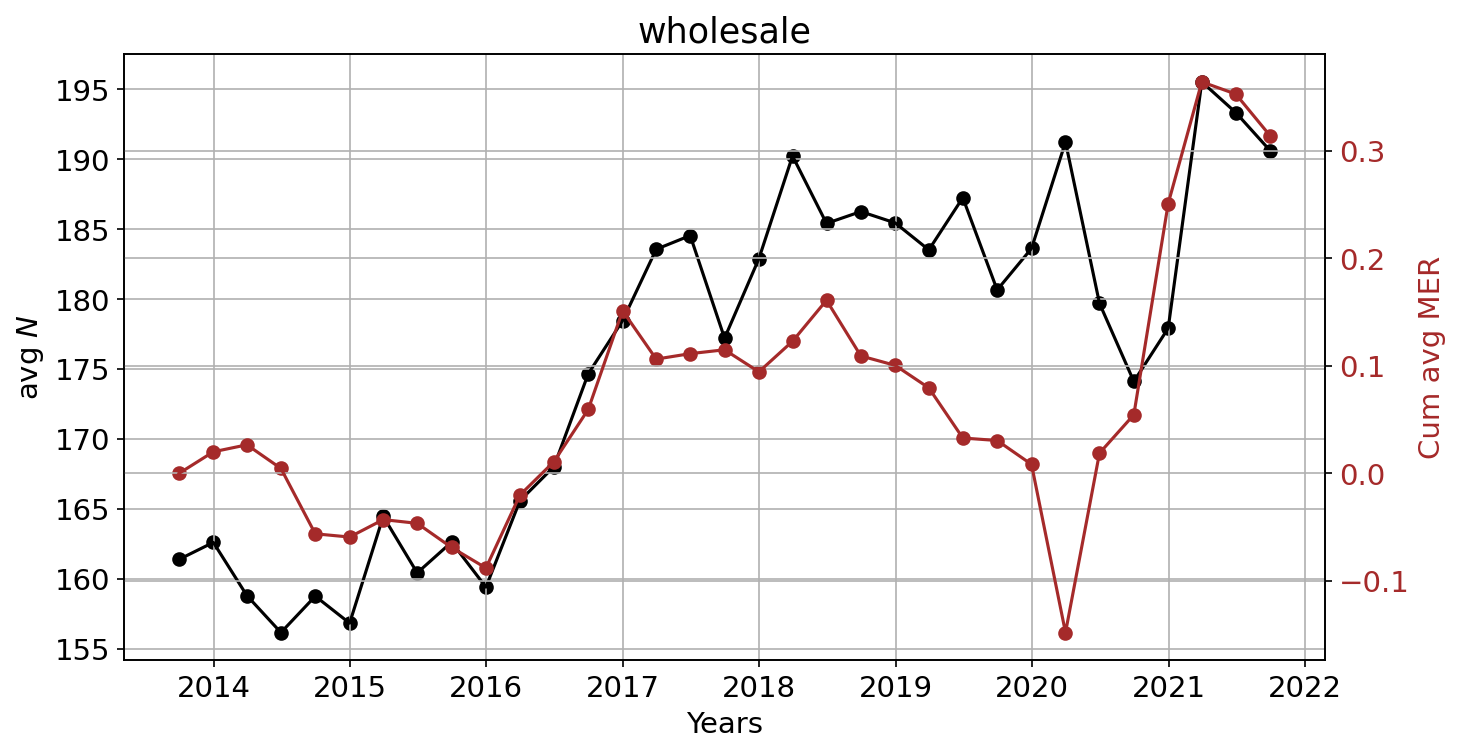}  
\end{subfigure}\newline

\begin{subfigure}{.45\textwidth}
  \centering
  \includegraphics[width=.99\linewidth]{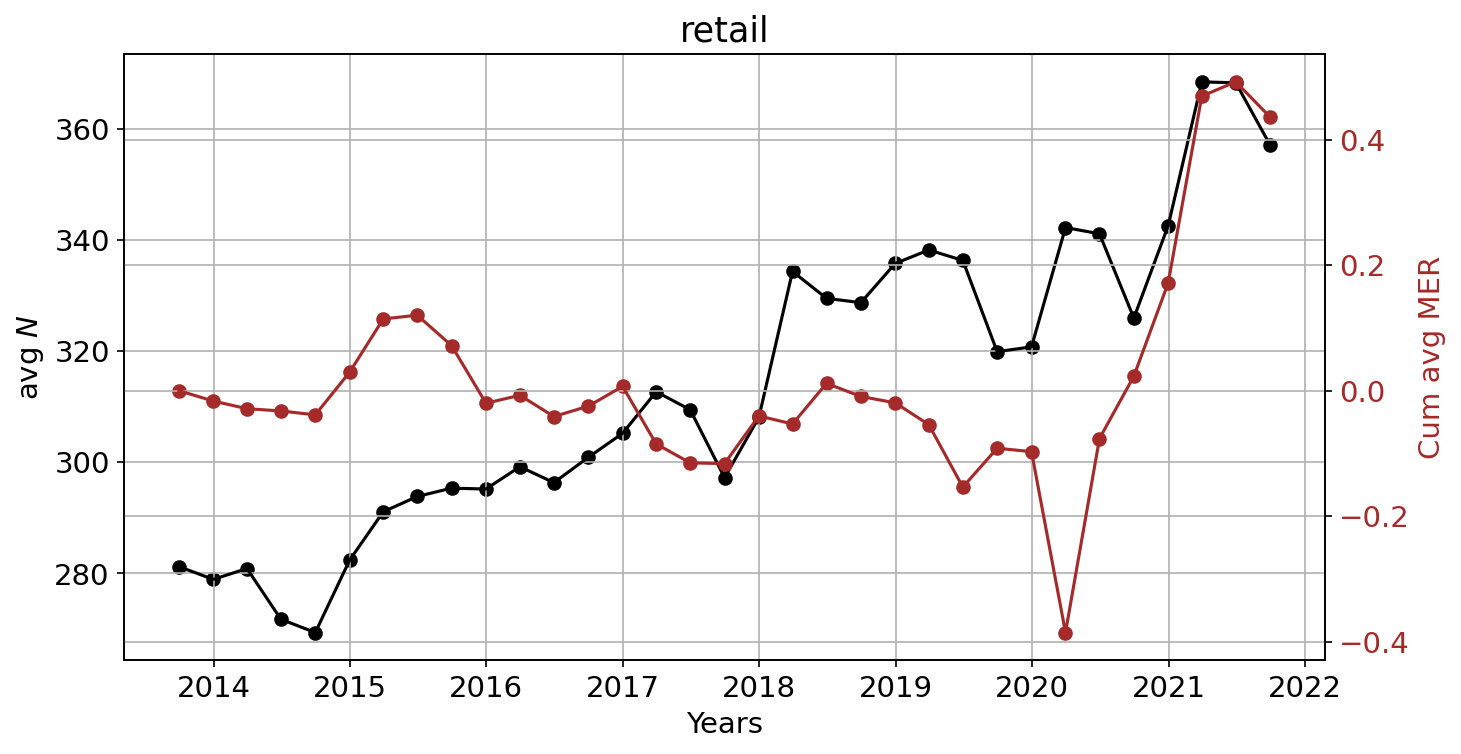}  
\end{subfigure}
\begin{subfigure}{.45\textwidth}
  \centering
  \includegraphics[width=.99\linewidth]{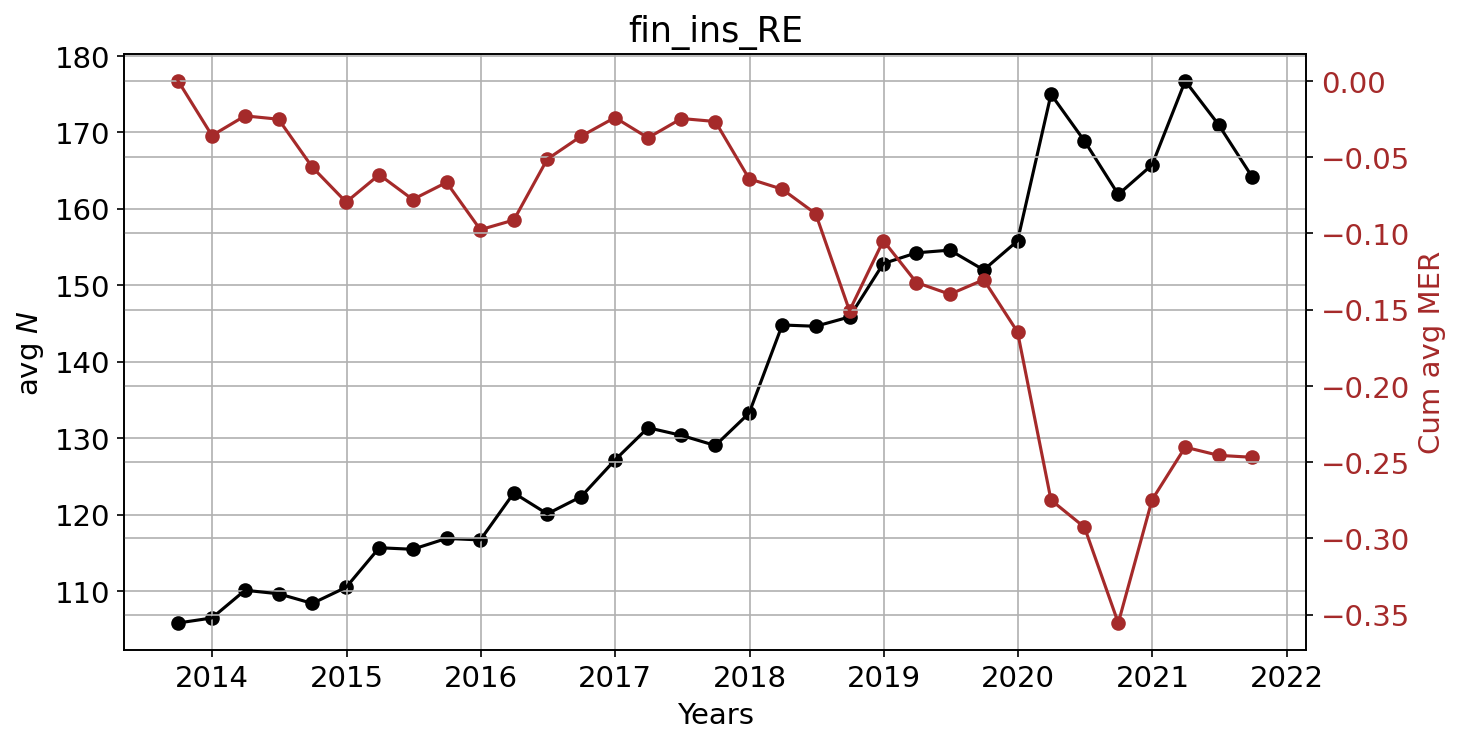}  
\end{subfigure}\newline

\begin{subfigure}{.45\textwidth}
  \centering
  \includegraphics[width=.99\linewidth]{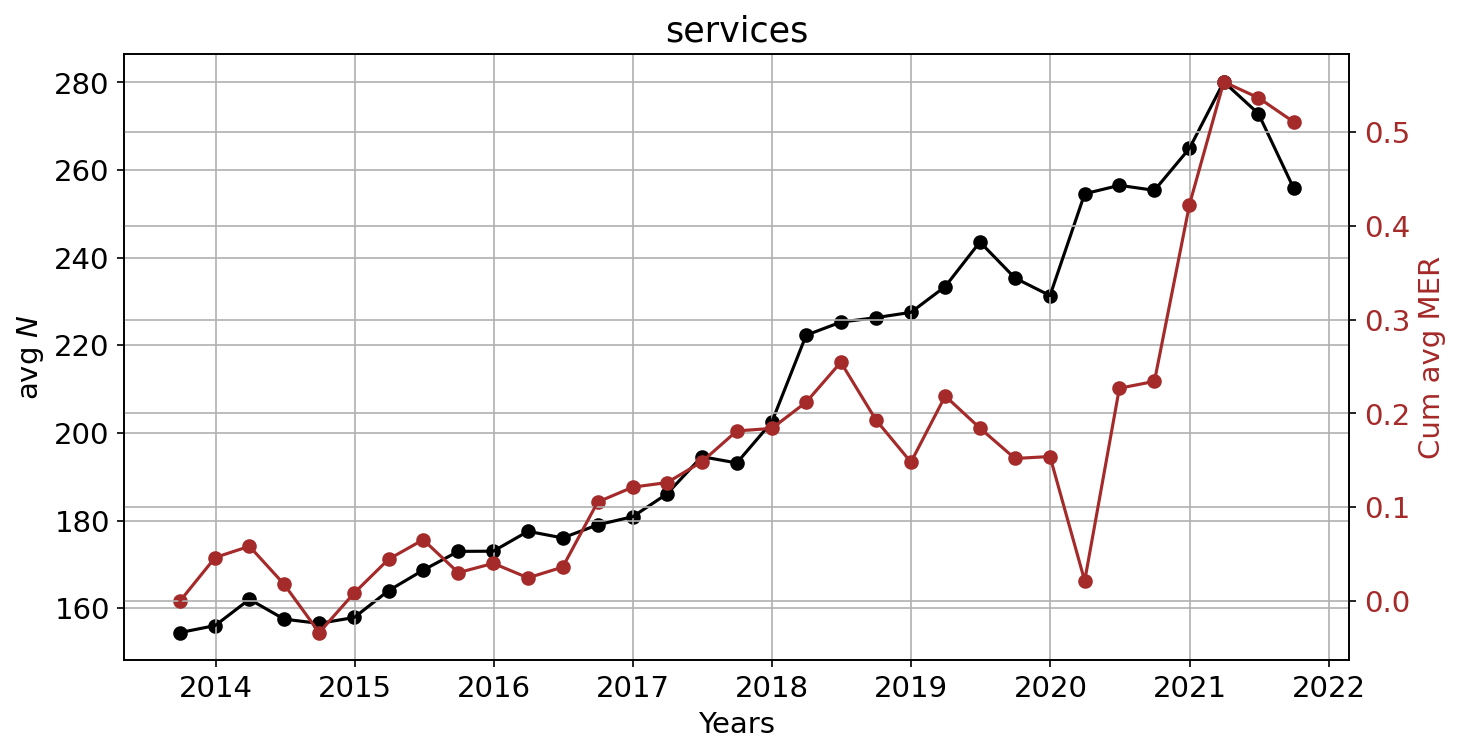}  
\end{subfigure}
\begin{subfigure}{.45\textwidth}
  \centering
  \includegraphics[width=.99\linewidth]{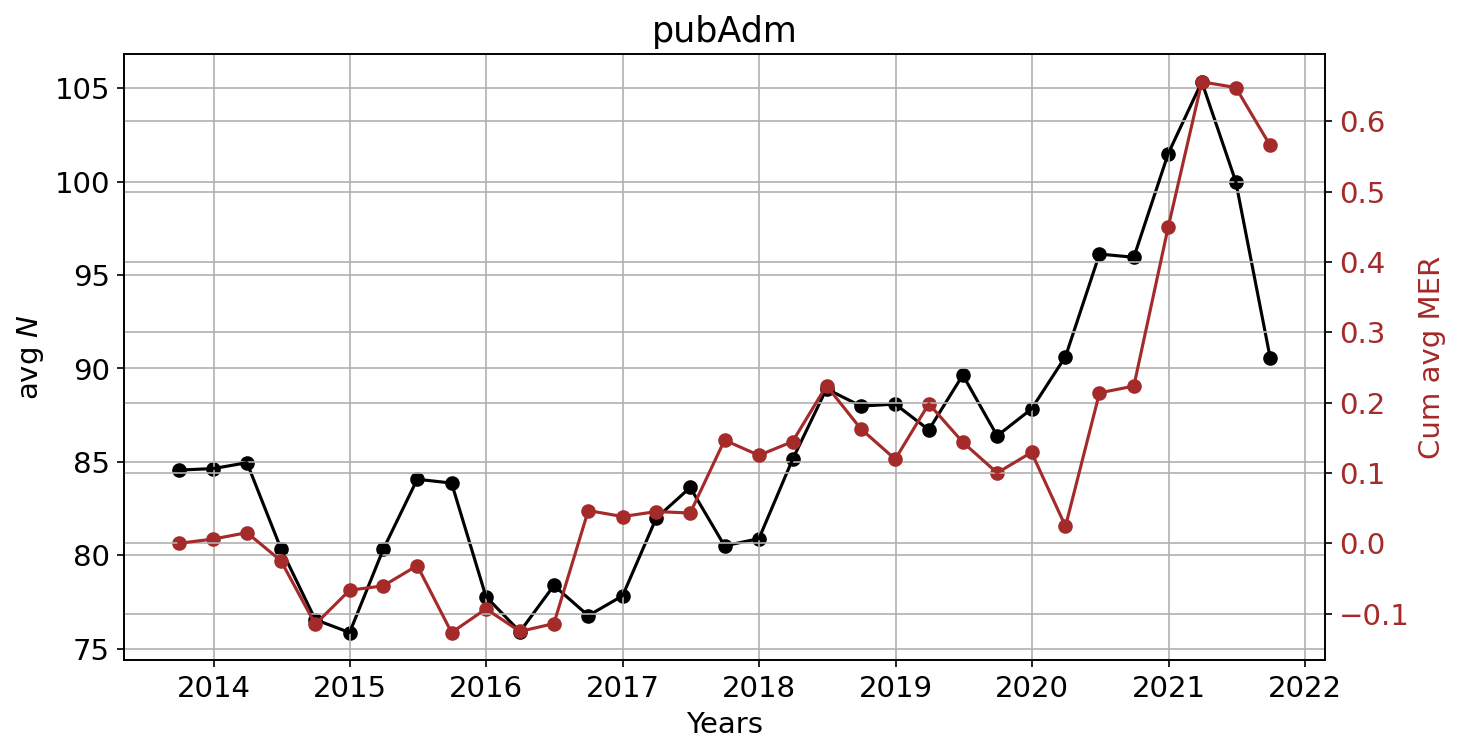}  
\end{subfigure}\newline

\caption[]{Evolution of the average popularity of stocks within sectors and related overall performance.}
\label{popularity}
\end{figure}

For each period $p$ and SIC sector, we compute the average number of active funds $N$ on the related stocks. Figure \ref{popularity} shows this evolution of \say{popularity} of sectors across time and compares it to the cumulative average MER of holding the related stocks.

\end{document}